\documentclass[%
reprint,
superscriptaddress,
 amsmath,amssymb,
aps,
pra,
]{revtex4-2}

\usepackage{graphicx}
\graphicspath{ {/} }

\usepackage{amsmath}

\usepackage{bm} 

\usepackage{dcolumn}

\usepackage{xcolor}
\usepackage[normalem]{ulem}
\usepackage{soul}

\newcommand{\replace}[2]{#2}
\newcommand{\edit}[1]{#1}

\begin{document}

\title{Distortions in Charged-Particle Images of Laser Direct-Drive Inertial Confinement Fusion Implosions}

\author{P. V. Heuer}
 \email{pheu@lle.rochester.edu}
\affiliation{Laboratory for Laser Energetics, University of Rochester, Rochester, NY 14623, USA}

\author{J. L. Peebles}
\affiliation{Laboratory for Laser Energetics, University of Rochester, Rochester, NY 14623, USA}

\author{J. Kunimune}
\affiliation{Massachusetts Institute of Technology Plasma Science and Fusion Center, Cambridge, Massachusetts 02139, USA }

\author{H. G. Rinderknecht}
\affiliation{Laboratory for Laser Energetics, University of Rochester, Rochester, NY 14623, USA}

\author{J. R. Davies}
\affiliation{Laboratory for Laser Energetics, University of Rochester, Rochester, NY 14623, USA}
\affiliation{Department of Mechanical Engineering, University of Rochester, Rochester, NY 14627, USA}

\author{V. Gopalaswamy}
\affiliation{Laboratory for Laser Energetics, University of Rochester, Rochester, NY 14623, USA}
\affiliation{Department of Mechanical Engineering, University of Rochester, Rochester, NY 14627, USA}

\author{J. Frelier}
\affiliation{Laboratory for Laser Energetics, University of Rochester, Rochester, NY 14623, USA}

\author{M. Scott}
\affiliation{Laboratory for Laser Energetics, University of Rochester, Rochester, NY 14623, USA}

\author{J. Roberts}
\affiliation{Laboratory for Laser Energetics, University of Rochester, Rochester, NY 14623, USA}

\author{R. B. Brannon}
\affiliation{Laboratory for Laser Energetics, University of Rochester, Rochester, NY 14623, USA}

\author{H. McClow}
\affiliation{Laboratory for Laser Energetics, University of Rochester, Rochester, NY 14623, USA}

\author{R. Fairbanks}
\affiliation{Laboratory for Laser Energetics, University of Rochester, Rochester, NY 14623, USA}

\author{S. P. Regan}
\affiliation{Laboratory for Laser Energetics, University of Rochester, Rochester, NY 14623, USA}
\affiliation{Department of Mechanical Engineering, University of Rochester, Rochester, NY 14627, USA}

\author{J. A. Frenje}
\affiliation{Massachusetts Institute of Technology Plasma Science and Fusion Center, Cambridge, Massachusetts 02139, USA }

\author{M. Gatu Johnson}
\affiliation{Massachusetts Institute of Technology Plasma Science and Fusion Center, Cambridge, Massachusetts 02139, USA }

\author{F. H. S\'eguin}
\affiliation{Massachusetts Institute of Technology Plasma Science and Fusion Center, Cambridge, Massachusetts 02139, USA }

\author{A. J. Crilly}
\affiliation{Centre for Inertial Fusion Studies, The Blackett Laboratory, Imperial College, London SW7 2AZ, United Kingdom}

\author{B. D. Appelbe}
\affiliation{Centre for Inertial Fusion Studies, The Blackett Laboratory, Imperial College, London SW7 2AZ, United Kingdom}

\author{M. Farrell}
\affiliation{General Atomics, San Diego, CA 92121, USA}

\author{J. Stutz}
\affiliation{General Atomics, San Diego, CA 92121, USA}

\date{\today}
\begin{abstract}
Energetic charged particles generated by inertial confinement fusion (ICF) implosions encode information about the spatial morphology of the hot-spot and dense fuel during the time of peak fusion reactions. The knock-on deuteron imager (KoDI) was developed at the Omega Laser Facility to image these particles in order to diagnose low-mode asymmetries in the hot-spot and dense fuel layer of cryogenic deuterium--tritium ICF implosions. However, the images collected are distorted in several ways that prevent reconstruction of the deuteron source. In this paper we describe these distortions and a series of attempts to mitigate or compensate for them. We present several potential mechanisms for the distortions, including a new model for scattering of charged particles in filamentary electric or magnetic fields surrounding the implosion. \edit{Particle-tracing is used} to create synthetic KoDI data based on the filamentary field model that reproduces the main experimentally observed image distortions. \edit{We conclude that the filamentary scattering model best matches the observed image distortions. Finally, we discuss potential impacts of filamentary fields on other charged-particle diagnostics.}

\end{abstract}

\maketitle

\section{Introduction}

Fusion reactions in inertial confinement fusion (ICF) implosions generate charged particles that encode information about the implosion. Some charged particles are direct products of primary or secondary fusion reactions, while other ``knock-on" particles are produced by collisions between fusion neutrons and ions in the surrounding plasma. The spectra of these particles have previously been used to diagnose the yield, areal density, ion temperature, and other implosion variables~\cite{Seguin2003spectrometry}. The spatial distribution of the source of both fusion products and knock-on particles encodes information about the hot-spot and the dense fuel surrounding it. Imaging these particles from multiple lines of sight would allow 3D reconstruction of the burn volume and the surrounding dense fuel, allowing the identification of low-mode asymmetries that reduce implosion performance~\cite{Regan2016demonstration, Igumenshchev2017three, GatuJohnson2019impact, Colaitis20223d}. 

\begin{figure}[htb]
\includegraphics[width=\columnwidth]{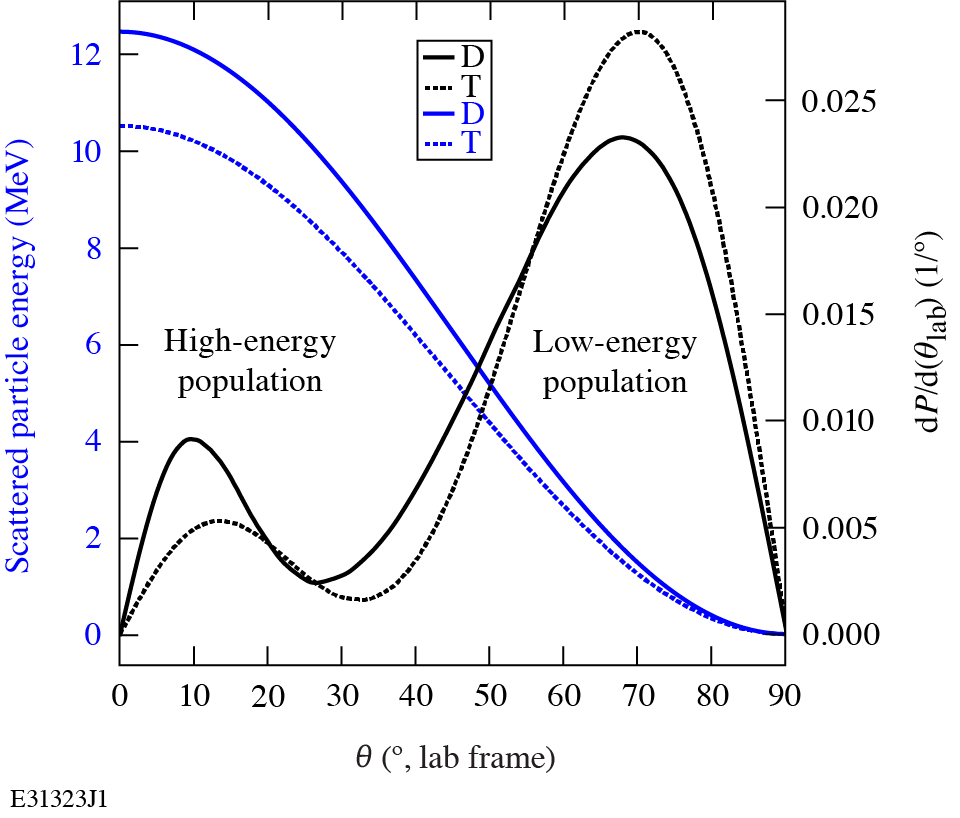}
\caption{\label{fig:scatter_prob}
DT knock-on deuteron (solid) and knock-on triton (dashed) scattering probability (black) and scattered particle energy (blue) as a function of scattering angle.}
\end{figure}

\begin{figure}[htb]
\includegraphics[width=\columnwidth]{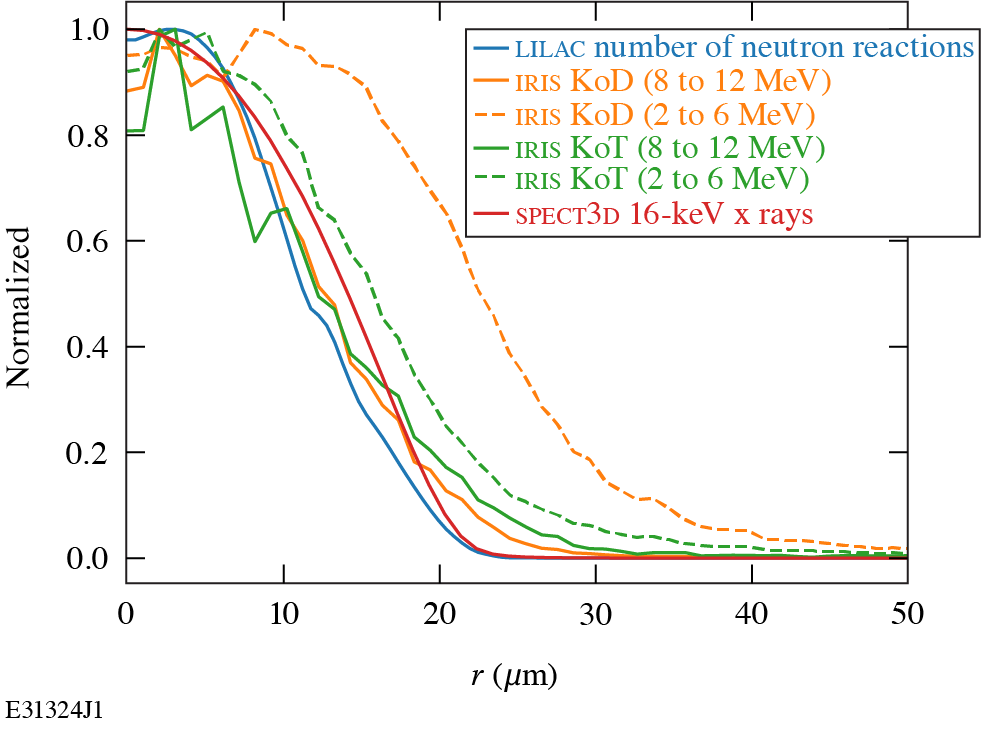}
\caption{\label{fig:1d_simulated_profiles}
A comparison of synthetic neutron, KoD, KoT, and time-integrated x-ray images generated by IRIS and SPECT3D, respectively using output from the 1D hydrodynamic code LILAC for OMEGA shot 109675. The hot-spot radius measured by high-energy KoD/KoT or x-ray imaging diagnostics is a good approximation of the time-integrated neutron-production profile. The x-ray image energy band that best matches the neutron hot-spot radius must be chosen for a given implosion based on simulations and synthetic diagnostics. }
\end{figure}

\begin{figure*}[tbh]
\includegraphics[width=\textwidth]{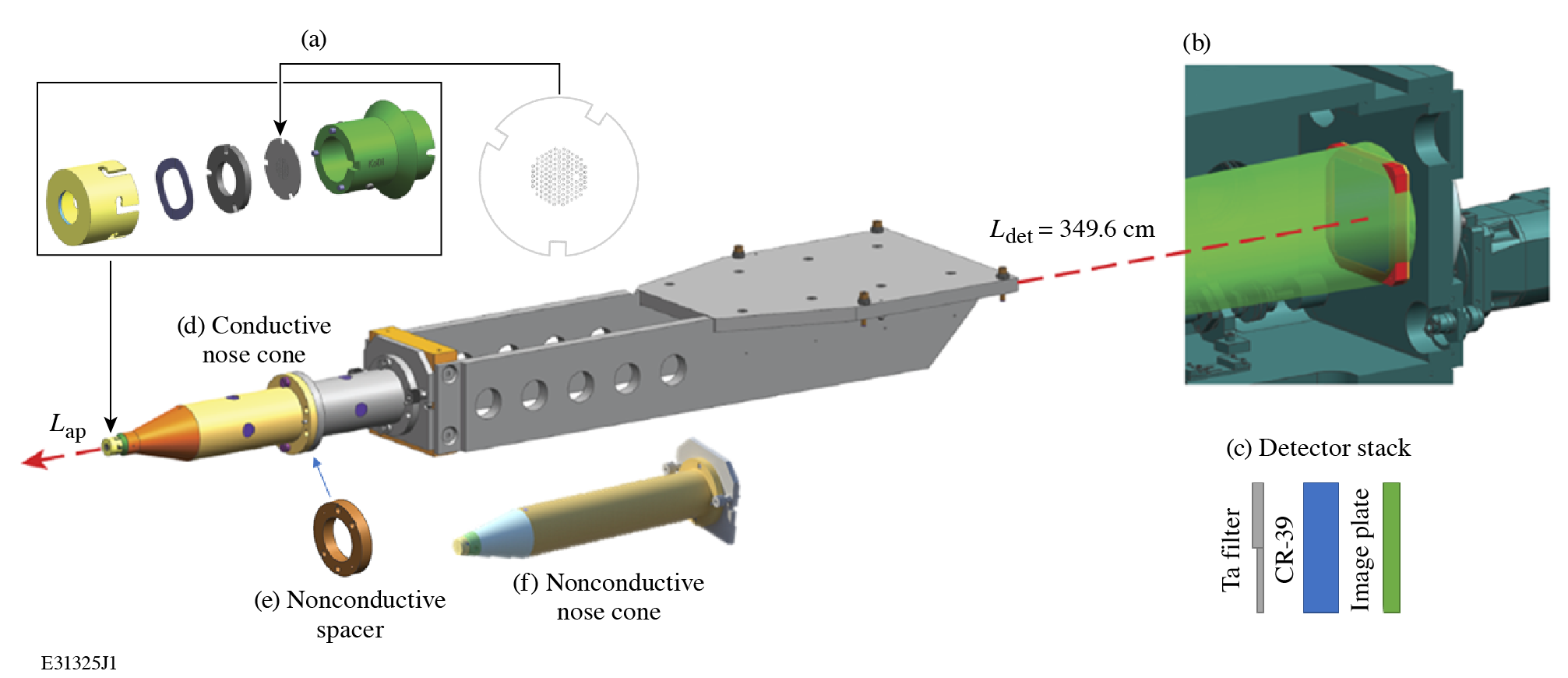}
\caption{\label{fig:hardware}
Diagram of the original KoDI hardware showing (a) the pinhole array, (b) the detector position, and (c) the detector stack. The original (d) conductive nose cone and alternative hardware including (e) the nonconductive spacer and (f) non-conductive nose cone are also shown.}
\end{figure*}

Imaging of charged-particle fusion products ($14.7$-MeV protons from D--$^3$He reactions) was initially performed at the Omega Laser Facility with the proton core imaging system (PCIS), which was used to diagnose the radius of the burn volume~\cite{Seguin2004d3he, DeCiantis2006proton}. This technique could not be extended to the deuterium--tritium (DT) fueled cryogenic implosions of greatest interest for ICF, which have sufficiently high areal densities to stop the lower-energy charged particles produced by the primary DT reactions. 

The knock-on deuteron imager (KoDI)~\cite{Rinderknecht2022knock, Kunimune2022knockon} was developed to image knock-on deuterons (KoDs) and knock-on tritons (KoTs) \edit{as well as x-rays} produced by DT cryogenic implosions on the OMEGA laser. These KoDs are divided into distinct high (9 to 12 MeV) and low (3 to 6 MeV) energy populations (Fig.~\ref{fig:scatter_prob}) resulting from small (${\sim}10^\circ$) and large (${\sim} 70^\circ$) angle scattering, respectively~\cite{Rinderknecht2022knock}. High-energy deuterons detected along a particular line-of-sight (LOS) resemble a direct neutron image along that LOS. Low-energy deuterons, which must be produced by neutrons initially moving quasi-perpendicular to the detector LOS, appear to come from the dense fuel where they scatter and therefore encode information about the fuel density profile and symmetry. \edit{The KoDI KoD/KoT, and x-ray measurements are time-integrated, but are effectively time-gated by the sharp ${\sim} 80$~ps duration peak in x-ray, neutron, and subsequently KoD/KoT production around the neutron bang time.}

Synthetic KoD and KoT images generated from 1D hydrodynamic simulations in LILAC~\cite{Delettrez1987effect} using the post-processing codes IRIS~\cite{Gopalaswamy2022analysis} show that the high-energy KoD image provides a good approximation of the profile of the neutron producing region and that the images produced by the high energy KoD and KoT have similar profiles (Fig.~\ref{fig:1d_simulated_profiles}). Synthetic x-ray images generated from the LILAC output using SPECT3D~\cite{MacFarlane2007spect3d} show that that a time-integrated x-ray image provides a similar profile to the high energy deuteron image, although the exact energy band of x-ray emission that corresponds to the hot-spot depends on the density and temperature profile of the implosion.

Mechanically, KoDI consists of an array of penumbral apertures (typical $R_\text{ap} = 150$ $\mu$m) positioned a distance $L_\text{ap}$ (3 to 20 cm) from the implosion at the target chamber center (TCC) (Fig.~\ref{fig:hardware}). The aperture array is $200$-$\mu$m-thick tantalum, and each aperture has an ${\sim}5^\circ$ cone angle positioned with the smaller side facing toward the implosion. A detector pack is positioned $L_\text{det}$  (${\sim}3.5$ m) from the aperture, and consists of a 10-cm $\times$ 10-cm piece of CR-39~\cite{Cartwright1978nuclear,Lahmann2020cr39} (for detecting deuterons) followed by an image plate (which detects x-rays that pass through the CR-39). Two regions of tantalum filtration are positioned in front of the detector media so the two halves of the CR-39 detect either high-energy or low-energy deuterons and tritons. Discriminating between KoD and KoT in the charged-particle images is an unsolved problem. For simplicity, since the KoD and KoT produce similar images (Fig.~\ref{fig:1d_simulated_profiles}), in this paper we will assume that all charged particles collected are deuterons. 

During the KoDI commissioning experiments, several distortions of the penumbral deuteron images were observed, as described in Sec.~\ref{sec:image_distortions}. Several possible explanations for the distortions are explored in Sec.~\ref{sec:explanations}. The initial hypothesis~\cite{Rinderknecht2022knock, Kunimune2022knockon} that electric fields produced by charging of the pinhole apertures were responsible provides a reasonable fit to the penumbral images, but is inconsistent with later experiments (Sec.~\ref{sec:aperture_charging}). Here, we propose an alternative model based on the scattering of knock-on charged particles in strong filamentary electric or magnetic fields around the imploded target (Sec.~\ref{sec:filament_scattering}). A particle tracer to generate synthetic proton radiographs and synthetic KoDI data including these filamentary fields is described in Sec.~\ref{sec:particle_pusher}. The particle trajectories through these fields and the synthetic KoD images are analyzed in Sec.~\ref{sec:synthetic_kodi_analysis}. Finally, we present conclusions based on this filamentary field model for KoDI and other charged-particle diagnostics in Sec.~\ref{sec:conclusion}. 

\edit{This paper presents data from around forty cryogenic DT directly driven implosions~\cite{Craxton2015direct} (some fielding multiple KoDI) conducted on the OMEGA Laser Facility in 2021-2023. In each experiment, a ${\sim}$800~$\mu$m OD spherical CD shell with an inner cryogenic DT ice layer and gas fill was imploded with a shaped laser pulse of ${\sim}$28~kJ, generating neutron yields of ${\sim} 10^{14}$ and hotspot temperatures of several keV. While exact target and laser parameters varied~\cite{Gopalaswamy2024demonstration, Williams2024demonstration} all of these implosions achieved sufficient neutron yields and compressed fuel areal densities to produce directly comparable KoDI measurements.} A discussion of how our results relate to previous work with warm DT implosions and direct imaging of charged-particle fusion products is included in Sec.~\ref{sec:conclusion}.

\begin{figure*}[!tbh]
\includegraphics[width=\textwidth]{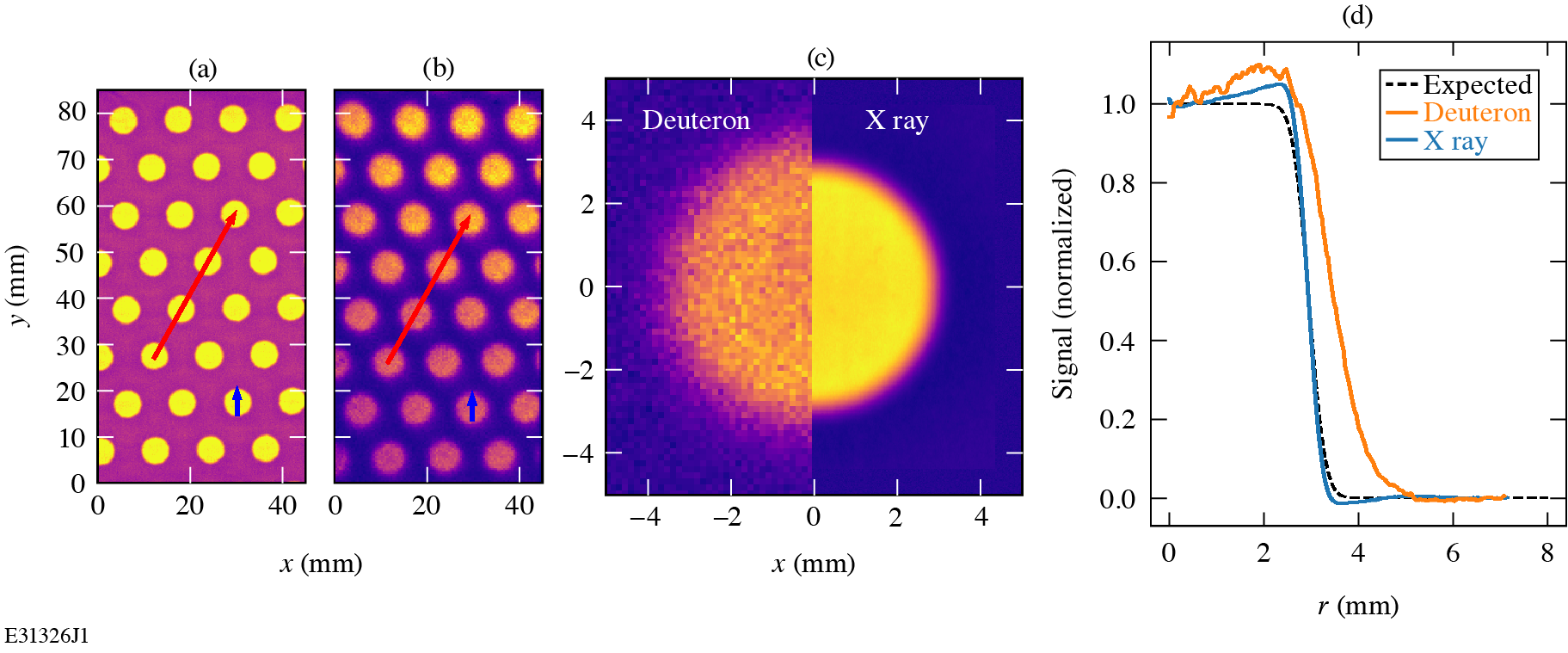}
\caption{\label{calc_mag}
(a) Image plate (x-ray) and (b) CR-39 (deuteron) data from a cryogenic DT implosion. Given the dimensions of the aperture array, the system magnification can be calculated from either dataset using either the diameter of each penumbral image (blue arrows) or the separation between any two pinhole images (red arrows). (c) When the individual penumbral images are averaged, the deuteron image is ${\sim}$20\% larger. (d) Radial profiles produced by averaging over the polar angle show the same discrepancy.}
\end{figure*}

\section{Image Distortions\label{sec:image_distortions}}

Two primary distortions are observed in the KoD images collected by KoDI. First, the penumbral images are larger than expected, which we describe as ``anomalous magnification'' (Sec.~\ref{sec:anomalous_magnification}). Second, low-energy KoD penumbral images are not round, but are ``smeared'' across the detector, making reconstruction of the penumbral images impossible (Sec.~\ref{sec:smearing}).  Nonuniformities in the umbra of the penumbral images are also observed (Sec.~\ref{sec:umbra_nonuniformity}). Since the reconstruction of penumbral images is extremely sensitive to changes in the effective instrument \edit{response} function, the presence of these distortions makes an accurate reconstruction of the deuteron source impossible in most cases.

\subsection{Anomalous Magnification\label{sec:anomalous_magnification}}

The 50\% radius of a penumbral image is theoretically independent of the source profile, and is set only by the hardware magnification and the aperture  radius. In experiments on the OMEGA Laser System, the hardware magnification $M=1+L_\text{det}/L_\text{ap}$ is accurate to $\leq 1$\%. Some additional variability in the aperture radius (${\sim}1$\%) also contributes to uncertainty in the expected size of the penumbral images. However, on KoDI, the simultaneous collection of x-ray images through the same apertures as the deuteron images allows for a precise calculation of the effective hardware magnification independent of the true aperture radius and $L_\text{ap}$ (the distance between the CR-39 and image plate, $< 0.1$\% of $L_\text{det}$, is negligible). Furthermore, since an array of pinhole apertures is used, the effective magnification of the system can be determined by using either the radius of the pinhole images or their separation. As shown in Figs.~\ref{calc_mag}(a) and ~\ref{calc_mag}(b), this process can be applied independently to both the deuteron and x-ray data.

Since the individual penumbral images have a low signal-to-noise ratio, the apparent magnification based on the radius of each image $M_\text{r}$ is determined by fitting the data with a 2D axisymmetric penumbral model function
\begin{equation}
I(r) = \frac{A}{2} \bigg  ( 1 - \text{erf}\bigg [ \frac{r - M R_\text{ap}}{\sqrt 2 \sigma(M-1) }
\bigg ] \bigg ) + B \text{,}
\label{eq:penumbral_model}
\end{equation}
where $r$ is the radius in the image plane and $\text{erf}$ is the error function. This model function gives the analytical profile of a penumbral image of a Gaussian object with standard deviation $\sigma$ through an aperture of radius $R_\text{ap}$ (assuming $R_\text{ap}\gg \sigma$) with radiographic magnification $M=1+ L_\text{det}/L_\text{ap}$. $A$ and $B$ are fitting constants representing the total signal and background level, respectively. This model function is fit to each penumbral image with a differential evolution algorithm~\cite{Storn1997differential, 2020SciPy-NMeth}. A Markov Chain Monte-Carlo (MCMC) sampler from the Python package \textit{emcee}~\cite{emcee2012} applied to a subset of images showed that the error in this fit is small relative to the variance of the fits between different penumbral images on a single shot, so the standard deviation of fit results is used as the error in the mean value of the fit. The apparent magnification based on the separation of penumbral images, $M_\text{sep}$, is determined by first finding the centers of each image, fitting a model of the known aperture array geometry with the differential evolution algorithm, and then again using the \textit{emcee} MCMC sampler to estimate the error in the fit. When comparing $M_\text{r}$ to $M_\text{sep}$, an additional 2\% error in the aperture radius and separation is included to account for the physical tolerances of the aperture arrays (resulting in the larger errors in brackets in Table.~\ref{table:mag}). 

\renewcommand{\arraystretch}{1.5}

\begin{table}[!tbh]
\begin{tabular}{c|c|c|}
\cline{2-3}
                                     & \textbf{M$_\text{sep}$} & \textbf{M$_r$}   \\ \hline 
\multicolumn{1}{|c|}{\textbf{X-ray}} & $19.72 \pm 0.04 [0.40]$   & $19.03 \pm 0.17 [0.42]$ \\ \hline
\multicolumn{1}{|c|}{\textbf{KoD}}   & $19.77 \pm 0.04 [0.40]$   & $22.75 \pm 1.11 [1.20]$ \\ \hline
\end{tabular}
\caption{\label{table:mag} Apparent magnification determined from the x-ray and knock-on deuteron data shown in Fig.~\ref{calc_mag} using the radius of the penumbral images and the separation between penumbral images. The fitting error (appropriate for comparing $M_\text{r,x-ray}$ with $M_\text{r,KoD}$ or $M_\text{sep,x-ray}$ with $M_\text{sep,KoD}$) is reported, followed by the fitting error combined with the 2\% error in the true aperture radius and separation in brackets (appropriate for comparing $M_\text{r}$ with $M_\text{sep}$).\replace{The separation magnifications agree within error. $M_\text{r,x-ray}$ disagrees slightly, which may be due to error in the aperture size. $M_\text{r,x-ray}$ and $M_\text{r,KoD}$ disagree significantly, indicating anomalous magnification of the deuteron images.}{}}
\end{table}

Table~\ref{table:mag} shows the apparent magnification calculated from the x-ray and knock-on deuteron data shown in Fig.~\ref{calc_mag} using  both techniques. The image separation technique is highly accurate, with both x-ray and KoD data establishing a true magnification of $19.74\times$, very close to the nominal hardware magnification of $19.75\times$. The magnification  estimated from the x-ray penumbra radius is slightly smaller, possibly showing that the apertures are slightly smaller than nominal, and has only a slightly higher error, which indicates that the image is well fit by Eq.~(\ref{eq:penumbral_model}). The apparent magnification from the KoD penumbra radius is significantly larger, and falls far outside of the error bars from $M_\text{r,x-ray}$. This result implies that some anomalous magnification of the KoD images is present that does not affect the x-ray images or the spacing of the KoD images. This effect is visually apparent in Fig.~\ref{calc_mag}(c-d), in which the deuteron image is clearly larger than the x-ray image. 

Throughout the remainder of this paper, the anomalous magnification ratio $m_\text{r} = M_\text{r}/M_\text{true} \approx M_\text{r} / M_\text{sep}$ is used as measure of the magnitude of this effect, where a value of $1$ represents no anomalous magnification. An anomalous magnification of more than a few percent creates a sufficient difference between the data and the nominal instrument point-spread function to preclude reconstruction of the source from the penumbral images.

\subsection{Smearing\label{sec:smearing}}

\begin{figure*}[tbh]
\includegraphics[width=\textwidth]{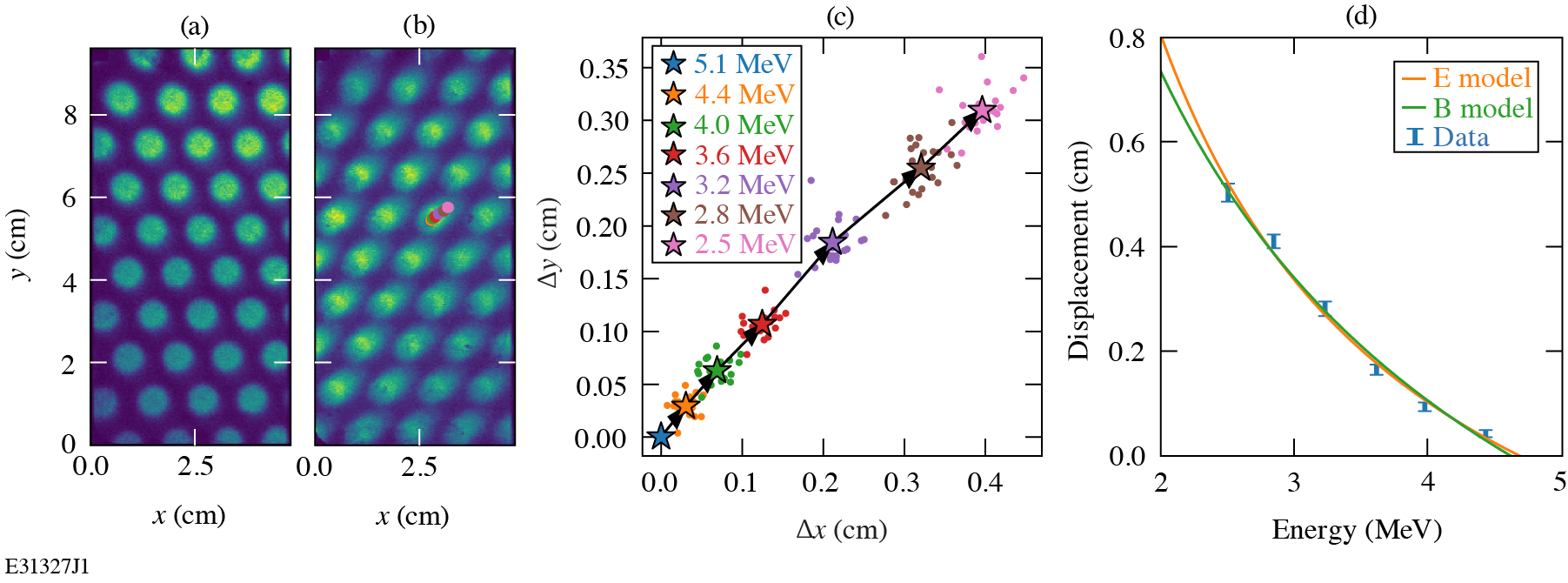}
\caption{\label{fig:smearing}
Low-energy KoD images from different shots exhibiting (a) minimal or (b) significant smearing. The points in (b) mark the center of one of the penumbral images in successive energy ranges. (c) The displacement of the center of each penumbral image (points) from the center of the highest energy bin image (stars)\replace{shows the drift clearly}{}. (d) The displacement as a function of deuteron energy is well fit by models for deflection by electric and magnetic fields.}
\end{figure*}

Penumbral images produced by a round aperture with $R_\text{ap} \gg R_\text{obj}$ are round even for sources that are far from round. However, the low-energy deuteron images collected by KoDI on many shots appear to be stretched along one axis, forming an elliptical and blurry image [Fig.~\ref{fig:smearing}(a) versus Fig.~\ref{fig:smearing}(b)]. We refer to this distortion of the image as ``smearing.''

A better understanding of this phenomenon is afforded by separating the low-energy deuteron image (${\sim}$ 2 to 6 MeV) into images over smaller ${\sim}0.5$-MeV energy ranges. This is possible because tracks left by charged particles in the CR-39 detector can be related to the energy of the particle~\cite{Lahmann2020cr39}. In the resulting images the penumbral images are circular, but the center of each image drifts as a function of the deuteron energy [Fig.~\ref{fig:smearing}(b), points]. We conclude that the ``smearing'' distortion is actually a result of a large-scale deflection of the deuteron images as a function of their energy, forming a smeared image when summed over a wider range of energies. Unfortunately, the signal level in the KoD images is too low to analyze without summing over a larger energy range, so taking narrow energy slices of the data does not allow for reconstruction of the smeared penumbral images. 

Figure~\ref{fig:smearing}(c) shows the displacement of the image centers in each energy bin relative to the image centers in the highest energy bin (which we assume to be the least perturbed). Within a single KoD image, the direction and magnitude of the deflection is approximately consistent across the array of penumbral images and the amount of deflection increases as the energy decreases. Between different diagnostic lines of sight on a single shot, or between different shots, the direction and magnitude of the smearing has no discernible pattern.

Electric and/or magnetic fields are a likely explanation for the deflection of a beam of charged particles. The deflection $\delta$ of a particle of charge $q$, mass $m$, and kinetic energy $W$ on a detector a distance $L$ away from a region of electric field $E$ or magnetic field $B$ is 
\begin{equation}
\delta / L = \tan \theta = \frac{q}{2W} \int E \text{d}l + \frac{q}{\sqrt{2mW}} \int B \text{d}l,
\end{equation}
where the integral is over the path of the particle through the field and $\theta$ is the deflection angle. Taking cases with $B=0, E\neq0$ and $E=0, B\neq0$, this provides two models for the deflection as different functions of energy, $\delta \propto 1/W$ and $\delta \propto 1/\sqrt{W}$, respectively, which in theory allow the two terms to be distinguished. Figure~\ref{fig:smearing}(d) shows the mean displacement of the image centers with error bars calculated as the standard deviation of the displacement across the population of penumbral images in the array. The data are fit almost to within error equally well by both the electric and magnetic field deflection models. A larger range of particle energies would be required to distinguish between the two models. We therefore conclude that the observed smearing is consistent with either an electric- or magnetic-field, but that we cannot distinguish between the two on this basis. 

The fits to the displacement curve in Fig.~\ref{fig:smearing}(d) do allow an approximate calculation of the range of path-integrated electric or magnetic field necessary to explain the results. The lowest field consistent with the fields is obtained if the fields responsible are at the aperture, where $L=L_\text{det}$, while a generous upper bound is given by the reasonable assumption that $\theta < 90^\circ$. These assumptions lead to ranges of
$15 $~kV $\leq \int E \text{d}l \leq 5500$~kV or $1.7$~T~mm $\leq \int B {d}l \leq 620$~T~mm for electric and magnetic fields respectively, assuming the deflection is due exclusively to one or the other. 

\subsection{Umbra Nonuniformity\label{sec:umbra_nonuniformity}} 
The umbra (central region) of an ideal penumbral image is uniform. KoDI data, however, frequently exhibit both radially symmetric and asymmetric features in the umbra. Early KoDI data collected exhibit a peak near the center of the umbra~\cite{Rinderknecht_thesis, Kunimune2022knockon} attributed to aperture charging, while more recent data [Figs.~\ref{calc_mag}(c) and \ref{calc_mag}(d)] are brighter near the edges of the umbra (``limb brightening"). A distinction of potential importance between these cases is that the former images were produced by warm DT implosions with large ($2000$-$\mu$m-OD) apertures close to the implosion ($4.2$ cm), while the latter are from cryogenic DT implosions with small apertures ($300$-$\mu$m OD) farther from the implosion ($17$ cm). Radially asymmetric variations in brightness across the umbra also occur. In penumbral imaging the umbra contains no information. Distortions at the edge of the umbra, however, can also affect the top edge of the penumbra, and therefore affect the reconstructed penumbral images.

\section{Possible Explanations\label{sec:explanations}}
Several possible explanations for these image distortions have been previously explored. The most common, electrical charging of the aperture array, is discussed in Sec.~\ref{sec:aperture_charging}, while scattering of KoD in the aperture is modeled in Sec.~\ref{sec:aperture_scattering}. \edit{Ultimately, we show that none of these models are satisfactory.} In Sec.~\ref{sec:filament_scattering} we propose a new explanation that we believe better fits the accumulated data. 

\subsection{Aperture Charging\label{sec:aperture_charging}}

\begin{figure}[tbh]
\includegraphics[width=\columnwidth]{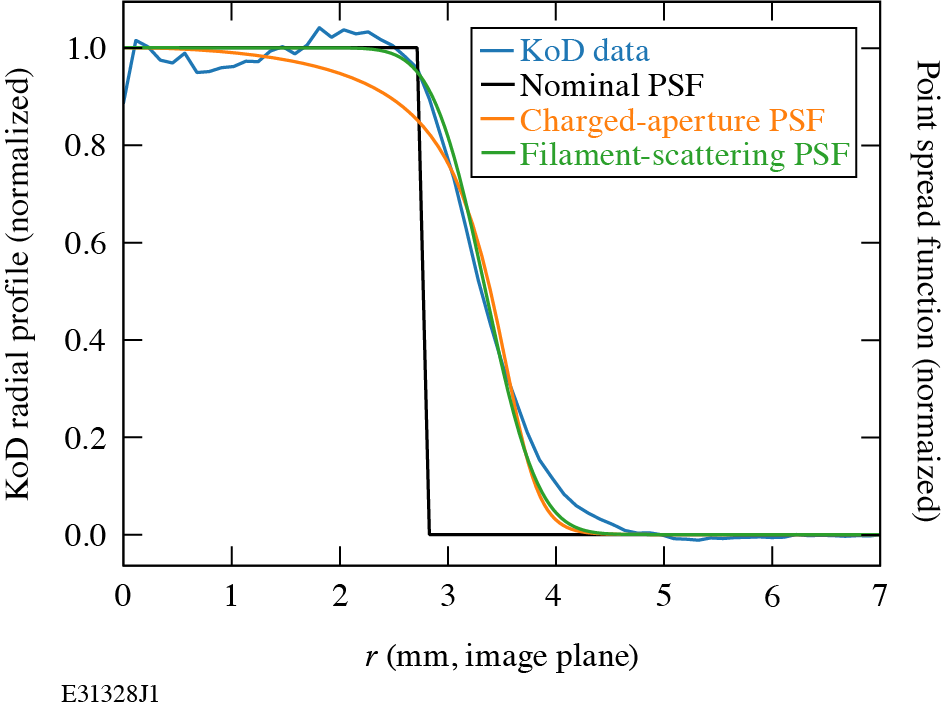}
\caption{\label{fig:psf}
Azimuthally averaged high-energy deuteron data (OMEGA shot 109675) compared to the nominal PSF and PSFs from the aperture-charging model (Sec.~\ref{sec:aperture_charging}) and the filament-scattering model (Sec.~\ref{sec:filament_scattering}). \replace{The 50\% radius of the data is significantly larger than the nominal PSF due to anomalous magnification.}{}}
\end{figure}

After the initial observations of anomalous magnification, it was proposed that the effect was due to a net-negative electric charge accumulated on the aperture prior to the arrival of the deuterons~\cite{Kunimune2022knockon, Rinderknecht2022knock}. This charge would produce a radial electric field within each aperture that would deflect positively charged particles away from the axis, acting as an electrostatic lens to\replace{anomalously}{} magnify the image. The radial dependence of the magnitude of the electric field could also explain some radially symmetric nonuniformities in the umbra. 

A modified analytical point-spread function (PSF) including this electric field was developed in previous work~\cite{Kunimune2022knockon, Rinderknecht2022knock}. The modified PSF predicts the anomalous magnification~\cite{Rinderknecht2022knock}, but depends on the magnitude of the charge on the aperture $Q_\text{ap}$ such that $M_\text{r}/M_\text{true} \propto Q_\text{ap}$. The aperture charge cannot be directly measured, but can be estimated by comparing the projected radius of the deuteron and x-ray images from KoDI. This is done by fitting the azimuthally-averaged penumbral image with the modified PSF convolved with a Gaussian kernel corresponding to the magnified radius of the source object. An example PSF fit in this manner is shown in Fig.~\ref{fig:psf}. Alternately, $Q_\text{ap}$ can be chosen such that the anomalous magnification of the PSF (measured as the radius of the 50\% contour) matches the measured ratio $M_\text{r,KoD}/M_\text{r,x-ray}$. Both methods compensate for the anomalous magnification by definition, allowing a source to be reconstructed from the penumbral image. A significant downside of either approach is that the radius of the reconstructed source is very sensitive to the chosen PSF, and therefore to the estimate of the aperture charge and associated error in the measured anomalous magnification ratio. Since the aperture charge cannot be independently measured, this introduces significant uncertainty in the validity of the reconstructed images. 

\begin{figure}[tbh]
\includegraphics[width=0.45\textwidth]{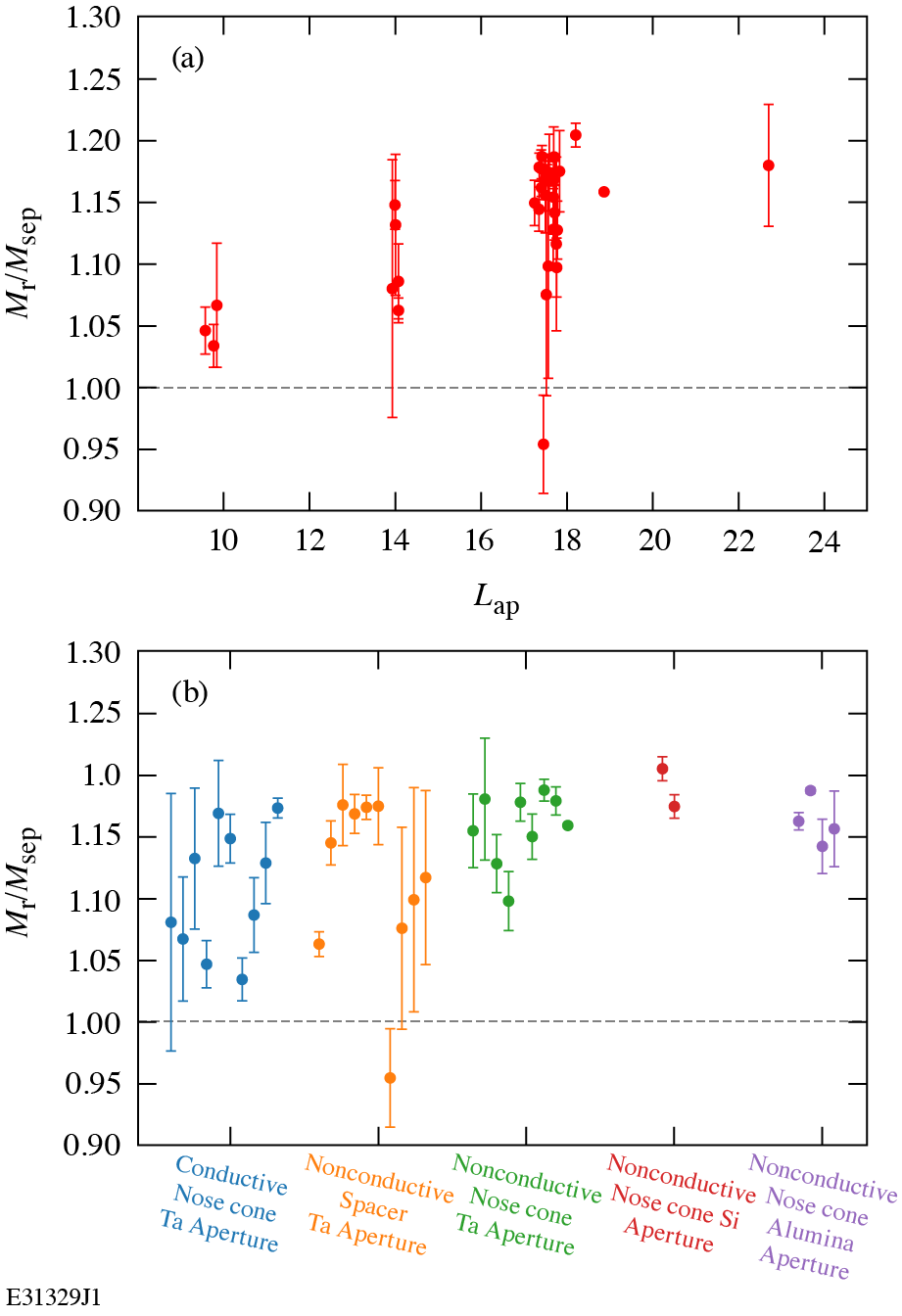}
\caption{\label{fig:mratio}
The high energy KoD anomalous magnification ratio for 35 shots as a function of a) the distance of the aperture from the implosion and b) the various iterations on the KoDI hardware fielded to attempt to mitigate aperture charging.}
\end{figure}

The simplest explanation for the source of the aperture charging is electrons ejected from the laser target. In this case, we expect $Q_\text{ap} \propto m_r \propto 1/L_\text{ap}^2$ by geometric arguments, so increasing $L_\text{ap}$ should mitigate the aperture charging effect. Observations [Fig.~\ref{fig:mratio}(a)] show, however, that the anomalous magnification ratio either increases linearly or is constant with respect to $L_\text{ap}$. Given the small number of samples at ${\sim}9$ cm and $> 20$ cm, it appears more likely that $Q_\text{ap}$ is independent of $L_\text{ap}$. 

Another explanation for the aperture charging posited that an electromagnetic pulse radiated from the target stalk during the implosion induces a potential in the body of KoDI, drawing a negative net charge to the apertures~\cite{Rinderknecht2022knock}. This model predicts a linear dependence $Q_\text{ap} \propto L_\text{ap}$, consistent with the measurements in Fig.~\ref{fig:mratio}(a). To mitigate this, several hardware modifications were developed to\replace{attempt to} prevent the flow of current from the diagnostic body to the nose tip (Fig.~\ref{fig:hardware}). First, a nonconductive spacer was placed between the nose cone and the body of the diagnostic. Next, the entire nose cone (except for the cap, which remained metal) was replaced with nonconductive components (plastic and ceramic). Finally, prototype apertures were manufactured out of semiconductive (silicon) and nonconductive (alumina, Al$_2$O$_3$) to prevent the movement of charge onto the aperture. These apertures had identical dimensions to the tantalum apertures but were thicker (${\sim}700$ $\mu$m for silicon and alumina compared to $200$ $\mu$m for tantalum) so that they remained sufficiently opaque to x-rays. The results of these experiments, shown in Fig.~\ref{fig:mratio}(b), show no reduction in the anomalous magnification with any of these hardware changes. 

An independent test of the validity of the charged-aperture model and the process of inferring the aperture charge is provided by comparing the reconstructed high-energy deuteron and x-ray images. As shown in Fig.~\ref{fig:1d_simulated_profiles}, a time-integrated x-ray image provides a similar profile to the high-energy deuteron image, provided the correct band of x-ray energies is selected using simulations and synthetic diagnostics. If the PSF applied to KoDI data is correct, we therefore expect approximate agreement between the reconstructed high energy KoD and x-ray images. In particular, for implosions with low-mode asymmetry, we expect the asymmetry to be observed in both the high-energy KoD and x-ray images. 

\begin{figure}[tbh]
\includegraphics[width=\columnwidth]{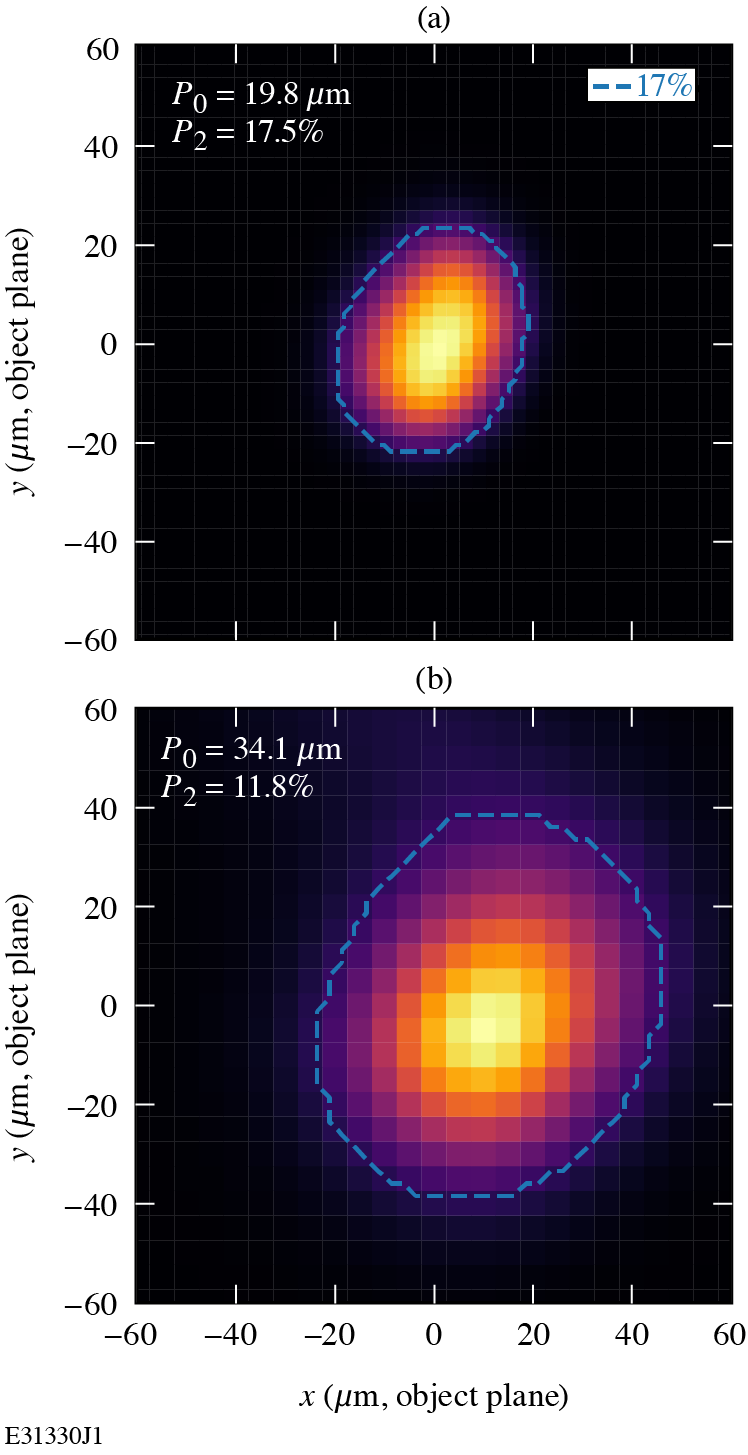}
\caption{\label{fig:kod_xray_comparison}
Reconstructed (a) x-ray and (b) high-energy KoD images of an implosion with a significant $P_2$ hot-spot asymmetry (OMEGA shot 109675).}
\end{figure}

Figure~\ref{fig:kod_xray_comparison} shows a comparison of reconstructed high-energy KoD and x-ray images (from the same LOS) of an implosion with a significant $P_2$ hot-spot asymmetry. Both images are reconstructed using an iterative maximum likelihood algorithm~\cite{Gelfgat1993programs, Rinderknecht2022knock}, and the KoD image is created using the charged-aperture PSF, with the inferred charge determined by the magnification ratio determined from the penumbra radius and the pinhole array separation. The clear disagreement between both the radius ($P_0$) and asymmetry ($P_2$), and the overall shift of the KoD image away from the center of the image indicates that the deuteron image is not an accurate reconstruction of the KoD source. Both images include a similarly-oriented $P_2$, although the disagreement in the magnitude (${\sim}60$\%) is substantial. In this case, either the aperture charging model or the method for estimating the charge on the aperture is insufficient. KoD reconstructions with a nominal PSF with no aperture charging are in even greater disagreement. 

Combining these observations, we conclude that, while the modified PSF derived for a charged aperture is consistent with the observation of anomalous magnification and provides an improved fit to the observed penumbra~\cite{Kunimune2022knockon}, the correction is not sufficient to accurately determine the hot-spot radius. Furthermore, the data are inconsistent with any of the mechanisms for explaining the presence of the aperture charge so far considered.

\subsection{Aperture Scattering\label{sec:aperture_scattering}}

\begin{figure}[tbh]
\includegraphics[width=\columnwidth]{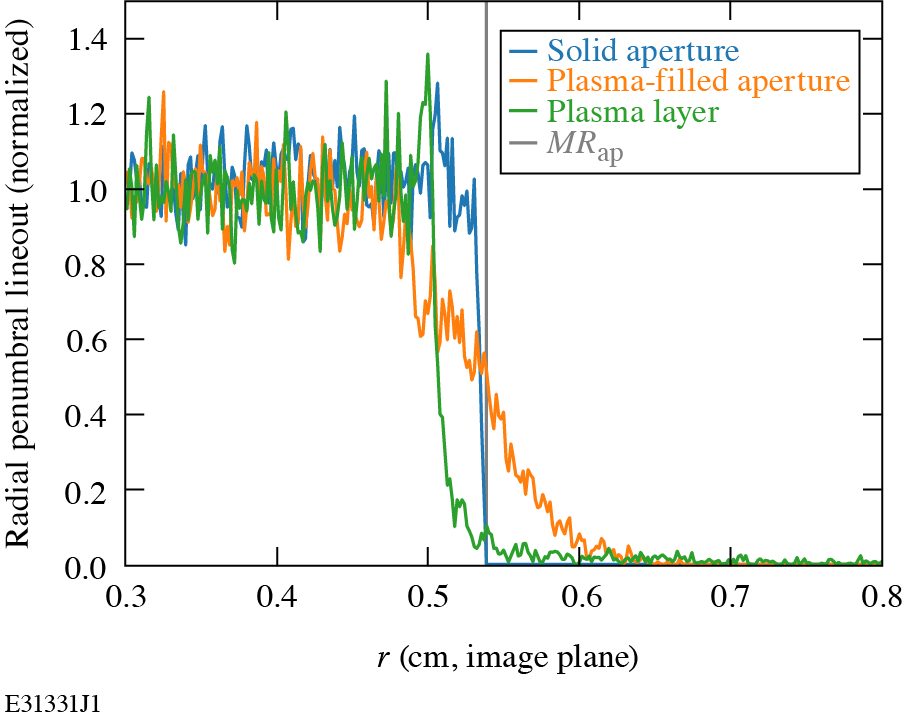}
\caption{\label{fig:aperture_scattering}
Radial profiles of synthetic penumbral images produced by a point source of 6-MeV KoD with a solid aperture (blue) and with the addition of a tenuous plasma either filling the aperture (orange) or just along the edges of the aperture (green). The nominal aperture radius is marked in gray.}
\end{figure}

Early work on the charged particle imaging of ICF implosions~\cite{DeCiantis2006proton} considered the possibility that particles could scatter in the edges of the apertures, altering the shape of the penumbral image. \replace{To investigate the possibility,}{}We conducted Monte Carlo charged-particle scattering simulations using a particle tracker developed as part of the PlasmaPy project~\cite{PlasmaPy2024_10_0}. A point source of $10^5$ 6-MeV deuterons is positioned at TCC, and a detector at the same location as in the experiments. The aperture is represented by a $2$-$\mu$m resolution grid, from which the mass density and temperature at each particle location are interpolated at each time step. The mean scattering angle is calculated for each particle using cross sections from the ENDF~\cite{Brown2018ENDF} database. Finally, a histogram of particle locations in the detector plane is made and azimuthally averaged to obtain the radial profile of the penumbral image. 

As shown in Fig.~\ref{fig:aperture_scattering}, scattering in the aperture does not move the edge of the penumbral image outward. This is because even a very small amount ($< 1$ $\mu$m) of solid Ta is sufficient to scatter a deuteron by an amount $\gg R_\text{ap}$ on the detector plane, so any deuterons that scatter off of the edge of the aperture effectively form a uniform background over the detector. As further evidence, we note that deuterons would experience less scattering in the lighter silicon apertures as compared to the tantalum apertures,  but Fig.~\ref{fig:mratio}(b) shows no change in the observed anomalous magnification between these cases.

Weaker scattering on the scale of the penumbral images might occur if the deuterons passed through a plasma in the aperture, such as might be ablated by x-rays from the implosion. We consider the case of a tenuous plasma uniformly filling the aperture, which we model as a Ta plasma with an ion density of $5 \times 10^{15}$ cm$^{-3}$. We also consider the effect of a dense layer of plasma formed along the edges of the solid aperture, which we model as an exponential density profile starting at the solid aperture density then dropping off with a $0.5$-$\mu$m scale length. In both cases the plasma temperature is set to $T_\text{i} = 10$ eV. We find that adding a tenuous plasma within the aperture does scatter some deuterons to larger radii, but a similar number are scattered to smaller radii such that the 50\% radius of the penumbral image is unchanged. This is expected since scattering in the plasma can be modeled as a Gaussian blur applied to the penumbral image, which does not shift the 50\% radius. Adding a plasma along the inside edge of the aperture actually decreases the size of the penumbral image since the effective radius of the aperture (for deuterons) is decreased. 

Another consideration is straggling of the deuterons in the tantalum filtration in front of the CR-39 detector~\cite{DeCiantis2006proton}, which effectively convolves a Gaussian kernel with the KoD images. SRIM~\cite{SRIM} simulations, however, indicate that the scale of this Gaussian is $\sigma < 20$ $\mu$m for the high-energy deuterons and $\sigma < 1$ $\mu$m for the low-energy deuterons, which is too small to explain the observed anomalous magnification. This effect would also be highly repeatable between shots, inconsistent with the variance in Fig.~\ref{fig:mratio}(b). 

We conclude that scattering in the aperture or in material ablated from the aperture cannot explain the observed anomalous magnification. It is also important to note that neither scattering in the aperture nor the filtration on the detector pack can explain the smearing effect that dominates the low-energy deuteron data. 

\subsection{Filamentary Fields\label{sec:filament_scattering}}

\begin{figure}[!tbh]
	\includegraphics[width=0.8\columnwidth]{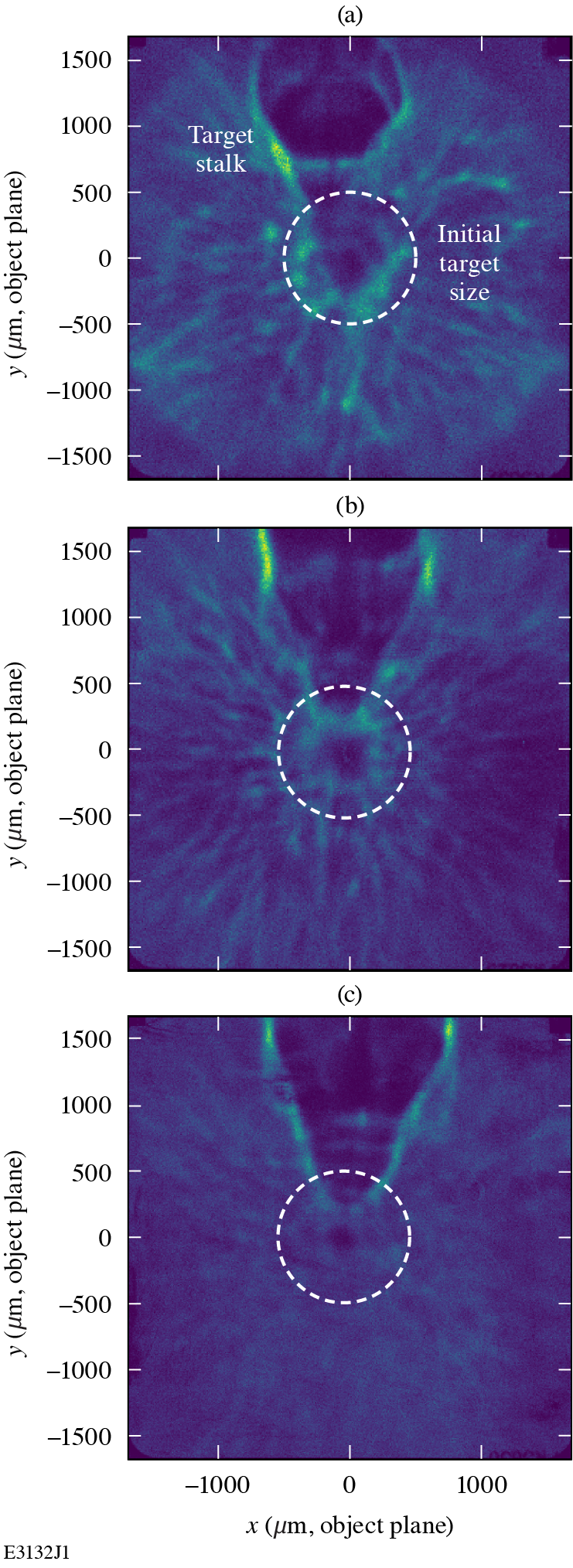}
	\caption{Experimental 15-MeV proton radiographs of a spherical target implosion driven by 40 beams with a 1-ns square-shaped pulse at (a) 0.39~ns, (b) 0.88~ns, and (c) 1.37~ns. Early in the pulse (a), the filamentary fields expand outward, filling the field of view later in time (b). About 300~ps after the end of the pulse (c), the filaments have largely dissipated. A dotted line shows the initial position of the target. The large feature at the top of the radiographs is caused by a current traveling down the target stalk.}
	\label{fig:spherical_prad}
\end{figure}

Early proton radiographs~\cite{Rygg2008proton} of spherical implosions on the OMEGA Laser System showed complex electric and/or magnetic structures surrounding the imploded target extending for at least a millimeter outside the original target radius. These filaments were observed to grow early in the drive laser pulse ($\leq 0.5$ ns) and persist for at least 1 ns after the end of the drive pulse~\cite{Seguin2012time}. Similar filamentary structures have been observed surrounding cylindrical implosions~\cite{Zylstra2016development,Heuer2022diagnosing}. Notably, while most of these measurements were performed with 15-MeV D$^3$He protons, the filaments are still clearly visible in some published data collected using 30-MeV protons generated by target normal sheath acceleration~\cite{Zylstra2012using}, indicating that the fields in the filaments must be quite strong. These features have been modeled as either filaments of electric charge or filaments surrounded by toroidal magnetic fields, and it was recognized that these fields might deflect charged particles emitted by implosions~\cite{Seguin2012time}. Identifying the mechanism(s) that generates these fields remains an open area of research~\cite{Sutcliffe2022observation} and will be discussed in future work. 

A recent campaign conducted on OMEGA investigated the time dependence of filamentary fields around spherical implosions. A D--$^3$He backlighter implosion, producing ${\sim}$3-MeV DD and ${\sim}$15-MeV D$^3$He protons, was placed 8~mm behind a primary spherical implosion of a plastic shell mass equivalent to a typical cryogenic implosion. The primary implosion was driven with 25~kJ over a 1~ns square pulse in 50 beams (the remaining ten OMEGA beams being used for the backlighter implosion). The time delay between the primary implosion and the backlighter was varied to collect proton radiographs at different times. Both 3-MeV (not shown) and 15-MeV radiographs (Fig.~\ref{fig:spherical_prad}) show a web of filamentary structure extending several millimeters away from the implosion. The filaments form early in time, and appear to initially expand spherically outward from the target [Fig.~\ref{fig:spherical_prad}(a)]. Later in the laser pulse, the fields fill a large volume around the target [Fig.~\ref{fig:spherical_prad}(b)]. Other experiments show that the filaments remain present while the laser remains on. Finally, by around 300~ps after the laser drive has ended, the filaments have dissipated enough to no longer be visible in the proton radiographs [Fig.~\ref{fig:spherical_prad}(c)]. In cryogenic implosion experiments on OMEGA, peak neutron production (bang time) occurs very close to the end of the laser drive and so the KoD generated by those neutrons pass through fields that are similar to those shown in Fig.~\ref{fig:spherical_prad}(b).

The ability of these fields to deflect $\geq 15$-MeV protons in proton radiography experiments is clear evidence that they are strong enough to significantly deflect KoD. We suggest that the scattering of KoD in these filaments may explain the image distortions reported in Sec.~\ref{sec:image_distortions}. Stochastic scattering off of the filaments imposes a Gaussian blur on the PSF. We suppose (and later show in Sec.~\ref{sec:synthetic_kodi_analysis}) that scattering in the radially oriented filaments around the implosion also tends on average to move the apparent source of the deuterons closer to the aperture, increasing the apparent image magnification. We convolve these effects in a heuristic model 
\begin{equation}
I(r) = K(\sigma) \ast H(M m_\text{r} R_\text{ap} - r)
\end{equation}
where $H$ is the Heaviside function and $K(\sigma)$ is a normalized Gaussian kernel with standard deviation $\sigma /M =\sigma_s + \sigma_b$ where $\sigma_s$ is the size of the source and $\sigma_b$ is a Gaussian blur imposed by scattering off of the filamentary fields. We see in Fig.~\ref{fig:psf} that this PSF fits experimental data better than the charged aperture PSF, and consequently can also be used to reconstruct a reasonable source from the penumbral image. As in the case of the charged aperture PSF, however, the reconstructed sources are highly dependent on the chosen PSF and the source size $\sigma_s$ is degenerate with the unknown amount of blurring $\sigma_b$, limiting the value of these reconstructions.

\section{Generating Synthetic Data with a Custom Particle Pusher\label{sec:particle_pusher}}

\begin{figure}[tbh]
	\includegraphics[width=\columnwidth]{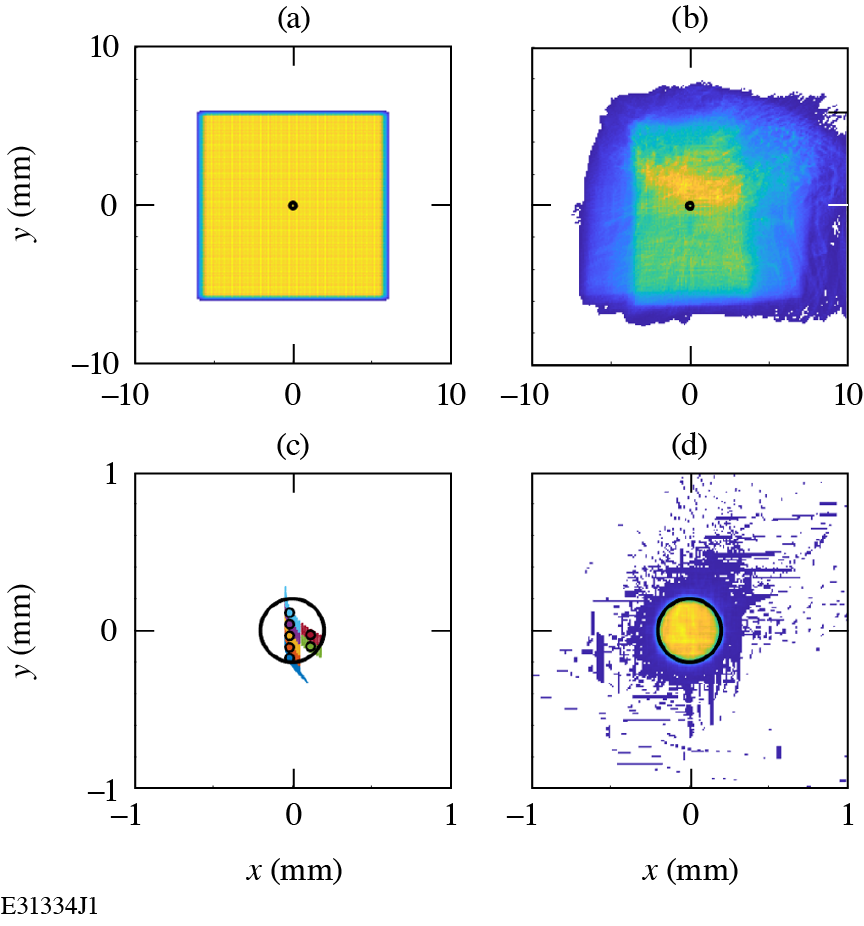}
	\caption{A histogram (a) shows the locations of the initial low resolution batch of deuterons at the aperture plane in the absence of any fields (the aperture indicated by small black circle). Deuterons are initialized as a perfect grid upon reaching the aperture plane when no field is present. When the field is introduced (b) the deuterons are scattered and only a small fraction make it to the aperture. (c) An example of 7 color coded particles that have reached the aperture (c, large circles) and the locations of the high resolution deuterons created around them (c, small dots) in the next higher resolution iteration. In the aperture plane histogram of the final high resolution iteration (d), nearly all particles make it through the aperture.}
	\label{Fig:Algorithm1}
\end{figure}

A particle-pushing algorithm was developed in Matlab 2023b  to generate both synthetic proton radiographs of spherical implosions and synthetic KoDI data. In both modes, the pusher explicitly calculates a grid of magnetic or electric fields from current or charge distributions using Biot--Savart's and Gauss's laws, respectively. Particles are pushed through the grid using the Boris leapfrog pusher algorithm~\cite{Birdsall2004plasma}. After leaving the field grid, particles are extrapolated to a detector or aperture plane in a single step.

In Sec.~\ref{sec:synthetic_radiographs}, synthetic proton radiographs are generated that correspond to experimental radiographs (Fig.~\ref{fig:spherical_prad}). These radiographs are used to
set a lower bound on the electric or magnetic field strength in the filaments surrounding an implosion. These field magnitudes are then used in the synthetic KoDI diagnostic developed in Sec.~\ref{sec:synthetic_kodi_tracker}. 

\subsection{Generating Synthetic Proton Radiographs\label{sec:synthetic_radiographs}}

\begin{figure}[tbh]
	\includegraphics[width=0.7\columnwidth]{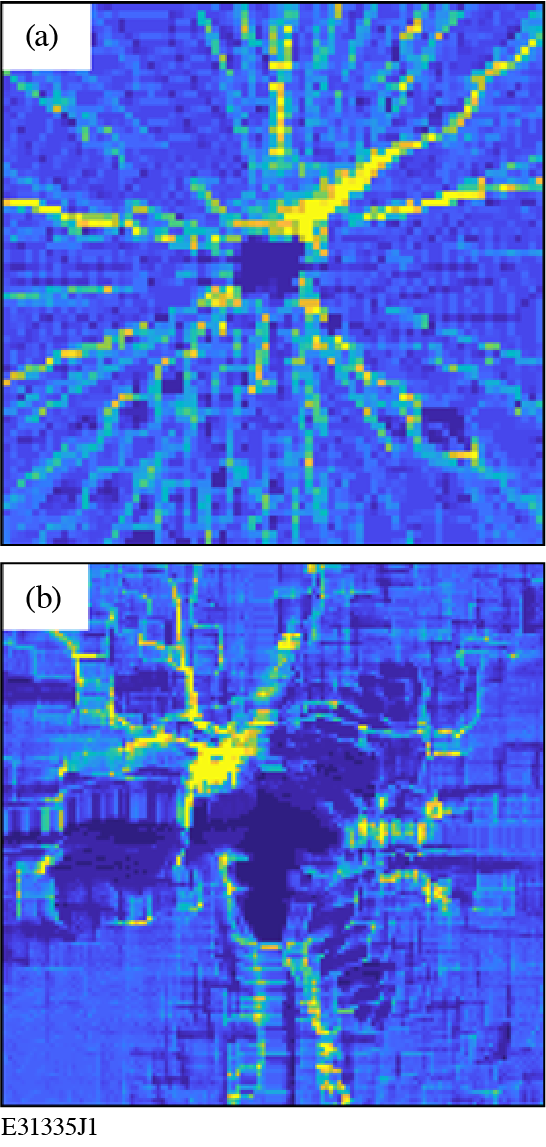}
	\caption{Synthetic radiographs corresponding to the experimental radiographs in Fig.~\ref{fig:spherical_prad} generated using models of (a) filamentary electric and (b) magnetic field models. The electric field model places a negative ($-64$-nC) charge in 120 radial filaments and a positive ($16$-nC) charge at the center of the implosion. The magnetic field model places a current ($0.5$~kA) in alternating directions along the filaments. The void structure in the magnetic field case is notably absent in the experimental radiographs.}
	\label{Fig:PRad}
\end{figure}

Synthetic proton radiographs are generated with a point source of 15-MeV protons placed 8~mm away from the center of the filaments (matching the experimental setup described in Sec.~\ref{sec:filament_scattering}) with a velocity distribution that would produce a perfect grid of evenly spaced protons in the detector plane in the absence of electric or magnetic fields. For the proton radiography simulations, the field grid is 6~mm $\times$ 6~mm $\times$ 12~mm or larger. Field sources are modeled as either currents or lines of charges placed radially outward from near the target center out to ${\sim}$3~mm. In the electric field case, a positive charge is placed in the center to represent the overall positive potential of the capsule~\cite{Seguin2003spectrometry}. For the magnetic field case, each filament alternates between the current traveling in and out of the target. Although these charge/current distributions do not completely reproduce the web-like structure, which is closer to a Voronoi lattice, they are close enough to be used for estimating the field strengths required to produce the experimental radiographs. Some grid-pattern artifacts in the synthetic radiographs are due to the proximity of the fields to the particle source; as in the experiment, the filamentary fields from the implosion extend almost to the proton source. 

The magnitude of the fields in the experiment is estimated by varying the magnitude of the charges and currents in the simulations across an ensemble of runs and comparing to the experimental radiographs. An upper bound on the field could be established if the filaments disappeared above a certain proton energy. Unfortunately, in all of the experimental radiographs the caustic filamentary features persists for all measured proton energies, so an upper bound on the field strength cannot be established from this data. 

A lower bound, however, can be provided by finding the lowest filamentary field strength required to reproduce the caustic features. We find that in the electric field model, a charge of at least $-32$~nC must be placed within a 3-mm$^3$ spherical volume around the target to reproduce these features, while for the magnetic field model the currents along the filaments must be at least 500~A. The field strength varies based on proximity to the filaments, but the average magnitudes between the filaments correspond roughly to electric fields of $\geq 10^8$~V/m and magnetic fields $\geq 1$~T. Over an interaction length of ${\sim}1$ mm, these correspond to line-integrated fields of $\int E \text{d}l \sim 100$~kV and $\int B \text{d}l\sim 1$~T~mm, consistent with the order of magnitude of fields estimated based on the smearing data in Sec.~\ref{sec:smearing}.

\subsection{Generating Synthetic KoD Images\label{sec:synthetic_kodi_tracker}}

To investigate the impact of filamentary fields on KoDI, the proton radiography particle-pushing algorithm was modified to produce synthetic KoDI data. An aperture is introduced between the deuteron source and the detector. After particles leave the field grid they are evolved to the aperture plane, and particles that miss the aperture are removed before the remaining particles are propagated to the detector plane. The deuteron source is modeled as a spherical shell with a range of velocities centered on the detector. Only one energy of deuterons is traced in a given simulation, but multiple deuteron energies from multiple simulations are combined to generate synthetic data with a range of deuteron energies comparable to the experimental images. Based on the KoD scattering probability (Fig.~\ref{fig:scatter_prob}), high-energy deuterons are primarily generated close to the center of the shell facing the detector, while low-energy deuterons are generated in a ring at large angles. The initial distribution of the deuteron source in the shell is weighted to reflect this. 

The introduction of an aperture presents a significant challenge. Reproducing the experimental KoDI data requires that millions of deuterons pass through the small solid angle of the aperture, but the presence of the filamentary fields around the capsule means that particles from a large range of initial angles must be sampled since they could potentially scatter into the aperture. Simulating all particles over this range of angles with the required resolution is computationally intractable. 

To address this issue we developed an algorithm that begins by tracing the trajectories of a conservative number (${\sim}10^6$) of deuterons distributed over a large solid angle centered on the KoDI aperture. Figure~\ref{Fig:Algorithm1}(a) shows that in the absence of fields around the target this initial batch of deuterons cover both the aperture and a large region around it. When an electric or magnetic field is added to the grid [Fig.~\ref{Fig:Algorithm1}(b)], the particles are scattered such that some particles that previously missed the aperture pass through it, and vice versa. Only a small fraction (typically $<1$\%) make it through the aperture. These particles are tagged, and a second-generation, higher-resolution population of deuterons is created with initial positions and velocities distributed around those of the tagged deuterons from the previous generation [Fig.~\ref{Fig:Algorithm1}(c)]. This process continues iteratively until $1.5 \times 10^6$ deuterons pass through the aperture [Fig.~\ref{Fig:Algorithm1}(d)]. The final group of particles that pass through the aperture are then propagated to the detector plane to form a penumbral image. 

\section{Analysis of Synthetic KoD Images\label{sec:synthetic_kodi_analysis}}

\begin{figure}[tbh]
	\includegraphics[width=\columnwidth]{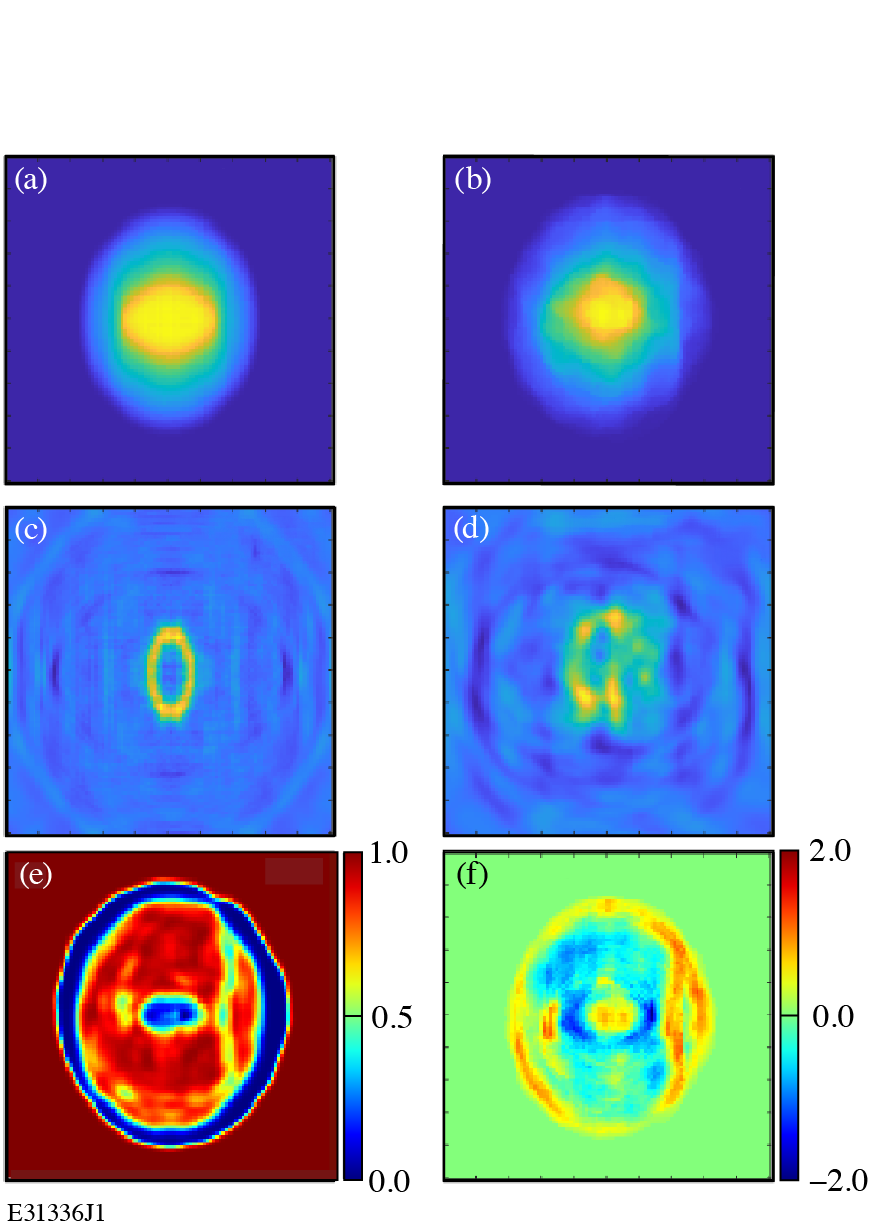}
	\caption{(a) Penumbral image of a hemi-ellipsoidal deuteron source with no deflecting fields. (b) A field is applied via filaments around the deuteron source, which deflects and skews the penumbral image. (c) The no-field penumbral image is deconvolved with the PSF, assuming high signal to noise, and reproduces the source shape. (d) The penumbral image with fields is deconvolved using the nominal (no fields) PSF and is consequently both skewed and distorted (e) The structural similarity index measure (SSIM) of each pixel when comparing the two penumbral images shows the loss of mode information and increase penumbral image size. (f) The absolute value of the difference between the two penumbral images (normalized). }
	\label{Fig:SynthKoDI1}
\end{figure}

\begin{figure*}[tbh]
	\includegraphics[width=\textwidth]{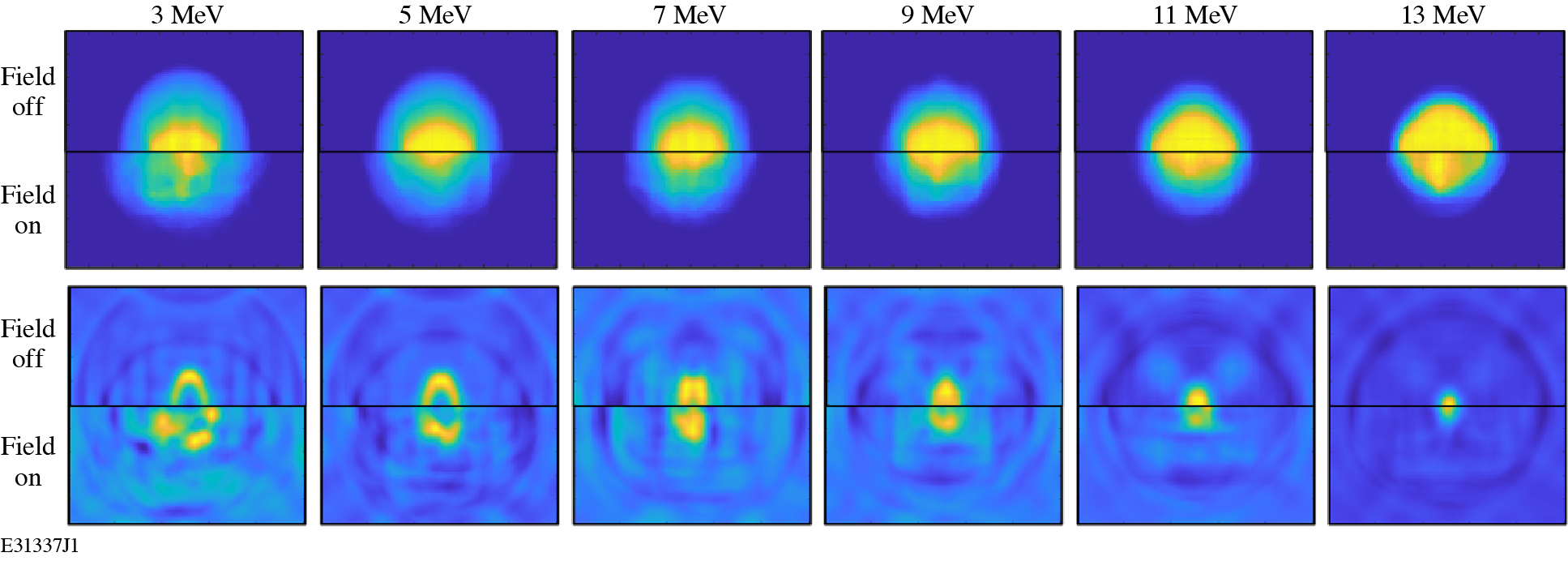}
	\caption{ (top row) Penumbral images and (bottom row) deconvolved images of the inferred deuteron source from synthetic KoDI for six different deuteron energy bins. The simulations for each energy bin are done with the same scattering field distribution. The angular distribution for knock on deuterons from Fig.~\ref{fig:scatter_prob} are taken into account in these analyses: low-energy deuterons are emitted and scattered at large angles, which result in a source with a larger apparent radius; high-energy deuterons are scattered at small angles, resulting in close to a point source.}
	\label{Fig:SynthKoDI2}
\end{figure*}

The synthetic KoDI data are analyzed in a similar manner to the experimental data (Sec.~\ref{sec:image_distortions}). The magnification ratio, $m_\text{r}$, is determined by fitting the 50\% radius of the penumbral image, which is equivalent to Eq.~(\ref{eq:penumbral_model}). The offset of the center of mass of each monoenergetic deuteron image is compared to the center of the corresponding no-field case to determine the effective offset, which corresponds to the experimentally observed smearing. In addition, since the no-field penumbral images are available for the synthetic data, we quantify the distortion of the penumbral images by computing the structural similarity index measure (SSIM) between the field-on and field-off images (after correcting for the overall offset of the field-on image). The SSIM performs three comparison measurements for each pixel between the two images with terms for luminescence, contrast and structure, and ranges from 0~(completely dissimilar) to 1~(identical).

A KoD source image is reconstructed from each synthetic penumbral image by first applying a median filter (to reduce noise from finite particle statistics) and then using the Wiener deconvolution algorithm to deconvolve the instrument PSF. The Wiener deconvolution requires an estimate of the signal-to-noise ratio, which was chosen to be $10^{-8}$ to $10^{-9}$ for the data presented here. In all cases, the nominal PSF calculated in the no-field case is used since this is\replace{what is} known for the experimental data. For penumbral images generated with no electric or magnetic fields around the source, this PSF is exact and the deconvolution reproduces the source except for some noise added by the median filter. For cases with nonzero electric or magnetic fields, the reconstructed source will be distorted. 

Since only a lower bound on the magnitude of the fields around the source could be established (Sec.~\ref{sec:synthetic_radiographs}), a variety of field strengths at or exceeding the minimum were tested. As in Sec.~\ref{sec:synthetic_radiographs}, fields were modeled as either filaments of charge or filaments of current. Electric field simulations were conducted with total charges of $-32$~nC (the minimum inferred from experiments), $-64$~nC, and $-96$~nC distributed over 120 filaments with 240 macrocharges in each filament. An additional positive charge of 32~nC was added in the center to represent the overall positive potential of the capsule~\cite{Seguin2003spectrometry}, although including this charge had little effect on the synthetic KoDI data. One magnetic field configuration was used in which radial filaments alternated direction in and out from the source, and each filament carried 500~A (the minimum inferred from the experiments). 

For every field configuration, as well as the no-field configuration, monoenergetic synthetic images were created over a range of deuteron energies from 3~MeV to 13~MeV in 1-MeV steps. Finally, for each combination of deuteron source profile, deuteron energy, and field configuration, each simulation was repeated three times with three different randomized filament configurations to quantify the variance due to the stochastic filament locations. 

\subsection{Validation\label{sec:synthetic_kodi_validation}}

The synthetic KoDI routine was validated by demonstrating that the expected penumbral image was generated for a known source in both the no-field and $-64$~nC electric field cases. An asymmetrical source with a large $P_2$ mode (an ellipsoidal source) was chosen as a simple but nontrivial object. The uniform distribution of the source was achieved by creating 150 source locations distributed via a Fibonacci lattice over an hemi-ellipsoid on the side facing the aperture. The maximum height of the initial source was 150~$\mathrm{\mu m}$, while the maximum width was 75~$\mathrm{\mu m}$. The resulting penumbral histogram image with 5-MeV deuterons is shown in Fig.~\ref{Fig:SynthKoDI1}(a), and the accurately reconstructed source profile is shown in Fig.~\ref{Fig:SynthKoDI1}(c).

Figure \ref{Fig:SynthKoDI1}(b) shows the penumbral image produced by the same hemi-ellipsoidal source with the field on, in which the image is significantly distorted. Since scattering in the fields has effectively altered the PSF, deconvolution of the penumbral image using the nominal (no field) PSF [Fig.\ref{Fig:SynthKoDI1}(d)] no longer reproduces the true source profile. The deconvolved image is now both larger (magnified) and distorted, although the large $P_2$ asymmetry is still recognizable. The SSIM, as well as the absolute difference between the field and no field images, are plotted in Figs.~\ref{Fig:SynthKoDI1}(e) and \ref{Fig:SynthKoDI1}(f), respectively, and show the significant distortion of the field-on image. 

Penumbral images and reconstructed source profiles for the field-on and field-off cases over the full range of deuteron energies for the ellipsoidal source are shown in Fig.~\ref{Fig:SynthKoDI2}. In the no-field cases, the source is accurately reconstructed and the size increases for lower deuteron energies, as expected based on the scattering probability. In nearly all cases with the field applied, however, both the penumbral images (not shown) and deconvolved sources are larger than in the field-off case. The structure of the reconstructed deuteron sources in the no-field cases and the increasing deleterious effect of the field on lower-energy deuterons conforms with our theoretical expectations, establishing that the synthetic KoDI algorithm is a reasonable surrogate for the real diagnostic.

\subsection{Comparison to Experimental Data\label{sec:synthetic_kodi_comparison}}

\begin{figure}[tbh]
	\includegraphics[width=\columnwidth]{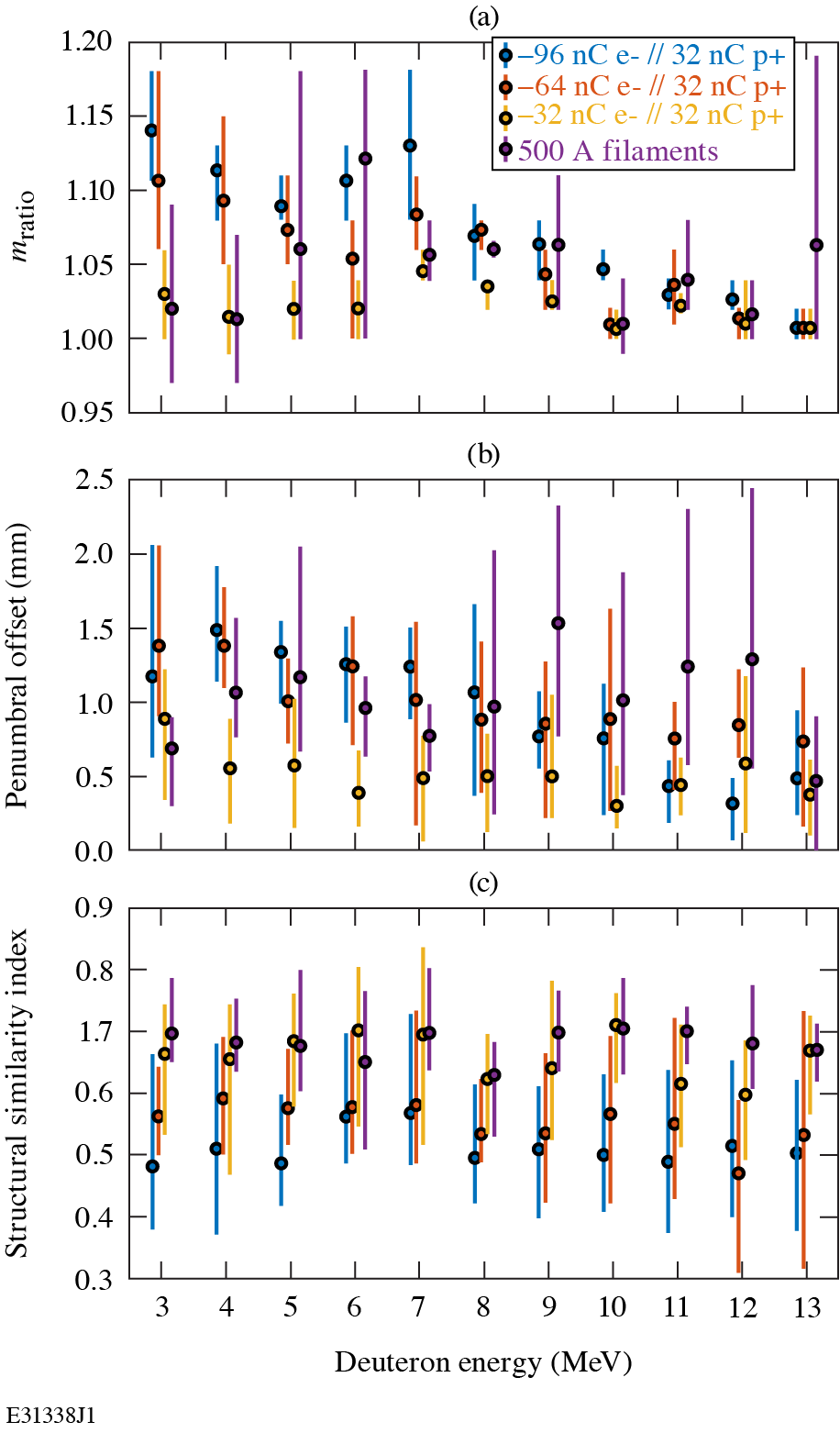}
	\caption{Plots of (a) magnification ratio, (b) penumbral image offset distance, and (c) structural similarity index from synthetic KoDI data generated with different deuteron energies and field configurations. The numbers are an average of three simulations for each field. Both image offset and magnification ratio depend on scattering field strength and deuteron energy, but even in the low-field case a significant image offset of roughly 400~$\mathrm{\mu m}$ was measured. Structural similarity showed that even in the low-field case, the penumbral image was not close to the same shape or pattern to the case with no scattering field.}
\label{Fig:SynthKoDI4}
\end{figure}

\begin{figure}[tbh]
	\includegraphics[width=0.45\textwidth]{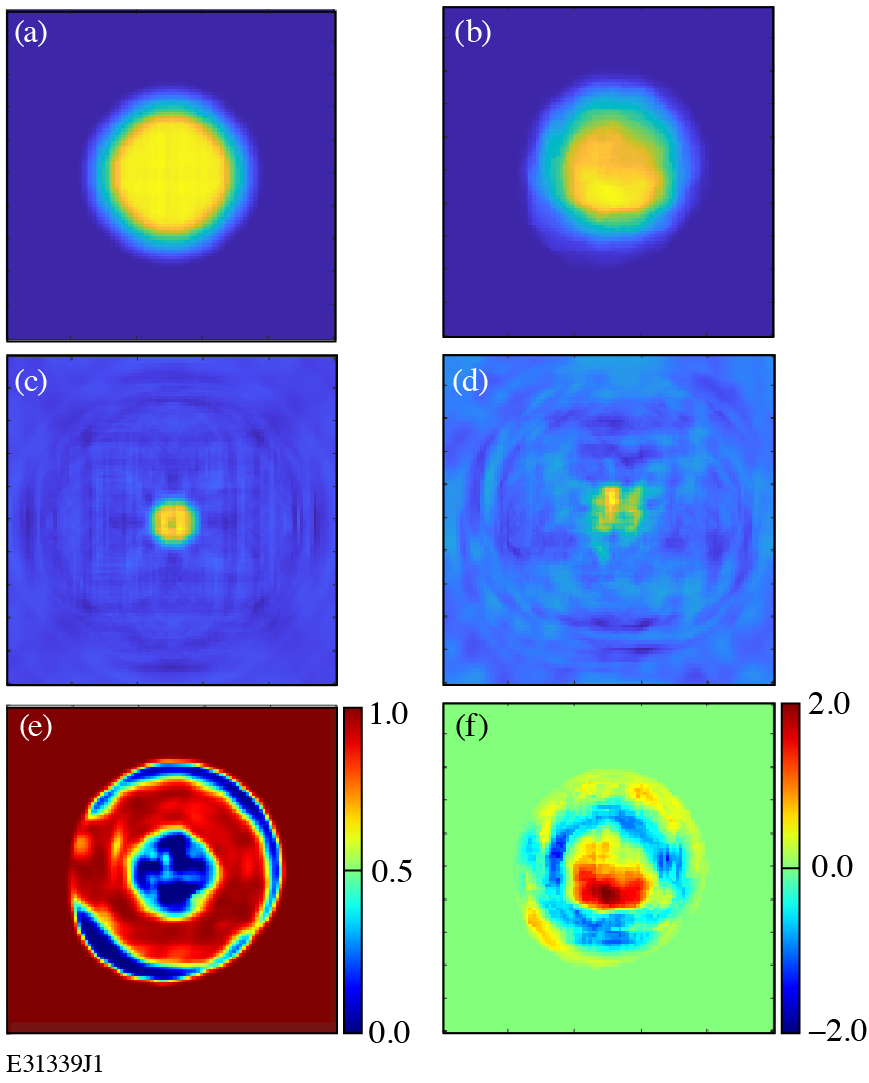}
	\caption{Penumbral images formed with a round 12-MeV deuteron source (a) with no filamentary fields and (b) with the lowest electric field case, $-32$~nC. The magnification ratio and offset of the penumbral image are minimal in this example, but the shape is still significantly distorted. (c) The deconvolved penumbral image of the no-field case reproduces the source, while (d) the reconstruction of the field-on case is substantially distorted. The (e) SSIM and (f) absolute difference between the two images shows that the field has scattered deuterons in a ring toward the center and edges of the image.}
	\label{Fig:Penum}
\end{figure}

\begin{figure}[tbh]
	\includegraphics[width=\columnwidth]{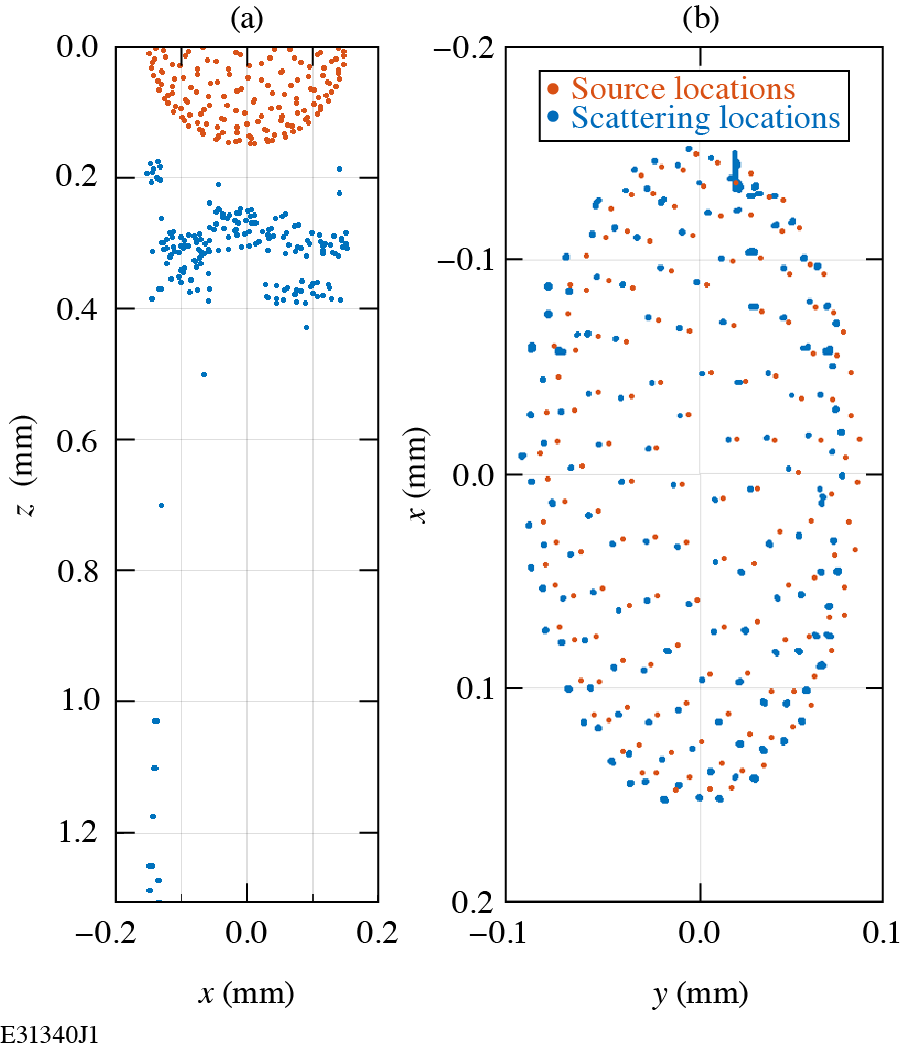}
	\caption{(a) Initial deuteron source locations are shown in red, while the locations where deuterons experience maximum deflection from surrounding fields are shown in blue. Deuterons travel from the top to bottom in this image. (b) The same plot but from the perspective of the detector looking at the source. The locations where the particles are scattered shows that the source has effectively been magnified and offset.}
	\label{Fig:SynthKoDI3}
\end{figure}

Implosions in the experimental dataset (Sec.~\ref{sec:image_distortions}) are known from other diagnostics to generally be much closer to round than the example in Sec.~\ref{sec:synthetic_kodi_validation}. Therefore, to quantitatively compare synthetic data to the experimentally observed anomalous magnification and smearing, an ensemble of simulations was run with a round deuteron source. These simulations were again run for the no-field case, and each of the four field-on cases described above. When analyzing the synthetic data over many simulations in a similar way to the experimental data (Fig. \ref{Fig:SynthKoDI4}), we find that the image offsets and magnification ratio depend both on the field strength surrounding the target and deuteron energy. 

The calculated magnification ratios range from 1 to 1.2, comparable to the experimental data in Fig.~\ref{fig:mratio}(a). The image offsets from the synthetic data are generally lower than the experimental data, indicating that the field strengths used for these simulations are likely an underestimate. Further simulations could be performed with higher field, but would require more computation power than currently available since, as scattering increases, a larger initial solid angle of deuterons is needed to capture particles deflected into the aperture. The SSIM shows that, for all field strengths, the images are significantly distorted by the fields. 

Even in low-field cases where the image offset and magnification ratio appear minimal, the fields have a significant impact on the structure of the penumbral image that prevents an accurate reconstruction of the source. In the lowest electric-field case ($-32$~nC) with 12-MeV deuterons, the magnification ratio approaches unity and the image offset is $\leq 0.3$~mm, but the penumbral image profiles are significantly distorted [Fig.~\ref{Fig:Penum}(b)]. Figures~\ref{Fig:Penum}(e) and~\ref{Fig:Penum}(f) show that even in this case, the fields have significantly altered the penumbra, preventing an accurate reconstruction. As a result, the SSIM plotted for these cases in Fig.~\ref{Fig:SynthKoDI4}(c) remains $< 0.8$ for all energies, even as the image offset decreases and the magnification ratio approaches unity. As shown in Fig.~\ref{Fig:Penum}(d), this level of SSIM still results in an inaccurate reconstructed source. This suggests that valid penumbral reconstructions may not be possible even for experimental data with minimal smearing or anomalous magnification. 

\subsection{Physical Interpretation\label{sec:synthetic_kodi_interpretation}}

Inspecting the locations where scattering occurs provides a physical picture for how scattering leads to anomalous magnification and image offset. Figure~\ref{Fig:SynthKoDI3}(a) shows the initial locations of a sample of deuterons that eventually pass through the aperture, and the locations along the trajectories of those deuterons where the largest deflection is experienced. The same particle locations are plotted from the perspective of the detector in Fig.~\ref{Fig:SynthKoDI3}(b). While deuterons are generated only 75 to 150 $\mathrm{\mu m}$ from the center of the target, they receive the largest deflections from fields up to a millimeter from the target (within the filaments). 

Since the scattering direction is effectively random, scattering at this distance has the effect of substantially increasing the apparent size of the deuteron source. The apparent change in source size is linearly dependent on where along the imaging axis the scattering happens. This is apparent when comparing two extreme cases: if a particle is scattered immediately, it has a very similar position to its original source, while a particle that is scattered after traveling 2~mm will likely also have traveled a significant distance perpendicular to the imaging axis away from its original source. As a result, particles scattered far from their initial positions substantially alter the apparent source of the penumbral image, leading to an apparent increase in magnification. We emphasize that any scattering field, electric or magnetic, can have this effect. Collisional scattering would produce a similar result, but KoD collisions are negligible in the plasma surrounding the implosion. 

Figure~\ref{Fig:SynthKoDI3}(b) shows that the scattering locations are also shifted uniformly in one direction. This is an effect of the slightly asymmetrical distribution of the filaments, which leads, on average, to the particles experiencing a larger deflection in one transverse direction. This component results in an overall offset of the penumbral image in the detector plane as a function of particle energy, consistent with the smearing observed in the experiments. 

\section{Conclusions\label{sec:conclusion}}

Energetic charged particles generated by ICF implosions encode information about the spatial morphology of the hot-spot and dense fuel during the time of peak fusion reactions. The KoDI was developed to image knock-on deuterons and tritons produced by implosions on the OMEGA Laser System, but the penumbral images collected are anomalously magnified, are smeared out across the detector, and exhibit other nonuniformities that together prevent the reconstruction of the deuteron source. A model based on charging of the penumbral aperture array predicts some of the distortions, but is inconsistent with other experimental data. Another model based on the scattering of deuterons in the aperture walls or within a plasma filling the aperture \replace{is found to}{does} not match the observed distortions.

We suggest that the image distortions\replace{may be}{are} better explained by\replace{a new model that describes}{} scattering of the KoD by filamentary electric or magnetic field structures known to exist around the implosion at stagnation. A heuristic PSF based on this model provides a better fit to experimental measurements than the charged-aperture model. Particle-tracing simulations in filamentary fields based on experimental measurements reproduce all of the image distortions observed in KoDI data. 

While the KoD data presented in this paper are restricted to cryogenic implosions, filamentary fields are known to exist around warm implosions and should have a similar effect on charged particle imaging in that case. Previous data with KoDI~\cite{Kunimune2022knockon} exhibited only minor anomalous magnification, requiring a modified PSF for reconstruction, but producing KoD images that were roughly consistent with x-ray images and hydrodynamic simulations. Similarly, minimal distortions were observed in previous experiments directly imaging charged-particle fusion products with PCIS~\cite{Seguin2004d3he,DeCiantis2006proton}. In both of these experiments, however, the target was driven with a $1$-ns square pulse, while the bang time (when KoD or fusion protons are produced) occurred at ${\sim} 1.8$~ns. As shown in Fig.~\ref{fig:spherical_prad}, this difference in time is sufficient for the filaments to largely dissipate, reducing their effect on the images. Some remnants of the filamentary fields may cause the minor image distortions interpreted as aperture charging in these implosions. Alternatively, some additional aperture charging may actually be present in these experiments, in which the aperture is much closer to the target ($L_\text{ap} \sim 3$ cm) than in the cryogenic implosions. Unfortunately, efficient cryogenic implosions require that energy be coupled into the target up to bang time, so filaments cannot be mitigated in this manner for cryogenic implosions.

Nonuniformities have frequently been observed in the fusion proton fluence from D--$^3$He exploding-pusher implosions used as backlighters for proton radiography. In most of these experiments, the laser drive continues up to or past the proton bang time. Therefore, we expect the protons to interact with filaments surrounding the backlighter capsule, and suggest that this interaction accounts for the fluence nonuniformities. We suggest that these nonuniformities could be mitigated by using backlighters with a coast phase between the end of the laser drive and bang time to allow time for the filaments to dissipate.

We conclude that the presence of these filamentary fields is a fundamental obstacle to charged particle imaging of ICF implosions, unless the charged particles of interest are produced long after the end of the laser drive, after the filaments have dissipated. No method of mitigating the formation of filaments for ICF-relevant target materials and laser intensities is known. We plan to conduct further experiments \edit{utilizing proton radiography and optical probing to investigate the mechanism that forms the filaments and to observe how the filaments evolve in time.}

\edit{The presence of filamentary fields in the corona should also be considered when evaluating other high energy density experiments, including other potential laser direct drive fusion approaches. For example, if charged particles in a fast ignition or laboratory astrophysics experiment pass through a coronal plasma, they may be dispersed by filamentary fields.}

The filaments do not prevent the use of non-imaging charged-particle diagnostics (e.g., spectrometers), but the presence of the filaments must be considered in the interpretation of measurements from those diagnostics. Some evidence from previous experiments suggests that scattering in filaments does not shift the energy distribution of charged particles measured by spectrometers~\cite{Zylstra2016development, Lahmann2023measuring}. Scattering from filaments, however, may change the effective field of view of a diagnostic or increase the effective source size. Filaments may also cause large-scale fluctuations in the otherwise isotropic distribution of particles emitted from implosions, leading to nonuniformity across a detector or departures of the measurement from the 4$\pi$ average of the signal. These effects will also be investigated in future work. 

\begin{acknowledgments} 
This material is based upon work supported by the Department of Energy [National Nuclear Security Administration] University of Rochester “National Inertial Confinement Fusion Program” under Award Number(s) DE-NA0004144. Work at General Atomics was conducted under the auspices of the Department of Energy [National Nuclear Security Administration] under contract 89233119CNA000063.

This report was prepared as an account of work sponsored by an agency of the U.S. Government. Neither the U.S. Government nor any agency thereof, nor any of their employees, makes any warranty, express or implied, or assumes any legal liability or responsibility for the accuracy, completeness, or usefulness of any information, apparatus, product, or process disclosed, or represents that its use would not infringe privately owned rights. Reference herein to any specific commercial product, process, or service by trade name, trademark, manufacturer, or otherwise does not necessarily constitute or imply its endorsement, recommendation, or favoring by the U.S. Government or any agency thereof. The views and opinions of authors expressed herein do not necessarily state or reflect those of the U.S. Government or any agency thereof.
\end{acknowledgments}


\begin{thebibliography}{33}%
\makeatletter
\providecommand \@ifxundefined [1]{%
 \@ifx{#1\undefined}
}%
\providecommand \@ifnum [1]{%
 \ifnum #1\expandafter \@firstoftwo
 \else \expandafter \@secondoftwo
 \fi
}%
\providecommand \@ifx [1]{%
 \ifx #1\expandafter \@firstoftwo
 \else \expandafter \@secondoftwo
 \fi
}%
\providecommand \natexlab [1]{#1}%
\providecommand \enquote  [1]{``#1''}%
\providecommand \bibnamefont  [1]{#1}%
\providecommand \bibfnamefont [1]{#1}%
\providecommand \citenamefont [1]{#1}%
\providecommand \href@noop [0]{\@secondoftwo}%
\providecommand \href [0]{\begingroup \@sanitize@url \@href}%
\providecommand \@href[1]{\@@startlink{#1}\@@href}%
\providecommand \@@href[1]{\endgroup#1\@@endlink}%
\providecommand \@sanitize@url [0]{\catcode `\\12\catcode `\$12\catcode
  `\&12\catcode `\#12\catcode `\^12\catcode `\_12\catcode `\%12\relax}%
\providecommand \@@startlink[1]{}%
\providecommand \@@endlink[0]{}%
\providecommand \url  [0]{\begingroup\@sanitize@url \@url }%
\providecommand \@url [1]{\endgroup\@href {#1}{\urlprefix }}%
\providecommand \urlprefix  [0]{URL }%
\providecommand \Eprint [0]{\href }%
\providecommand \doibase [0]{https://doi.org/}%
\providecommand \selectlanguage [0]{\@gobble}%
\providecommand \bibinfo  [0]{\@secondoftwo}%
\providecommand \bibfield  [0]{\@secondoftwo}%
\providecommand \translation [1]{[#1]}%
\providecommand \BibitemOpen [0]{}%
\providecommand \bibitemStop [0]{}%
\providecommand \bibitemNoStop [0]{.\EOS\space}%
\providecommand \EOS [0]{\spacefactor3000\relax}%
\providecommand \BibitemShut  [1]{\csname bibitem#1\endcsname}%
\let\auto@bib@innerbib\@empty
\bibitem [{\citenamefont {S{\'{e}}guin}\ \emph {et~al.}(2003)\citenamefont
  {S{\'{e}}guin}, \citenamefont {Frenje}, \citenamefont {Li}, \citenamefont
  {Hicks}, \citenamefont {Kurebayashi}, \citenamefont {Rygg}, \citenamefont
  {Schwartz}, \citenamefont {Petrasso}, \citenamefont {Roberts}, \citenamefont
  {Soures}, \citenamefont {Meyerhofer}, \citenamefont {Sangster}, \citenamefont
  {Knauer}, \citenamefont {Sorce}, \citenamefont {Glebov}, \citenamefont
  {Stoeckl}, \citenamefont {Phillips}, \citenamefont {Leeper}, \citenamefont
  {Fletcher},\ and\ \citenamefont {Padalino}}]{Seguin2003spectrometry}%
  \BibitemOpen
  \bibfield  {author} {\bibinfo {author} {\bibfnamefont {F.~H.}\ \bibnamefont
  {S{\'{e}}guin}}, \bibinfo {author} {\bibfnamefont {J.~A.}\ \bibnamefont
  {Frenje}}, \bibinfo {author} {\bibfnamefont {C.~K.}\ \bibnamefont {Li}},
  \bibinfo {author} {\bibfnamefont {D.~G.}\ \bibnamefont {Hicks}}, \bibinfo
  {author} {\bibfnamefont {S.}~\bibnamefont {Kurebayashi}}, \bibinfo {author}
  {\bibfnamefont {J.~R.}\ \bibnamefont {Rygg}}, \bibinfo {author}
  {\bibfnamefont {B.-E.}\ \bibnamefont {Schwartz}}, \bibinfo {author}
  {\bibfnamefont {R.~D.}\ \bibnamefont {Petrasso}}, \bibinfo {author}
  {\bibfnamefont {S.}~\bibnamefont {Roberts}}, \bibinfo {author} {\bibfnamefont
  {J.~M.}\ \bibnamefont {Soures}}, \bibinfo {author} {\bibfnamefont {D.~D.}\
  \bibnamefont {Meyerhofer}}, \bibinfo {author} {\bibfnamefont {T.~C.}\
  \bibnamefont {Sangster}}, \bibinfo {author} {\bibfnamefont {J.~P.}\
  \bibnamefont {Knauer}}, \bibinfo {author} {\bibfnamefont {C.}~\bibnamefont
  {Sorce}}, \bibinfo {author} {\bibfnamefont {V.~Y.}\ \bibnamefont {Glebov}},
  \bibinfo {author} {\bibfnamefont {C.}~\bibnamefont {Stoeckl}}, \bibinfo
  {author} {\bibfnamefont {T.~W.}\ \bibnamefont {Phillips}}, \bibinfo {author}
  {\bibfnamefont {R.~J.}\ \bibnamefont {Leeper}}, \bibinfo {author}
  {\bibfnamefont {K.}~\bibnamefont {Fletcher}},\ and\ \bibinfo {author}
  {\bibfnamefont {S.}~\bibnamefont {Padalino}},\ }\bibfield  {title} {\bibinfo
  {title} {Spectrometry of charged particles from inertial-confinement-fusion
  plasmas},\ }\href {https://doi.org/10.1063/1.1518141} {\bibfield  {journal}
  {\bibinfo  {journal} {Rev. Sci. Instrum.}\ }\textbf {\bibinfo {volume}
  {74}},\ \bibinfo {pages} {975} (\bibinfo {year} {2003})}\BibitemShut
  {NoStop}%
\bibitem [{\citenamefont {Regan}\ \emph {et~al.}(2016)\citenamefont {Regan},
  \citenamefont {Goncharov}, \citenamefont {Igumenshchev}, \citenamefont
  {Sangster}, \citenamefont {Betti}, \citenamefont {Bose}, \citenamefont
  {Boehly}, \citenamefont {Bonino}, \citenamefont {Campbell}, \citenamefont
  {Cao}, \citenamefont {Collins}, \citenamefont {Craxton}, \citenamefont
  {Davis}, \citenamefont {Delettrez}, \citenamefont {Edgell}, \citenamefont
  {Epstein}, \citenamefont {Forrest}, \citenamefont {Frenje}, \citenamefont
  {Froula}, \citenamefont {Gatu~Johnson}, \citenamefont {Glebov}, \citenamefont
  {Harding}, \citenamefont {Hohenberger}, \citenamefont {Hu}, \citenamefont
  {Jacobs-Perkins}, \citenamefont {Janezic}, \citenamefont {Karasik},
  \citenamefont {Keck}, \citenamefont {Kelly}, \citenamefont {Kessler},
  \citenamefont {Knauer}, \citenamefont {Kosc}, \citenamefont {Loucks},
  \citenamefont {Marozas}, \citenamefont {Marshall}, \citenamefont {McCrory},
  \citenamefont {McKenty}, \citenamefont {Meyerhofer}, \citenamefont {Michel},
  \citenamefont {Myatt}, \citenamefont {Obenschain}, \citenamefont {Petrasso},
  \citenamefont {Radha}, \citenamefont {Rice}, \citenamefont {Rosenberg},
  \citenamefont {Schmitt}, \citenamefont {Schmitt}, \citenamefont {Seka},
  \citenamefont {Shmayda}, \citenamefont {Shoup}, \citenamefont {Shvydky},
  \citenamefont {Skupsky}, \citenamefont {Solodov}, \citenamefont {Stoeckl},
  \citenamefont {Theobald}, \citenamefont {Ulreich}, \citenamefont {Wittman},
  \citenamefont {Woo}, \citenamefont {Yaakobi},\ and\ \citenamefont
  {Zuegel}}]{Regan2016demonstration}%
  \BibitemOpen
  \bibfield  {author} {\bibinfo {author} {\bibfnamefont {S.}~\bibnamefont
  {Regan}}, \bibinfo {author} {\bibfnamefont {V.}~\bibnamefont {Goncharov}},
  \bibinfo {author} {\bibfnamefont {I.}~\bibnamefont {Igumenshchev}}, \bibinfo
  {author} {\bibfnamefont {T.}~\bibnamefont {Sangster}}, \bibinfo {author}
  {\bibfnamefont {R.}~\bibnamefont {Betti}}, \bibinfo {author} {\bibfnamefont
  {A.}~\bibnamefont {Bose}}, \bibinfo {author} {\bibfnamefont {T.}~\bibnamefont
  {Boehly}}, \bibinfo {author} {\bibfnamefont {M.}~\bibnamefont {Bonino}},
  \bibinfo {author} {\bibfnamefont {E.}~\bibnamefont {Campbell}}, \bibinfo
  {author} {\bibfnamefont {D.}~\bibnamefont {Cao}}, \bibinfo {author}
  {\bibfnamefont {T.}~\bibnamefont {Collins}}, \bibinfo {author} {\bibfnamefont
  {R.}~\bibnamefont {Craxton}}, \bibinfo {author} {\bibfnamefont
  {A.}~\bibnamefont {Davis}}, \bibinfo {author} {\bibfnamefont
  {J.}~\bibnamefont {Delettrez}}, \bibinfo {author} {\bibfnamefont
  {D.}~\bibnamefont {Edgell}}, \bibinfo {author} {\bibfnamefont
  {R.}~\bibnamefont {Epstein}}, \bibinfo {author} {\bibfnamefont
  {C.}~\bibnamefont {Forrest}}, \bibinfo {author} {\bibfnamefont
  {J.}~\bibnamefont {Frenje}}, \bibinfo {author} {\bibfnamefont
  {D.}~\bibnamefont {Froula}}, \bibinfo {author} {\bibfnamefont
  {M.}~\bibnamefont {Gatu~Johnson}}, \bibinfo {author} {\bibfnamefont
  {V.}~\bibnamefont {Glebov}}, \bibinfo {author} {\bibfnamefont
  {D.}~\bibnamefont {Harding}}, \bibinfo {author} {\bibfnamefont
  {M.}~\bibnamefont {Hohenberger}}, \bibinfo {author} {\bibfnamefont
  {S.}~\bibnamefont {Hu}}, \bibinfo {author} {\bibfnamefont {D.}~\bibnamefont
  {Jacobs-Perkins}}, \bibinfo {author} {\bibfnamefont {R.}~\bibnamefont
  {Janezic}}, \bibinfo {author} {\bibfnamefont {M.}~\bibnamefont {Karasik}},
  \bibinfo {author} {\bibfnamefont {R.}~\bibnamefont {Keck}}, \bibinfo {author}
  {\bibfnamefont {J.}~\bibnamefont {Kelly}}, \bibinfo {author} {\bibfnamefont
  {T.}~\bibnamefont {Kessler}}, \bibinfo {author} {\bibfnamefont
  {J.}~\bibnamefont {Knauer}}, \bibinfo {author} {\bibfnamefont
  {T.}~\bibnamefont {Kosc}}, \bibinfo {author} {\bibfnamefont {S.}~\bibnamefont
  {Loucks}}, \bibinfo {author} {\bibfnamefont {J.}~\bibnamefont {Marozas}},
  \bibinfo {author} {\bibfnamefont {F.}~\bibnamefont {Marshall}}, \bibinfo
  {author} {\bibfnamefont {R.}~\bibnamefont {McCrory}}, \bibinfo {author}
  {\bibfnamefont {P.}~\bibnamefont {McKenty}}, \bibinfo {author} {\bibfnamefont
  {D.}~\bibnamefont {Meyerhofer}}, \bibinfo {author} {\bibfnamefont
  {D.}~\bibnamefont {Michel}}, \bibinfo {author} {\bibfnamefont
  {J.}~\bibnamefont {Myatt}}, \bibinfo {author} {\bibfnamefont
  {S.}~\bibnamefont {Obenschain}}, \bibinfo {author} {\bibfnamefont
  {R.}~\bibnamefont {Petrasso}}, \bibinfo {author} {\bibfnamefont
  {P.}~\bibnamefont {Radha}}, \bibinfo {author} {\bibfnamefont
  {B.}~\bibnamefont {Rice}}, \bibinfo {author} {\bibfnamefont {M.}~\bibnamefont
  {Rosenberg}}, \bibinfo {author} {\bibfnamefont {A.}~\bibnamefont {Schmitt}},
  \bibinfo {author} {\bibfnamefont {M.}~\bibnamefont {Schmitt}}, \bibinfo
  {author} {\bibfnamefont {W.}~\bibnamefont {Seka}}, \bibinfo {author}
  {\bibfnamefont {W.}~\bibnamefont {Shmayda}}, \bibinfo {author} {\bibfnamefont
  {M.}~\bibnamefont {Shoup}}, \bibinfo {author} {\bibfnamefont
  {A.}~\bibnamefont {Shvydky}}, \bibinfo {author} {\bibfnamefont
  {S.}~\bibnamefont {Skupsky}}, \bibinfo {author} {\bibfnamefont
  {A.}~\bibnamefont {Solodov}}, \bibinfo {author} {\bibfnamefont
  {C.}~\bibnamefont {Stoeckl}}, \bibinfo {author} {\bibfnamefont
  {W.}~\bibnamefont {Theobald}}, \bibinfo {author} {\bibfnamefont
  {J.}~\bibnamefont {Ulreich}}, \bibinfo {author} {\bibfnamefont
  {M.}~\bibnamefont {Wittman}}, \bibinfo {author} {\bibfnamefont
  {K.}~\bibnamefont {Woo}}, \bibinfo {author} {\bibfnamefont {B.}~\bibnamefont
  {Yaakobi}},\ and\ \bibinfo {author} {\bibfnamefont {J.}~\bibnamefont
  {Zuegel}},\ }\bibfield  {title} {\bibinfo {title} {Demonstration of fuel
  hot-spot pressure in excess of 50 gbar for direct-drive, layered
  deuterium-tritium implosions on omega},\ }\href
  {https://doi.org/10.1103/physrevlett.117.025001} {\bibfield  {journal}
  {\bibinfo  {journal} {Physical Review Letters}\ }\textbf {\bibinfo {volume}
  {117}},\ \bibinfo {pages} {025001} (\bibinfo {year} {2016})}\BibitemShut
  {NoStop}%
\bibitem [{\citenamefont {Igumenshchev}\ \emph {et~al.}(2017)\citenamefont
  {Igumenshchev}, \citenamefont {Michel}, \citenamefont {Shah}, \citenamefont
  {Campbell}, \citenamefont {Epstein}, \citenamefont {Forrest}, \citenamefont
  {Glebov}, \citenamefont {Goncharov}, \citenamefont {Knauer}, \citenamefont
  {Marshall}, \citenamefont {McCrory}, \citenamefont {Regan}, \citenamefont
  {Sangster}, \citenamefont {Stoeckl}, \citenamefont {Schmitt},\ and\
  \citenamefont {Obenschain}}]{Igumenshchev2017three}%
  \BibitemOpen
  \bibfield  {author} {\bibinfo {author} {\bibfnamefont {I.~V.}\ \bibnamefont
  {Igumenshchev}}, \bibinfo {author} {\bibfnamefont {D.~T.}\ \bibnamefont
  {Michel}}, \bibinfo {author} {\bibfnamefont {R.~C.}\ \bibnamefont {Shah}},
  \bibinfo {author} {\bibfnamefont {E.~M.}\ \bibnamefont {Campbell}}, \bibinfo
  {author} {\bibfnamefont {R.}~\bibnamefont {Epstein}}, \bibinfo {author}
  {\bibfnamefont {C.~J.}\ \bibnamefont {Forrest}}, \bibinfo {author}
  {\bibfnamefont {V.~Y.}\ \bibnamefont {Glebov}}, \bibinfo {author}
  {\bibfnamefont {V.~N.}\ \bibnamefont {Goncharov}}, \bibinfo {author}
  {\bibfnamefont {J.~P.}\ \bibnamefont {Knauer}}, \bibinfo {author}
  {\bibfnamefont {F.~J.}\ \bibnamefont {Marshall}}, \bibinfo {author}
  {\bibfnamefont {R.~L.}\ \bibnamefont {McCrory}}, \bibinfo {author}
  {\bibfnamefont {S.~P.}\ \bibnamefont {Regan}}, \bibinfo {author}
  {\bibfnamefont {T.~C.}\ \bibnamefont {Sangster}}, \bibinfo {author}
  {\bibfnamefont {C.}~\bibnamefont {Stoeckl}}, \bibinfo {author} {\bibfnamefont
  {A.~J.}\ \bibnamefont {Schmitt}},\ and\ \bibinfo {author} {\bibfnamefont
  {S.}~\bibnamefont {Obenschain}},\ }\bibfield  {title} {\bibinfo {title}
  {Three-dimensional hydrodynamic simulations of omega implosions},\ }\bibfield
   {journal} {\bibinfo  {journal} {Physics of Plasmas}\ }\textbf {\bibinfo
  {volume} {24}},\ \href {https://doi.org/10.1063/1.4979195}
  {10.1063/1.4979195} (\bibinfo {year} {2017})\BibitemShut {NoStop}%
\bibitem [{\citenamefont {Gatu~Johnson}\ \emph {et~al.}(2019)\citenamefont
  {Gatu~Johnson}, \citenamefont {Appelbe}, \citenamefont {Chittenden},
  \citenamefont {Crilly}, \citenamefont {Delettrez}, \citenamefont {Forrest},
  \citenamefont {Frenje}, \citenamefont {Glebov}, \citenamefont {Grimble},
  \citenamefont {Haines}, \citenamefont {Igumenshchev}, \citenamefont
  {Janezic}, \citenamefont {Knauer}, \citenamefont {Lahmann}, \citenamefont
  {Marshall}, \citenamefont {Michel}, \citenamefont {Séguin}, \citenamefont
  {Stoeckl}, \citenamefont {Walsh}, \citenamefont {Zylstra},\ and\
  \citenamefont {Petrasso}}]{GatuJohnson2019impact}%
  \BibitemOpen
  \bibfield  {author} {\bibinfo {author} {\bibfnamefont {M.}~\bibnamefont
  {Gatu~Johnson}}, \bibinfo {author} {\bibfnamefont {B.~D.}\ \bibnamefont
  {Appelbe}}, \bibinfo {author} {\bibfnamefont {J.~P.}\ \bibnamefont
  {Chittenden}}, \bibinfo {author} {\bibfnamefont {A.}~\bibnamefont {Crilly}},
  \bibinfo {author} {\bibfnamefont {J.}~\bibnamefont {Delettrez}}, \bibinfo
  {author} {\bibfnamefont {C.}~\bibnamefont {Forrest}}, \bibinfo {author}
  {\bibfnamefont {J.~A.}\ \bibnamefont {Frenje}}, \bibinfo {author}
  {\bibfnamefont {V.~Y.}\ \bibnamefont {Glebov}}, \bibinfo {author}
  {\bibfnamefont {W.}~\bibnamefont {Grimble}}, \bibinfo {author} {\bibfnamefont
  {B.~M.}\ \bibnamefont {Haines}}, \bibinfo {author} {\bibfnamefont {I.~V.}\
  \bibnamefont {Igumenshchev}}, \bibinfo {author} {\bibfnamefont
  {R.}~\bibnamefont {Janezic}}, \bibinfo {author} {\bibfnamefont {J.~P.}\
  \bibnamefont {Knauer}}, \bibinfo {author} {\bibfnamefont {B.}~\bibnamefont
  {Lahmann}}, \bibinfo {author} {\bibfnamefont {F.~J.}\ \bibnamefont
  {Marshall}}, \bibinfo {author} {\bibfnamefont {T.}~\bibnamefont {Michel}},
  \bibinfo {author} {\bibfnamefont {F.~H.}\ \bibnamefont {Séguin}}, \bibinfo
  {author} {\bibfnamefont {C.}~\bibnamefont {Stoeckl}}, \bibinfo {author}
  {\bibfnamefont {C.}~\bibnamefont {Walsh}}, \bibinfo {author} {\bibfnamefont
  {A.~B.}\ \bibnamefont {Zylstra}},\ and\ \bibinfo {author} {\bibfnamefont
  {R.~D.}\ \bibnamefont {Petrasso}},\ }\bibfield  {title} {\bibinfo {title}
  {Impact of imposed mode 2 laser drive asymmetry on inertial confinement
  fusion implosions},\ }\bibfield  {journal} {\bibinfo  {journal} {Physics of
  Plasmas}\ }\textbf {\bibinfo {volume} {26}},\ \href
  {https://doi.org/10.1063/1.5066435} {10.1063/1.5066435} (\bibinfo {year}
  {2019})\BibitemShut {NoStop}%
\bibitem [{\citenamefont {Colaïtis}\ \emph {et~al.}(2022)\citenamefont
  {Colaïtis}, \citenamefont {Turnbull}, \citenamefont {Igumenschev},
  \citenamefont {Edgell}, \citenamefont {Shah}, \citenamefont {Mannion},
  \citenamefont {Stoeckl}, \citenamefont {Jacob-Perkins}, \citenamefont
  {Shvydky}, \citenamefont {Janezic}, \citenamefont {Kalb}, \citenamefont
  {Cao}, \citenamefont {Forrest}, \citenamefont {Kwiatkowski}, \citenamefont
  {Regan}, \citenamefont {Theobald}, \citenamefont {Goncharov},\ and\
  \citenamefont {Froula}}]{Colaitis20223d}%
  \BibitemOpen
  \bibfield  {author} {\bibinfo {author} {\bibfnamefont {A.}~\bibnamefont
  {Colaïtis}}, \bibinfo {author} {\bibfnamefont {D.}~\bibnamefont {Turnbull}},
  \bibinfo {author} {\bibfnamefont {I.}~\bibnamefont {Igumenschev}}, \bibinfo
  {author} {\bibfnamefont {D.}~\bibnamefont {Edgell}}, \bibinfo {author}
  {\bibfnamefont {R.}~\bibnamefont {Shah}}, \bibinfo {author} {\bibfnamefont
  {O.}~\bibnamefont {Mannion}}, \bibinfo {author} {\bibfnamefont
  {C.}~\bibnamefont {Stoeckl}}, \bibinfo {author} {\bibfnamefont
  {D.}~\bibnamefont {Jacob-Perkins}}, \bibinfo {author} {\bibfnamefont
  {A.}~\bibnamefont {Shvydky}}, \bibinfo {author} {\bibfnamefont
  {R.}~\bibnamefont {Janezic}}, \bibinfo {author} {\bibfnamefont
  {A.}~\bibnamefont {Kalb}}, \bibinfo {author} {\bibfnamefont {D.}~\bibnamefont
  {Cao}}, \bibinfo {author} {\bibfnamefont {C.}~\bibnamefont {Forrest}},
  \bibinfo {author} {\bibfnamefont {J.}~\bibnamefont {Kwiatkowski}}, \bibinfo
  {author} {\bibfnamefont {S.}~\bibnamefont {Regan}}, \bibinfo {author}
  {\bibfnamefont {W.}~\bibnamefont {Theobald}}, \bibinfo {author}
  {\bibfnamefont {V.}~\bibnamefont {Goncharov}},\ and\ \bibinfo {author}
  {\bibfnamefont {D.}~\bibnamefont {Froula}},\ }\bibfield  {title} {\bibinfo
  {title} {3d simulations capture the persistent low-mode asymmetries evident
  in laser-direct-drive implosions on {OMEGA}},\ }\href
  {https://doi.org/10.1103/physrevlett.129.095001} {\bibfield  {journal}
  {\bibinfo  {journal} {Phys. Rev. Lett.}\ }\textbf {\bibinfo {volume} {129}},\
  \bibinfo {pages} {095001} (\bibinfo {year} {2022})}\BibitemShut {NoStop}%
\bibitem [{\citenamefont {S{\'{e}}guin}\ \emph {et~al.}(2004)\citenamefont
  {S{\'{e}}guin}, \citenamefont {DeCiantis}, \citenamefont {Frenje},
  \citenamefont {Kurebayashi}, \citenamefont {Li}, \citenamefont {Rygg},
  \citenamefont {Chen}, \citenamefont {Berube}, \citenamefont {Schwartz},
  \citenamefont {Petrasso}, \citenamefont {Smalyuk}, \citenamefont {Marshall},
  \citenamefont {Knauer}, \citenamefont {Delettrez}, \citenamefont {McKenty},
  \citenamefont {Meyerhofer}, \citenamefont {Roberts}, \citenamefont
  {Sangster}, \citenamefont {Mikaelian},\ and\ \citenamefont
  {Park}}]{Seguin2004d3he}%
  \BibitemOpen
  \bibfield  {author} {\bibinfo {author} {\bibfnamefont {F.~H.}\ \bibnamefont
  {S{\'{e}}guin}}, \bibinfo {author} {\bibfnamefont {J.~L.}\ \bibnamefont
  {DeCiantis}}, \bibinfo {author} {\bibfnamefont {J.~A.}\ \bibnamefont
  {Frenje}}, \bibinfo {author} {\bibfnamefont {S.}~\bibnamefont {Kurebayashi}},
  \bibinfo {author} {\bibfnamefont {C.~K.}\ \bibnamefont {Li}}, \bibinfo
  {author} {\bibfnamefont {J.~R.}\ \bibnamefont {Rygg}}, \bibinfo {author}
  {\bibfnamefont {C.}~\bibnamefont {Chen}}, \bibinfo {author} {\bibfnamefont
  {V.}~\bibnamefont {Berube}}, \bibinfo {author} {\bibfnamefont {B.~E.}\
  \bibnamefont {Schwartz}}, \bibinfo {author} {\bibfnamefont {R.~D.}\
  \bibnamefont {Petrasso}}, \bibinfo {author} {\bibfnamefont {V.~A.}\
  \bibnamefont {Smalyuk}}, \bibinfo {author} {\bibfnamefont {F.~J.}\
  \bibnamefont {Marshall}}, \bibinfo {author} {\bibfnamefont {J.~P.}\
  \bibnamefont {Knauer}}, \bibinfo {author} {\bibfnamefont {J.~A.}\
  \bibnamefont {Delettrez}}, \bibinfo {author} {\bibfnamefont {P.~W.}\
  \bibnamefont {McKenty}}, \bibinfo {author} {\bibfnamefont {D.~D.}\
  \bibnamefont {Meyerhofer}}, \bibinfo {author} {\bibfnamefont
  {S.}~\bibnamefont {Roberts}}, \bibinfo {author} {\bibfnamefont {T.~C.}\
  \bibnamefont {Sangster}}, \bibinfo {author} {\bibfnamefont {K.}~\bibnamefont
  {Mikaelian}},\ and\ \bibinfo {author} {\bibfnamefont {H.~S.}\ \bibnamefont
  {Park}},\ }\bibfield  {title} {\bibinfo {title} {D3he-proton emission imaging
  for inertial-confinement-fusion experiments (invited)},\ }\href
  {https://doi.org/10.1063/1.1788892} {\bibfield  {journal} {\bibinfo
  {journal} {Rev. Sci. Instrum.}\ }\textbf {\bibinfo {volume} {75}},\ \bibinfo
  {pages} {3520} (\bibinfo {year} {2004})}\BibitemShut {NoStop}%
\bibitem [{\citenamefont {DeCiantis}\ \emph {et~al.}(2006)\citenamefont
  {DeCiantis}, \citenamefont {Séguin}, \citenamefont {Frenje}, \citenamefont
  {Berube}, \citenamefont {Canavan}, \citenamefont {Chen}, \citenamefont
  {Kurebayashi}, \citenamefont {Li}, \citenamefont {Rygg}, \citenamefont
  {Schwartz}, \citenamefont {Petrasso}, \citenamefont {Delettrez},
  \citenamefont {Regan}, \citenamefont {Smalyuk}, \citenamefont {Knauer},
  \citenamefont {Marshall}, \citenamefont {Meyerhofer}, \citenamefont
  {Roberts}, \citenamefont {Sangster}, \citenamefont {Stoeckl}, \citenamefont
  {Mikaelian}, \citenamefont {Park},\ and\ \citenamefont
  {Robey}}]{DeCiantis2006proton}%
  \BibitemOpen
  \bibfield  {author} {\bibinfo {author} {\bibfnamefont {J.~L.}\ \bibnamefont
  {DeCiantis}}, \bibinfo {author} {\bibfnamefont {F.~H.}\ \bibnamefont
  {Séguin}}, \bibinfo {author} {\bibfnamefont {J.~A.}\ \bibnamefont {Frenje}},
  \bibinfo {author} {\bibfnamefont {V.}~\bibnamefont {Berube}}, \bibinfo
  {author} {\bibfnamefont {M.~J.}\ \bibnamefont {Canavan}}, \bibinfo {author}
  {\bibfnamefont {C.~D.}\ \bibnamefont {Chen}}, \bibinfo {author}
  {\bibfnamefont {S.}~\bibnamefont {Kurebayashi}}, \bibinfo {author}
  {\bibfnamefont {C.~K.}\ \bibnamefont {Li}}, \bibinfo {author} {\bibfnamefont
  {J.~R.}\ \bibnamefont {Rygg}}, \bibinfo {author} {\bibfnamefont {B.~E.}\
  \bibnamefont {Schwartz}}, \bibinfo {author} {\bibfnamefont {R.~D.}\
  \bibnamefont {Petrasso}}, \bibinfo {author} {\bibfnamefont {J.~A.}\
  \bibnamefont {Delettrez}}, \bibinfo {author} {\bibfnamefont {S.~P.}\
  \bibnamefont {Regan}}, \bibinfo {author} {\bibfnamefont {V.~A.}\ \bibnamefont
  {Smalyuk}}, \bibinfo {author} {\bibfnamefont {J.~P.}\ \bibnamefont {Knauer}},
  \bibinfo {author} {\bibfnamefont {F.~J.}\ \bibnamefont {Marshall}}, \bibinfo
  {author} {\bibfnamefont {D.~D.}\ \bibnamefont {Meyerhofer}}, \bibinfo
  {author} {\bibfnamefont {S.}~\bibnamefont {Roberts}}, \bibinfo {author}
  {\bibfnamefont {T.~C.}\ \bibnamefont {Sangster}}, \bibinfo {author}
  {\bibfnamefont {C.}~\bibnamefont {Stoeckl}}, \bibinfo {author} {\bibfnamefont
  {K.}~\bibnamefont {Mikaelian}}, \bibinfo {author} {\bibfnamefont {H.~S.}\
  \bibnamefont {Park}},\ and\ \bibinfo {author} {\bibfnamefont {H.~F.}\
  \bibnamefont {Robey}},\ }\bibfield  {title} {\bibinfo {title} {Proton core
  imaging of the nuclear burn in inertial confinement fusion implosions},\
  }\bibfield  {journal} {\bibinfo  {journal} {Review of Scientific
  Instruments}\ }\textbf {\bibinfo {volume} {77}},\ \href
  {https://doi.org/10.1063/1.2173788} {10.1063/1.2173788} (\bibinfo {year}
  {2006})\BibitemShut {NoStop}%
\bibitem [{\citenamefont {Rinderknecht}\ \emph {et~al.}(2022)\citenamefont
  {Rinderknecht}, \citenamefont {Heuer}, \citenamefont {Kunimune},
  \citenamefont {Adrian}, \citenamefont {Knauer}, \citenamefont {Theobald},
  \citenamefont {Fairbanks}, \citenamefont {Brannon}, \citenamefont
  {Ceurvorst}, \citenamefont {Gopalaswamy}, \citenamefont {Williams},
  \citenamefont {Radha}, \citenamefont {Regan}, \citenamefont {Johnson},
  \citenamefont {S{\'{e}}guin},\ and\ \citenamefont
  {Frenje}}]{Rinderknecht2022knock}%
  \BibitemOpen
  \bibfield  {author} {\bibinfo {author} {\bibfnamefont {H.~G.}\ \bibnamefont
  {Rinderknecht}}, \bibinfo {author} {\bibfnamefont {P.~V.}\ \bibnamefont
  {Heuer}}, \bibinfo {author} {\bibfnamefont {J.}~\bibnamefont {Kunimune}},
  \bibinfo {author} {\bibfnamefont {P.~J.}\ \bibnamefont {Adrian}}, \bibinfo
  {author} {\bibfnamefont {J.~P.}\ \bibnamefont {Knauer}}, \bibinfo {author}
  {\bibfnamefont {W.}~\bibnamefont {Theobald}}, \bibinfo {author}
  {\bibfnamefont {R.}~\bibnamefont {Fairbanks}}, \bibinfo {author}
  {\bibfnamefont {B.}~\bibnamefont {Brannon}}, \bibinfo {author} {\bibfnamefont
  {L.}~\bibnamefont {Ceurvorst}}, \bibinfo {author} {\bibfnamefont
  {V.}~\bibnamefont {Gopalaswamy}}, \bibinfo {author} {\bibfnamefont {C.~A.}\
  \bibnamefont {Williams}}, \bibinfo {author} {\bibfnamefont {P.~B.}\
  \bibnamefont {Radha}}, \bibinfo {author} {\bibfnamefont {S.~P.}\ \bibnamefont
  {Regan}}, \bibinfo {author} {\bibfnamefont {M.~G.}\ \bibnamefont {Johnson}},
  \bibinfo {author} {\bibfnamefont {F.~H.}\ \bibnamefont {S{\'{e}}guin}},\ and\
  \bibinfo {author} {\bibfnamefont {J.~A.}\ \bibnamefont {Frenje}},\ }\bibfield
   {title} {\bibinfo {title} {A knock-on deuteron imager for measurements of
  fuel and hotspot asymmetry in direct-drive inertial confinement fusion
  implosions (invited)},\ }\href {https://doi.org/10.1063/5.0099301} {\bibfield
   {journal} {\bibinfo  {journal} {Rev. Sci. Instrum.}\ }\textbf {\bibinfo
  {volume} {93}},\ \bibinfo {pages} {093507} (\bibinfo {year}
  {2022})}\BibitemShut {NoStop}%
\bibitem [{\citenamefont {Kunimune}\ \emph {et~al.}(2022)\citenamefont
  {Kunimune}, \citenamefont {Rinderknecht}, \citenamefont {Adrian},
  \citenamefont {Heuer}, \citenamefont {Regan}, \citenamefont {S{\'{e}}guin},
  \citenamefont {Johnson}, \citenamefont {Bahukutumbi}, \citenamefont {Knauer},
  \citenamefont {Bachmann},\ and\ \citenamefont
  {Frenje}}]{Kunimune2022knockon}%
  \BibitemOpen
  \bibfield  {author} {\bibinfo {author} {\bibfnamefont {J.~H.}\ \bibnamefont
  {Kunimune}}, \bibinfo {author} {\bibfnamefont {H.~G.}\ \bibnamefont
  {Rinderknecht}}, \bibinfo {author} {\bibfnamefont {P.~J.}\ \bibnamefont
  {Adrian}}, \bibinfo {author} {\bibfnamefont {P.~V.}\ \bibnamefont {Heuer}},
  \bibinfo {author} {\bibfnamefont {S.~P.}\ \bibnamefont {Regan}}, \bibinfo
  {author} {\bibfnamefont {F.~H.}\ \bibnamefont {S{\'{e}}guin}}, \bibinfo
  {author} {\bibfnamefont {M.~G.}\ \bibnamefont {Johnson}}, \bibinfo {author}
  {\bibfnamefont {R.~P.}\ \bibnamefont {Bahukutumbi}}, \bibinfo {author}
  {\bibfnamefont {J.~P.}\ \bibnamefont {Knauer}}, \bibinfo {author}
  {\bibfnamefont {B.~L.}\ \bibnamefont {Bachmann}},\ and\ \bibinfo {author}
  {\bibfnamefont {J.~A.}\ \bibnamefont {Frenje}},\ }\bibfield  {title}
  {\bibinfo {title} {Knock-on deuteron imaging for diagnosing the morphology of
  an {ICF} implosion at {OMEGA}},\ }\href {https://doi.org/10.1063/5.0096786}
  {\bibfield  {journal} {\bibinfo  {journal} {Phys. Plasmas}\ }\textbf
  {\bibinfo {volume} {29}},\ \bibinfo {pages} {072711} (\bibinfo {year}
  {2022})}\BibitemShut {NoStop}%
\bibitem [{\citenamefont {Delettrez}\ \emph {et~al.}(1987)\citenamefont
  {Delettrez}, \citenamefont {Epstein}, \citenamefont {Richardson},
  \citenamefont {Jaanimagi},\ and\ \citenamefont
  {Henke}}]{Delettrez1987effect}%
  \BibitemOpen
  \bibfield  {author} {\bibinfo {author} {\bibfnamefont {J.}~\bibnamefont
  {Delettrez}}, \bibinfo {author} {\bibfnamefont {R.}~\bibnamefont {Epstein}},
  \bibinfo {author} {\bibfnamefont {M.~C.}\ \bibnamefont {Richardson}},
  \bibinfo {author} {\bibfnamefont {P.~A.}\ \bibnamefont {Jaanimagi}},\ and\
  \bibinfo {author} {\bibfnamefont {B.~L.}\ \bibnamefont {Henke}},\ }\bibfield
  {title} {\bibinfo {title} {Effect of laser illumination nonuniformity on the
  analysis of time-resolved x-ray measurements in uv spherical transport
  experiments},\ }\href {https://doi.org/10.1103/physreva.36.3926} {\bibfield
  {journal} {\bibinfo  {journal} {Physical Review A}\ }\textbf {\bibinfo
  {volume} {36}},\ \bibinfo {pages} {3926} (\bibinfo {year}
  {1987})}\BibitemShut {NoStop}%
\bibitem [{\citenamefont {Gopalaswamy}\ \emph {et~al.}(2022)\citenamefont
  {Gopalaswamy}, \citenamefont {Betti}, \citenamefont {Radha}, \citenamefont
  {Crilly}, \citenamefont {Woo}, \citenamefont {Lees}, \citenamefont {Thomas},
  \citenamefont {Igumenshchev}, \citenamefont {Miller}, \citenamefont {Knauer},
  \citenamefont {Stoeckl}, \citenamefont {Forrest}, \citenamefont {Mannion},
  \citenamefont {Mohamed}, \citenamefont {Rinderknecht},\ and\ \citenamefont
  {Heuer}}]{Gopalaswamy2022analysis}%
  \BibitemOpen
  \bibfield  {author} {\bibinfo {author} {\bibfnamefont {V.}~\bibnamefont
  {Gopalaswamy}}, \bibinfo {author} {\bibfnamefont {R.}~\bibnamefont {Betti}},
  \bibinfo {author} {\bibfnamefont {P.~B.}\ \bibnamefont {Radha}}, \bibinfo
  {author} {\bibfnamefont {A.~J.}\ \bibnamefont {Crilly}}, \bibinfo {author}
  {\bibfnamefont {K.~M.}\ \bibnamefont {Woo}}, \bibinfo {author} {\bibfnamefont
  {A.}~\bibnamefont {Lees}}, \bibinfo {author} {\bibfnamefont {C.}~\bibnamefont
  {Thomas}}, \bibinfo {author} {\bibfnamefont {I.~V.}\ \bibnamefont
  {Igumenshchev}}, \bibinfo {author} {\bibfnamefont {S.~C.}\ \bibnamefont
  {Miller}}, \bibinfo {author} {\bibfnamefont {J.~P.}\ \bibnamefont {Knauer}},
  \bibinfo {author} {\bibfnamefont {C.}~\bibnamefont {Stoeckl}}, \bibinfo
  {author} {\bibfnamefont {C.~J.}\ \bibnamefont {Forrest}}, \bibinfo {author}
  {\bibfnamefont {O.~M.}\ \bibnamefont {Mannion}}, \bibinfo {author}
  {\bibfnamefont {Z.~L.}\ \bibnamefont {Mohamed}}, \bibinfo {author}
  {\bibfnamefont {H.~G.}\ \bibnamefont {Rinderknecht}},\ and\ \bibinfo {author}
  {\bibfnamefont {P.~V.}\ \bibnamefont {Heuer}},\ }\bibfield  {title} {\bibinfo
  {title} {Analysis of limited coverage effects on areal density measurements
  in inertial confinement fusion implosions},\ }\href
  {https://doi.org/10.1063/5.0085942} {\bibfield  {journal} {\bibinfo
  {journal} {Phys. Plasmas}\ }\textbf {\bibinfo {volume} {29}},\ \bibinfo
  {pages} {072706} (\bibinfo {year} {2022})}\BibitemShut {NoStop}%
\bibitem [{\citenamefont {MacFarlane}\ \emph {et~al.}(2007)\citenamefont
  {MacFarlane}, \citenamefont {Golovkin}, \citenamefont {Wang}, \citenamefont
  {Woodruff},\ and\ \citenamefont {Pereyra}}]{MacFarlane2007spect3d}%
  \BibitemOpen
  \bibfield  {author} {\bibinfo {author} {\bibfnamefont {J.}~\bibnamefont
  {MacFarlane}}, \bibinfo {author} {\bibfnamefont {I.}~\bibnamefont
  {Golovkin}}, \bibinfo {author} {\bibfnamefont {P.}~\bibnamefont {Wang}},
  \bibinfo {author} {\bibfnamefont {P.}~\bibnamefont {Woodruff}},\ and\
  \bibinfo {author} {\bibfnamefont {N.}~\bibnamefont {Pereyra}},\ }\bibfield
  {title} {\bibinfo {title} {Spect3d – a multi-dimensional
  collisional-radiative code for generating diagnostic signatures based on
  hydrodynamics and pic simulation output},\ }\href
  {https://doi.org/10.1016/j.hedp.2007.02.016} {\bibfield  {journal} {\bibinfo
  {journal} {High Energy Density Physics}\ }\textbf {\bibinfo {volume} {3}},\
  \bibinfo {pages} {181} (\bibinfo {year} {2007})}\BibitemShut {NoStop}%
\bibitem [{\citenamefont {Cartwright}\ \emph {et~al.}(1978)\citenamefont
  {Cartwright}, \citenamefont {Shirk},\ and\ \citenamefont
  {Price}}]{Cartwright1978nuclear}%
  \BibitemOpen
  \bibfield  {author} {\bibinfo {author} {\bibfnamefont {B.}~\bibnamefont
  {Cartwright}}, \bibinfo {author} {\bibfnamefont {E.}~\bibnamefont {Shirk}},\
  and\ \bibinfo {author} {\bibfnamefont {P.}~\bibnamefont {Price}},\ }\bibfield
   {title} {\bibinfo {title} {A nuclear-track-recording polymer of unique
  sensitivity and resolution},\ }\href
  {https://doi.org/10.1016/0029-554x(78)90989-8} {\bibfield  {journal}
  {\bibinfo  {journal} {Nuclear Instruments and Methods}\ }\textbf {\bibinfo
  {volume} {153}},\ \bibinfo {pages} {457} (\bibinfo {year}
  {1978})}\BibitemShut {NoStop}%
\bibitem [{\citenamefont {Lahmann}\ \emph {et~al.}(2020)\citenamefont
  {Lahmann}, \citenamefont {Johnson}, \citenamefont {Frenje}, \citenamefont
  {Glebov}, \citenamefont {Rinderknecht}, \citenamefont {S{\'{e}}guin},
  \citenamefont {Sutcliffe},\ and\ \citenamefont {Petrasso}}]{Lahmann2020cr39}%
  \BibitemOpen
  \bibfield  {author} {\bibinfo {author} {\bibfnamefont {B.}~\bibnamefont
  {Lahmann}}, \bibinfo {author} {\bibfnamefont {M.~G.}\ \bibnamefont
  {Johnson}}, \bibinfo {author} {\bibfnamefont {J.~A.}\ \bibnamefont {Frenje}},
  \bibinfo {author} {\bibfnamefont {Y.~Y.}\ \bibnamefont {Glebov}}, \bibinfo
  {author} {\bibfnamefont {H.~G.}\ \bibnamefont {Rinderknecht}}, \bibinfo
  {author} {\bibfnamefont {F.~H.}\ \bibnamefont {S{\'{e}}guin}}, \bibinfo
  {author} {\bibfnamefont {G.}~\bibnamefont {Sutcliffe}},\ and\ \bibinfo
  {author} {\bibfnamefont {R.~D.}\ \bibnamefont {Petrasso}},\ }\bibfield
  {title} {\bibinfo {title} {{CR}-39 nuclear track detector response to
  inertial confinement fusion relevant ions},\ }\href
  {https://doi.org/10.1063/5.0004129} {\bibfield  {journal} {\bibinfo
  {journal} {Review of Scientific Instruments}\ }\textbf {\bibinfo {volume}
  {91}},\ \bibinfo {pages} {053502} (\bibinfo {year} {2020})}\BibitemShut
  {NoStop}%
\bibitem [{\citenamefont {Craxton}\ \emph {et~al.}(2015)\citenamefont
  {Craxton}, \citenamefont {Anderson}, \citenamefont {Boehly}, \citenamefont
  {Goncharov}, \citenamefont {Harding}, \citenamefont {Knauer}, \citenamefont
  {McCrory}, \citenamefont {McKenty}, \citenamefont {Meyerhofer}, \citenamefont
  {Myatt}, \citenamefont {Schmitt}, \citenamefont {Sethian}, \citenamefont
  {Short}, \citenamefont {Skupsky}, \citenamefont {Theobald}, \citenamefont
  {Kruer}, \citenamefont {Tanaka}, \citenamefont {Betti}, \citenamefont
  {Collins}, \citenamefont {Delettrez}, \citenamefont {Hu}, \citenamefont
  {Marozas}, \citenamefont {Maximov}, \citenamefont {Michel}, \citenamefont
  {Radha}, \citenamefont {Regan}, \citenamefont {Sangster}, \citenamefont
  {Seka}, \citenamefont {Solodov}, \citenamefont {Soures}, \citenamefont
  {Stoeckl},\ and\ \citenamefont {Zuegel}}]{Craxton2015direct}%
  \BibitemOpen
  \bibfield  {author} {\bibinfo {author} {\bibfnamefont {R.~S.}\ \bibnamefont
  {Craxton}}, \bibinfo {author} {\bibfnamefont {K.~S.}\ \bibnamefont
  {Anderson}}, \bibinfo {author} {\bibfnamefont {T.~R.}\ \bibnamefont
  {Boehly}}, \bibinfo {author} {\bibfnamefont {V.~N.}\ \bibnamefont
  {Goncharov}}, \bibinfo {author} {\bibfnamefont {D.~R.}\ \bibnamefont
  {Harding}}, \bibinfo {author} {\bibfnamefont {J.~P.}\ \bibnamefont {Knauer}},
  \bibinfo {author} {\bibfnamefont {R.~L.}\ \bibnamefont {McCrory}}, \bibinfo
  {author} {\bibfnamefont {P.~W.}\ \bibnamefont {McKenty}}, \bibinfo {author}
  {\bibfnamefont {D.~D.}\ \bibnamefont {Meyerhofer}}, \bibinfo {author}
  {\bibfnamefont {J.~F.}\ \bibnamefont {Myatt}}, \bibinfo {author}
  {\bibfnamefont {A.~J.}\ \bibnamefont {Schmitt}}, \bibinfo {author}
  {\bibfnamefont {J.~D.}\ \bibnamefont {Sethian}}, \bibinfo {author}
  {\bibfnamefont {R.~W.}\ \bibnamefont {Short}}, \bibinfo {author}
  {\bibfnamefont {S.}~\bibnamefont {Skupsky}}, \bibinfo {author} {\bibfnamefont
  {W.}~\bibnamefont {Theobald}}, \bibinfo {author} {\bibfnamefont {W.~L.}\
  \bibnamefont {Kruer}}, \bibinfo {author} {\bibfnamefont {K.}~\bibnamefont
  {Tanaka}}, \bibinfo {author} {\bibfnamefont {R.}~\bibnamefont {Betti}},
  \bibinfo {author} {\bibfnamefont {T.~J.~B.}\ \bibnamefont {Collins}},
  \bibinfo {author} {\bibfnamefont {J.~A.}\ \bibnamefont {Delettrez}}, \bibinfo
  {author} {\bibfnamefont {S.~X.}\ \bibnamefont {Hu}}, \bibinfo {author}
  {\bibfnamefont {J.~A.}\ \bibnamefont {Marozas}}, \bibinfo {author}
  {\bibfnamefont {A.~V.}\ \bibnamefont {Maximov}}, \bibinfo {author}
  {\bibfnamefont {D.~T.}\ \bibnamefont {Michel}}, \bibinfo {author}
  {\bibfnamefont {P.~B.}\ \bibnamefont {Radha}}, \bibinfo {author}
  {\bibfnamefont {S.~P.}\ \bibnamefont {Regan}}, \bibinfo {author}
  {\bibfnamefont {T.~C.}\ \bibnamefont {Sangster}}, \bibinfo {author}
  {\bibfnamefont {W.}~\bibnamefont {Seka}}, \bibinfo {author} {\bibfnamefont
  {A.~A.}\ \bibnamefont {Solodov}}, \bibinfo {author} {\bibfnamefont {J.~M.}\
  \bibnamefont {Soures}}, \bibinfo {author} {\bibfnamefont {C.}~\bibnamefont
  {Stoeckl}},\ and\ \bibinfo {author} {\bibfnamefont {J.~D.}\ \bibnamefont
  {Zuegel}},\ }\bibfield  {title} {\bibinfo {title} {Direct-drive inertial
  confinement fusion: A review},\ }\href {https://doi.org/10.1063/1.4934714}
  {\bibfield  {journal} {\bibinfo  {journal} {Phys. Plasmas}\ }\textbf
  {\bibinfo {volume} {22}},\ \bibinfo {pages} {110501} (\bibinfo {year}
  {2015})},\ \Eprint {https://arxiv.org/abs/https://doi.org/10.1063/1.4934714}
  {https://doi.org/10.1063/1.4934714} \BibitemShut {NoStop}%
\bibitem [{\citenamefont {Gopalaswamy}\ \emph {et~al.}(2024)\citenamefont
  {Gopalaswamy}, \citenamefont {Williams}, \citenamefont {Betti}, \citenamefont
  {Patel}, \citenamefont {Knauer}, \citenamefont {Lees}, \citenamefont {Cao},
  \citenamefont {Campbell}, \citenamefont {Farmakis}, \citenamefont {Ejaz},
  \citenamefont {Anderson}, \citenamefont {Epstein}, \citenamefont
  {Carroll-Nellenbeck}, \citenamefont {Igumenshchev}, \citenamefont {Marozas},
  \citenamefont {Radha}, \citenamefont {Solodov}, \citenamefont {Thomas},
  \citenamefont {Woo}, \citenamefont {Collins}, \citenamefont {Hu},
  \citenamefont {Scullin}, \citenamefont {Turnbull}, \citenamefont {Goncharov},
  \citenamefont {Churnetski}, \citenamefont {Forrest}, \citenamefont {Glebov},
  \citenamefont {Heuer}, \citenamefont {McClow}, \citenamefont {Shah},
  \citenamefont {Stoeckl}, \citenamefont {Theobald}, \citenamefont {Edgell},
  \citenamefont {Ivancic}, \citenamefont {Rosenberg}, \citenamefont {Regan},
  \citenamefont {Bredesen}, \citenamefont {Fella}, \citenamefont {Koch},
  \citenamefont {Janezic}, \citenamefont {Bonino}, \citenamefont {Harding},
  \citenamefont {Bauer}, \citenamefont {Sampat}, \citenamefont {Waxer},
  \citenamefont {Labuzeta}, \citenamefont {Morse}, \citenamefont
  {Gatu-Johnson}, \citenamefont {Petrasso}, \citenamefont {Frenje},
  \citenamefont {Murray}, \citenamefont {Serrato}, \citenamefont {Guzman},
  \citenamefont {Shuldberg}, \citenamefont {Farrell},\ and\ \citenamefont
  {Deeney}}]{Gopalaswamy2024demonstration}%
  \BibitemOpen
  \bibfield  {author} {\bibinfo {author} {\bibfnamefont {V.}~\bibnamefont
  {Gopalaswamy}}, \bibinfo {author} {\bibfnamefont {C.~A.}\ \bibnamefont
  {Williams}}, \bibinfo {author} {\bibfnamefont {R.}~\bibnamefont {Betti}},
  \bibinfo {author} {\bibfnamefont {D.}~\bibnamefont {Patel}}, \bibinfo
  {author} {\bibfnamefont {J.~P.}\ \bibnamefont {Knauer}}, \bibinfo {author}
  {\bibfnamefont {A.}~\bibnamefont {Lees}}, \bibinfo {author} {\bibfnamefont
  {D.}~\bibnamefont {Cao}}, \bibinfo {author} {\bibfnamefont {E.~M.}\
  \bibnamefont {Campbell}}, \bibinfo {author} {\bibfnamefont {P.}~\bibnamefont
  {Farmakis}}, \bibinfo {author} {\bibfnamefont {R.}~\bibnamefont {Ejaz}},
  \bibinfo {author} {\bibfnamefont {K.~S.}\ \bibnamefont {Anderson}}, \bibinfo
  {author} {\bibfnamefont {R.}~\bibnamefont {Epstein}}, \bibinfo {author}
  {\bibfnamefont {J.}~\bibnamefont {Carroll-Nellenbeck}}, \bibinfo {author}
  {\bibfnamefont {I.~V.}\ \bibnamefont {Igumenshchev}}, \bibinfo {author}
  {\bibfnamefont {J.~A.}\ \bibnamefont {Marozas}}, \bibinfo {author}
  {\bibfnamefont {P.~B.}\ \bibnamefont {Radha}}, \bibinfo {author}
  {\bibfnamefont {A.~A.}\ \bibnamefont {Solodov}}, \bibinfo {author}
  {\bibfnamefont {C.~A.}\ \bibnamefont {Thomas}}, \bibinfo {author}
  {\bibfnamefont {K.~M.}\ \bibnamefont {Woo}}, \bibinfo {author} {\bibfnamefont
  {T.~J.~B.}\ \bibnamefont {Collins}}, \bibinfo {author} {\bibfnamefont
  {S.~X.}\ \bibnamefont {Hu}}, \bibinfo {author} {\bibfnamefont
  {W.}~\bibnamefont {Scullin}}, \bibinfo {author} {\bibfnamefont
  {D.}~\bibnamefont {Turnbull}}, \bibinfo {author} {\bibfnamefont {V.~N.}\
  \bibnamefont {Goncharov}}, \bibinfo {author} {\bibfnamefont {K.}~\bibnamefont
  {Churnetski}}, \bibinfo {author} {\bibfnamefont {C.~J.}\ \bibnamefont
  {Forrest}}, \bibinfo {author} {\bibfnamefont {V.~Y.}\ \bibnamefont {Glebov}},
  \bibinfo {author} {\bibfnamefont {P.~V.}\ \bibnamefont {Heuer}}, \bibinfo
  {author} {\bibfnamefont {H.}~\bibnamefont {McClow}}, \bibinfo {author}
  {\bibfnamefont {R.~C.}\ \bibnamefont {Shah}}, \bibinfo {author}
  {\bibfnamefont {C.}~\bibnamefont {Stoeckl}}, \bibinfo {author} {\bibfnamefont
  {W.}~\bibnamefont {Theobald}}, \bibinfo {author} {\bibfnamefont {D.~H.}\
  \bibnamefont {Edgell}}, \bibinfo {author} {\bibfnamefont {S.}~\bibnamefont
  {Ivancic}}, \bibinfo {author} {\bibfnamefont {M.~J.}\ \bibnamefont
  {Rosenberg}}, \bibinfo {author} {\bibfnamefont {S.~P.}\ \bibnamefont
  {Regan}}, \bibinfo {author} {\bibfnamefont {D.}~\bibnamefont {Bredesen}},
  \bibinfo {author} {\bibfnamefont {C.}~\bibnamefont {Fella}}, \bibinfo
  {author} {\bibfnamefont {M.}~\bibnamefont {Koch}}, \bibinfo {author}
  {\bibfnamefont {R.~T.}\ \bibnamefont {Janezic}}, \bibinfo {author}
  {\bibfnamefont {M.~J.}\ \bibnamefont {Bonino}}, \bibinfo {author}
  {\bibfnamefont {D.~R.}\ \bibnamefont {Harding}}, \bibinfo {author}
  {\bibfnamefont {K.~A.}\ \bibnamefont {Bauer}}, \bibinfo {author}
  {\bibfnamefont {S.}~\bibnamefont {Sampat}}, \bibinfo {author} {\bibfnamefont
  {L.~J.}\ \bibnamefont {Waxer}}, \bibinfo {author} {\bibfnamefont
  {M.}~\bibnamefont {Labuzeta}}, \bibinfo {author} {\bibfnamefont {S.~F.~B.}\
  \bibnamefont {Morse}}, \bibinfo {author} {\bibfnamefont {M.}~\bibnamefont
  {Gatu-Johnson}}, \bibinfo {author} {\bibfnamefont {R.~D.}\ \bibnamefont
  {Petrasso}}, \bibinfo {author} {\bibfnamefont {J.~A.}\ \bibnamefont
  {Frenje}}, \bibinfo {author} {\bibfnamefont {J.}~\bibnamefont {Murray}},
  \bibinfo {author} {\bibfnamefont {B.}~\bibnamefont {Serrato}}, \bibinfo
  {author} {\bibfnamefont {D.}~\bibnamefont {Guzman}}, \bibinfo {author}
  {\bibfnamefont {C.}~\bibnamefont {Shuldberg}}, \bibinfo {author}
  {\bibfnamefont {M.}~\bibnamefont {Farrell}},\ and\ \bibinfo {author}
  {\bibfnamefont {C.}~\bibnamefont {Deeney}},\ }\bibfield  {title} {\bibinfo
  {title} {Demonstration of a hydrodynamically equivalent burning plasma in
  direct-drive inertial confinement fusion},\ }\href
  {https://doi.org/10.1038/s41567-023-02361-4} {\bibfield  {journal} {\bibinfo
  {journal} {Nature Physics}\ }\textbf {\bibinfo {volume} {20}},\ \bibinfo
  {pages} {751} (\bibinfo {year} {2024})}\BibitemShut {NoStop}%
\bibitem [{\citenamefont {Williams}\ \emph {et~al.}(2024)\citenamefont
  {Williams}, \citenamefont {Betti}, \citenamefont {Gopalaswamy}, \citenamefont
  {Knauer}, \citenamefont {Forrest}, \citenamefont {Lees}, \citenamefont
  {Ejaz}, \citenamefont {Farmakis}, \citenamefont {Cao}, \citenamefont {Radha},
  \citenamefont {Anderson}, \citenamefont {Regan}, \citenamefont {Glebov},
  \citenamefont {Shah}, \citenamefont {Stoeckl}, \citenamefont {Ivancic},
  \citenamefont {Churnetski}, \citenamefont {Janezic}, \citenamefont {Fella},
  \citenamefont {Rosenberg}, \citenamefont {Bonino}, \citenamefont {Harding},
  \citenamefont {Shmayda}, \citenamefont {Carroll-Nellenback}, \citenamefont
  {Hu}, \citenamefont {Epstein}, \citenamefont {Collins}, \citenamefont
  {Thomas}, \citenamefont {Igumenshchev}, \citenamefont {Goncharov},
  \citenamefont {Theobald}, \citenamefont {Woo}, \citenamefont {Marozas},
  \citenamefont {Bauer}, \citenamefont {Sampat}, \citenamefont {Waxer},
  \citenamefont {Turnbull}, \citenamefont {Heuer}, \citenamefont {McClow},
  \citenamefont {Ceurvorst}, \citenamefont {Scullin}, \citenamefont {Edgell},
  \citenamefont {Koch}, \citenamefont {Bredesen}, \citenamefont {Johnson},
  \citenamefont {Frenje}, \citenamefont {Petrasso}, \citenamefont {Shuldberg},
  \citenamefont {Farrell}, \citenamefont {Murray}, \citenamefont {Guzman},
  \citenamefont {Serrato}, \citenamefont {Morse}, \citenamefont {Labuzeta},
  \citenamefont {Deeney},\ and\ \citenamefont
  {Campbell}}]{Williams2024demonstration}%
  \BibitemOpen
  \bibfield  {author} {\bibinfo {author} {\bibfnamefont {C.~A.}\ \bibnamefont
  {Williams}}, \bibinfo {author} {\bibfnamefont {R.}~\bibnamefont {Betti}},
  \bibinfo {author} {\bibfnamefont {V.}~\bibnamefont {Gopalaswamy}}, \bibinfo
  {author} {\bibfnamefont {J.~P.}\ \bibnamefont {Knauer}}, \bibinfo {author}
  {\bibfnamefont {C.~J.}\ \bibnamefont {Forrest}}, \bibinfo {author}
  {\bibfnamefont {A.}~\bibnamefont {Lees}}, \bibinfo {author} {\bibfnamefont
  {R.}~\bibnamefont {Ejaz}}, \bibinfo {author} {\bibfnamefont {P.~S.}\
  \bibnamefont {Farmakis}}, \bibinfo {author} {\bibfnamefont {D.}~\bibnamefont
  {Cao}}, \bibinfo {author} {\bibfnamefont {P.~B.}\ \bibnamefont {Radha}},
  \bibinfo {author} {\bibfnamefont {K.~S.}\ \bibnamefont {Anderson}}, \bibinfo
  {author} {\bibfnamefont {S.~P.}\ \bibnamefont {Regan}}, \bibinfo {author}
  {\bibfnamefont {V.~Y.}\ \bibnamefont {Glebov}}, \bibinfo {author}
  {\bibfnamefont {R.~C.}\ \bibnamefont {Shah}}, \bibinfo {author}
  {\bibfnamefont {C.}~\bibnamefont {Stoeckl}}, \bibinfo {author} {\bibfnamefont
  {S.}~\bibnamefont {Ivancic}}, \bibinfo {author} {\bibfnamefont
  {K.}~\bibnamefont {Churnetski}}, \bibinfo {author} {\bibfnamefont {R.~T.}\
  \bibnamefont {Janezic}}, \bibinfo {author} {\bibfnamefont {C.}~\bibnamefont
  {Fella}}, \bibinfo {author} {\bibfnamefont {M.~J.}\ \bibnamefont
  {Rosenberg}}, \bibinfo {author} {\bibfnamefont {M.~J.}\ \bibnamefont
  {Bonino}}, \bibinfo {author} {\bibfnamefont {D.~R.}\ \bibnamefont {Harding}},
  \bibinfo {author} {\bibfnamefont {W.~T.}\ \bibnamefont {Shmayda}}, \bibinfo
  {author} {\bibfnamefont {J.}~\bibnamefont {Carroll-Nellenback}}, \bibinfo
  {author} {\bibfnamefont {S.~X.}\ \bibnamefont {Hu}}, \bibinfo {author}
  {\bibfnamefont {R.}~\bibnamefont {Epstein}}, \bibinfo {author} {\bibfnamefont
  {T.~J.~B.}\ \bibnamefont {Collins}}, \bibinfo {author} {\bibfnamefont
  {C.~A.}\ \bibnamefont {Thomas}}, \bibinfo {author} {\bibfnamefont {I.~V.}\
  \bibnamefont {Igumenshchev}}, \bibinfo {author} {\bibfnamefont {V.~N.}\
  \bibnamefont {Goncharov}}, \bibinfo {author} {\bibfnamefont {W.}~\bibnamefont
  {Theobald}}, \bibinfo {author} {\bibfnamefont {K.~M.}\ \bibnamefont {Woo}},
  \bibinfo {author} {\bibfnamefont {J.~A.}\ \bibnamefont {Marozas}}, \bibinfo
  {author} {\bibfnamefont {K.~A.}\ \bibnamefont {Bauer}}, \bibinfo {author}
  {\bibfnamefont {S.}~\bibnamefont {Sampat}}, \bibinfo {author} {\bibfnamefont
  {L.~J.}\ \bibnamefont {Waxer}}, \bibinfo {author} {\bibfnamefont
  {D.}~\bibnamefont {Turnbull}}, \bibinfo {author} {\bibfnamefont {P.~V.}\
  \bibnamefont {Heuer}}, \bibinfo {author} {\bibfnamefont {H.}~\bibnamefont
  {McClow}}, \bibinfo {author} {\bibfnamefont {L.}~\bibnamefont {Ceurvorst}},
  \bibinfo {author} {\bibfnamefont {W.}~\bibnamefont {Scullin}}, \bibinfo
  {author} {\bibfnamefont {D.~H.}\ \bibnamefont {Edgell}}, \bibinfo {author}
  {\bibfnamefont {M.}~\bibnamefont {Koch}}, \bibinfo {author} {\bibfnamefont
  {D.}~\bibnamefont {Bredesen}}, \bibinfo {author} {\bibfnamefont {M.~G.}\
  \bibnamefont {Johnson}}, \bibinfo {author} {\bibfnamefont {J.~A.}\
  \bibnamefont {Frenje}}, \bibinfo {author} {\bibfnamefont {R.~D.}\
  \bibnamefont {Petrasso}}, \bibinfo {author} {\bibfnamefont {C.}~\bibnamefont
  {Shuldberg}}, \bibinfo {author} {\bibfnamefont {M.}~\bibnamefont {Farrell}},
  \bibinfo {author} {\bibfnamefont {J.}~\bibnamefont {Murray}}, \bibinfo
  {author} {\bibfnamefont {D.}~\bibnamefont {Guzman}}, \bibinfo {author}
  {\bibfnamefont {B.}~\bibnamefont {Serrato}}, \bibinfo {author} {\bibfnamefont
  {S.~F.~B.}\ \bibnamefont {Morse}}, \bibinfo {author} {\bibfnamefont
  {M.}~\bibnamefont {Labuzeta}}, \bibinfo {author} {\bibfnamefont
  {C.}~\bibnamefont {Deeney}},\ and\ \bibinfo {author} {\bibfnamefont {E.~M.}\
  \bibnamefont {Campbell}},\ }\bibfield  {title} {\bibinfo {title}
  {Demonstration of hot-spot fuel gain exceeding unity in direct-drive inertial
  confinement fusion implosions},\ }\href
  {https://doi.org/10.1038/s41567-023-02363-2} {\bibfield  {journal} {\bibinfo
  {journal} {Nature Physics}\ }\textbf {\bibinfo {volume} {20}},\ \bibinfo
  {pages} {758} (\bibinfo {year} {2024})}\BibitemShut {NoStop}%
\bibitem [{\citenamefont {Storn}\ and\ \citenamefont
  {Price}(1997)}]{Storn1997differential}%
  \BibitemOpen
  \bibfield  {author} {\bibinfo {author} {\bibfnamefont {R.}~\bibnamefont
  {Storn}}\ and\ \bibinfo {author} {\bibfnamefont {K.}~\bibnamefont {Price}},\
  }\bibfield  {title} {\bibinfo {title} {Differential evolution – a simple
  and efficient heuristic for global optimization over continuous spaces},\
  }\href {https://doi.org/10.1023/a:1008202821328} {\bibfield  {journal}
  {\bibinfo  {journal} {J. Global Optim.}\ }\textbf {\bibinfo {volume} {11}},\
  \bibinfo {pages} {341} (\bibinfo {year} {1997})}\BibitemShut {NoStop}%
\bibitem [{\citenamefont {Virtanen}\ \emph {et~al.}(2020)\citenamefont
  {Virtanen}, \citenamefont {Gommers}, \citenamefont {Oliphant}, \citenamefont
  {Haberland}, \citenamefont {Reddy}, \citenamefont {Cournapeau}, \citenamefont
  {Burovski}, \citenamefont {Peterson}, \citenamefont {Weckesser},
  \citenamefont {Bright}, \citenamefont {{van der Walt}}, \citenamefont
  {Brett}, \citenamefont {Wilson}, \citenamefont {Millman}, \citenamefont
  {Mayorov}, \citenamefont {Nelson}, \citenamefont {Jones}, \citenamefont
  {Kern}, \citenamefont {Larson}, \citenamefont {Carey}, \citenamefont {Polat},
  \citenamefont {Feng}, \citenamefont {Moore}, \citenamefont {{VanderPlas}},
  \citenamefont {Laxalde}, \citenamefont {Perktold}, \citenamefont {Cimrman},
  \citenamefont {Henriksen}, \citenamefont {Quintero}, \citenamefont {Harris},
  \citenamefont {Archibald}, \citenamefont {Ribeiro}, \citenamefont
  {Pedregosa}, \citenamefont {{van Mulbregt}},\ and\ \citenamefont {{SciPy 1.0
  Contributors}}}]{2020SciPy-NMeth}%
  \BibitemOpen
  \bibfield  {author} {\bibinfo {author} {\bibfnamefont {P.}~\bibnamefont
  {Virtanen}}, \bibinfo {author} {\bibfnamefont {R.}~\bibnamefont {Gommers}},
  \bibinfo {author} {\bibfnamefont {T.~E.}\ \bibnamefont {Oliphant}}, \bibinfo
  {author} {\bibfnamefont {M.}~\bibnamefont {Haberland}}, \bibinfo {author}
  {\bibfnamefont {T.}~\bibnamefont {Reddy}}, \bibinfo {author} {\bibfnamefont
  {D.}~\bibnamefont {Cournapeau}}, \bibinfo {author} {\bibfnamefont
  {E.}~\bibnamefont {Burovski}}, \bibinfo {author} {\bibfnamefont
  {P.}~\bibnamefont {Peterson}}, \bibinfo {author} {\bibfnamefont
  {W.}~\bibnamefont {Weckesser}}, \bibinfo {author} {\bibfnamefont
  {J.}~\bibnamefont {Bright}}, \bibinfo {author} {\bibfnamefont {S.~J.}\
  \bibnamefont {{van der Walt}}}, \bibinfo {author} {\bibfnamefont
  {M.}~\bibnamefont {Brett}}, \bibinfo {author} {\bibfnamefont
  {J.}~\bibnamefont {Wilson}}, \bibinfo {author} {\bibfnamefont {K.~J.}\
  \bibnamefont {Millman}}, \bibinfo {author} {\bibfnamefont {N.}~\bibnamefont
  {Mayorov}}, \bibinfo {author} {\bibfnamefont {A.~R.~J.}\ \bibnamefont
  {Nelson}}, \bibinfo {author} {\bibfnamefont {E.}~\bibnamefont {Jones}},
  \bibinfo {author} {\bibfnamefont {R.}~\bibnamefont {Kern}}, \bibinfo {author}
  {\bibfnamefont {E.}~\bibnamefont {Larson}}, \bibinfo {author} {\bibfnamefont
  {C.~J.}\ \bibnamefont {Carey}}, \bibinfo {author} {\bibfnamefont
  {{\.I}.}~\bibnamefont {Polat}}, \bibinfo {author} {\bibfnamefont
  {Y.}~\bibnamefont {Feng}}, \bibinfo {author} {\bibfnamefont {E.~W.}\
  \bibnamefont {Moore}}, \bibinfo {author} {\bibfnamefont {J.}~\bibnamefont
  {{VanderPlas}}}, \bibinfo {author} {\bibfnamefont {D.}~\bibnamefont
  {Laxalde}}, \bibinfo {author} {\bibfnamefont {J.}~\bibnamefont {Perktold}},
  \bibinfo {author} {\bibfnamefont {R.}~\bibnamefont {Cimrman}}, \bibinfo
  {author} {\bibfnamefont {I.}~\bibnamefont {Henriksen}}, \bibinfo {author}
  {\bibfnamefont {E.~A.}\ \bibnamefont {Quintero}}, \bibinfo {author}
  {\bibfnamefont {C.~R.}\ \bibnamefont {Harris}}, \bibinfo {author}
  {\bibfnamefont {A.~M.}\ \bibnamefont {Archibald}}, \bibinfo {author}
  {\bibfnamefont {A.~H.}\ \bibnamefont {Ribeiro}}, \bibinfo {author}
  {\bibfnamefont {F.}~\bibnamefont {Pedregosa}}, \bibinfo {author}
  {\bibfnamefont {P.}~\bibnamefont {{van Mulbregt}}},\ and\ \bibinfo {author}
  {\bibnamefont {{SciPy 1.0 Contributors}}},\ }\bibfield  {title} {\bibinfo
  {title} {{{SciPy} 1.0: Fundamental Algorithms for Scientific Computing in
  Python}},\ }\href {https://doi.org/10.1038/s41592-019-0686-2} {\bibfield
  {journal} {\bibinfo  {journal} {Nature Methods}\ }\textbf {\bibinfo {volume}
  {17}},\ \bibinfo {pages} {261} (\bibinfo {year} {2020})}\BibitemShut
  {NoStop}%
\bibitem [{\citenamefont {{Foreman-Mackey}}\ \emph {et~al.}(2013)\citenamefont
  {{Foreman-Mackey}}, \citenamefont {{Hogg}}, \citenamefont {{Lang}},\ and\
  \citenamefont {{Goodman}}}]{emcee2012}%
  \BibitemOpen
  \bibfield  {author} {\bibinfo {author} {\bibfnamefont {D.}~\bibnamefont
  {{Foreman-Mackey}}}, \bibinfo {author} {\bibfnamefont {D.~W.}\ \bibnamefont
  {{Hogg}}}, \bibinfo {author} {\bibfnamefont {D.}~\bibnamefont {{Lang}}},\
  and\ \bibinfo {author} {\bibfnamefont {J.}~\bibnamefont {{Goodman}}},\
  }\bibfield  {title} {\bibinfo {title} {{emcee: The MCMC Hammer}},\ }\href
  {https://doi.org/10.1086/670067} {\bibfield  {journal} {\bibinfo  {journal}
  {arXiv}\ }\textbf {\bibinfo {volume} {125}},\ \bibinfo {pages} {306}
  (\bibinfo {year} {2013})},\ \Eprint {https://arxiv.org/abs/1202.3665}
  {arXiv:1202.3665 [astro-ph.IM]} \BibitemShut {NoStop}%
\bibitem [{\citenamefont {Rinderknecht}(2015)}]{Rinderknecht_thesis}%
  \BibitemOpen
  \bibfield  {author} {\bibinfo {author} {\bibfnamefont {H.~G.}\ \bibnamefont
  {Rinderknecht}},\ }\emph {\bibinfo {title} {Studies of non-hydrodynamic
  processes in ICF implosionson OMEGA and the National Ignition Facility}},\
  \href
  {https://www.lle.rochester.edu/media/publications/documents/theses/Rinderknecht.pdf}
  {Ph.D. thesis},\ \bibinfo  {school} {Massachusetts Institute of Technology}
  (\bibinfo {year} {2015})\BibitemShut {NoStop}%
\bibitem [{\citenamefont {Gelfgat}\ \emph {et~al.}(1993)\citenamefont
  {Gelfgat}, \citenamefont {Kosarev},\ and\ \citenamefont
  {Podolyak}}]{Gelfgat1993programs}%
  \BibitemOpen
  \bibfield  {author} {\bibinfo {author} {\bibfnamefont {V.}~\bibnamefont
  {Gelfgat}}, \bibinfo {author} {\bibfnamefont {E.}~\bibnamefont {Kosarev}},\
  and\ \bibinfo {author} {\bibfnamefont {E.}~\bibnamefont {Podolyak}},\
  }\bibfield  {title} {\bibinfo {title} {Programs for signal recovery from
  noisy data using the maximum likelihood principle},\ }\href
  {https://doi.org/10.1016/0010-4655(93)90017-7} {\bibfield  {journal}
  {\bibinfo  {journal} {Comput. Phys. Commun.}\ }\textbf {\bibinfo {volume}
  {74}},\ \bibinfo {pages} {335} (\bibinfo {year} {1993})}\BibitemShut
  {NoStop}%
\bibitem [{\citenamefont {{PlasmaPy Community}}\ \emph
  {et~al.}(2024)\citenamefont {{PlasmaPy Community}}, \citenamefont {Murphy},
  \citenamefont {Everson}, \citenamefont {Stańczak-Marikin}, \citenamefont
  {Heuer}, \citenamefont {Kozlowski}, \citenamefont {Johnson}, \citenamefont
  {Malhotra}, \citenamefont {Schaffner}, \citenamefont {Vincena}, \citenamefont
  {Abler}, \citenamefont {Addison}, \citenamefont {Ahamed}, \citenamefont
  {Alarcon}, \citenamefont {Antognetti}, \citenamefont {Arran}, \citenamefont
  {Bagherianlemraski}, \citenamefont {Beckers}, \citenamefont {Bedmutha},
  \citenamefont {Bedoya-Lopez}, \citenamefont {Bergeron}, \citenamefont
  {Bessi}, \citenamefont {Britten}, \citenamefont {Brown}, \citenamefont
  {Bryant}, \citenamefont {Carroll}, \citenamefont {Cartagena-Sanchez},
  \citenamefont {Chambers}, \citenamefont {Chattopadhyay}, \citenamefont
  {Choubey}, \citenamefont {Choudhary}, \citenamefont {Clauss}, \citenamefont
  {Colom}, \citenamefont {Davies}, \citenamefont {Deal}, \citenamefont
  {Decristoforo}, \citenamefont {Diaz~Riega}, \citenamefont {Dover},
  \citenamefont {Drozdov}, \citenamefont {Du}, \citenamefont {Einhorn},
  \citenamefont {Ervin}, \citenamefont {Fan}, \citenamefont {Farid},
  \citenamefont {Fischer}, \citenamefont {Foo}, \citenamefont {Fütterer},
  \citenamefont {Gangadharan}, \citenamefont {Gerow}, \citenamefont {Gonzalez},
  \citenamefont {Goodall}, \citenamefont {Gorelli}, \citenamefont
  {Gordon-McKeon}, \citenamefont {Goudeau}, \citenamefont {Guidoni},
  \citenamefont {Guimiot}, \citenamefont {Haggerty}, \citenamefont {Hansen},
  \citenamefont {Haque}, \citenamefont {Hillairet}, \citenamefont {Hoang},
  \citenamefont {How}, \citenamefont {Huang}, \citenamefont {Humphrey},
  \citenamefont {Isupova}, \citenamefont {Jeandet}, \citenamefont {Jones},
  \citenamefont {Kastek}, \citenamefont {Kent}, \citenamefont {Klima},
  \citenamefont {Köhn-Seemann}, \citenamefont {Kulshrestha}, \citenamefont
  {Kumar}, \citenamefont {Kuszaj}, \citenamefont {Langendorf}, \citenamefont
  {Lanteri}, \citenamefont {Lee}, \citenamefont {Leonard}, \citenamefont
  {Lequette}, \citenamefont {Lim}, \citenamefont {Magarde}, \citenamefont
  {Martinelli}, \citenamefont {Masood}, \citenamefont {McHardy}, \citenamefont
  {Modi}, \citenamefont {Montes}, \citenamefont {Mumford}, \citenamefont
  {Munn}, \citenamefont {Murphy}, \citenamefont {Nie}, \citenamefont {Ortiz},
  \citenamefont {Panda}, \citenamefont {Pannala}, \citenamefont {Parashar},
  \citenamefont {Patel}, \citenamefont {Pavon}, \citenamefont {Pérez},
  \citenamefont {Pitzer}, \citenamefont {Polak}, \citenamefont {Qudsi},
  \citenamefont {Raj}, \citenamefont {Rajashekar}, \citenamefont {Rao},
  \citenamefont {Reep}, \citenamefont {Richardson}, \citenamefont {Roberts},
  \citenamefont {Rodriguez}, \citenamefont {Rojas~Zelaya}, \citenamefont
  {Salcido}, \citenamefont {Savcheva}, \citenamefont {Schneck}, \citenamefont
  {Shen}, \citenamefont {Sheng}, \citenamefont {Sherpa}, \citenamefont
  {Silvestri}, \citenamefont {Simon}, \citenamefont {Singh}, \citenamefont
  {Singh}, \citenamefont {Sipőcz}, \citenamefont {Skinner}, \citenamefont
  {Skrzypczak}, \citenamefont {Smirnov}, \citenamefont {Smith}, \citenamefont
  {Sobeske}, \citenamefont {Spedicato}, \citenamefont {Stansby}, \citenamefont
  {Stinson}, \citenamefont {Sugiharto}, \citenamefont {Švancarová},
  \citenamefont {Tavant}, \citenamefont {Tranquilino}, \citenamefont {Ulrich},
  \citenamefont {Valle}, \citenamefont {Varnish}, \citenamefont {Vo},
  \citenamefont {Wu}, \citenamefont {Xu}, \citenamefont {Yip},\ and\
  \citenamefont {Zhang}}]{PlasmaPy2024_10_0}%
  \BibitemOpen
  \bibfield  {author} {\bibinfo {author} {\bibnamefont {{PlasmaPy Community}}},
  \bibinfo {author} {\bibfnamefont {N.~A.}\ \bibnamefont {Murphy}}, \bibinfo
  {author} {\bibfnamefont {E.~T.}\ \bibnamefont {Everson}}, \bibinfo {author}
  {\bibfnamefont {D.}~\bibnamefont {Stańczak-Marikin}}, \bibinfo {author}
  {\bibfnamefont {P.~V.}\ \bibnamefont {Heuer}}, \bibinfo {author}
  {\bibfnamefont {P.~M.}\ \bibnamefont {Kozlowski}}, \bibinfo {author}
  {\bibfnamefont {E.~T.}\ \bibnamefont {Johnson}}, \bibinfo {author}
  {\bibfnamefont {R.}~\bibnamefont {Malhotra}}, \bibinfo {author}
  {\bibfnamefont {D.~A.}\ \bibnamefont {Schaffner}}, \bibinfo {author}
  {\bibfnamefont {S.~T.}\ \bibnamefont {Vincena}}, \bibinfo {author}
  {\bibfnamefont {M.}~\bibnamefont {Abler}}, \bibinfo {author} {\bibfnamefont
  {J.}~\bibnamefont {Addison}}, \bibinfo {author} {\bibfnamefont {A.~F.}\
  \bibnamefont {Ahamed}}, \bibinfo {author} {\bibfnamefont {P.~V.}\
  \bibnamefont {Alarcon}}, \bibinfo {author} {\bibfnamefont {B.}~\bibnamefont
  {Antognetti}}, \bibinfo {author} {\bibfnamefont {C.}~\bibnamefont {Arran}},
  \bibinfo {author} {\bibfnamefont {H.}~\bibnamefont {Bagherianlemraski}},
  \bibinfo {author} {\bibfnamefont {J.}~\bibnamefont {Beckers}}, \bibinfo
  {author} {\bibfnamefont {M.}~\bibnamefont {Bedmutha}}, \bibinfo {author}
  {\bibfnamefont {C.}~\bibnamefont {Bedoya-Lopez}}, \bibinfo {author}
  {\bibfnamefont {J.}~\bibnamefont {Bergeron}}, \bibinfo {author}
  {\bibfnamefont {L.}~\bibnamefont {Bessi}}, \bibinfo {author} {\bibfnamefont
  {R.}~\bibnamefont {Britten}}, \bibinfo {author} {\bibfnamefont
  {S.}~\bibnamefont {Brown}}, \bibinfo {author} {\bibfnamefont
  {K.}~\bibnamefont {Bryant}}, \bibinfo {author} {\bibfnamefont
  {S.}~\bibnamefont {Carroll}}, \bibinfo {author} {\bibfnamefont
  {C.}~\bibnamefont {Cartagena-Sanchez}}, \bibinfo {author} {\bibfnamefont
  {S.}~\bibnamefont {Chambers}}, \bibinfo {author} {\bibfnamefont
  {A.}~\bibnamefont {Chattopadhyay}}, \bibinfo {author} {\bibfnamefont
  {A.}~\bibnamefont {Choubey}}, \bibinfo {author} {\bibfnamefont
  {S.}~\bibnamefont {Choudhary}}, \bibinfo {author} {\bibfnamefont
  {C.}~\bibnamefont {Clauss}}, \bibinfo {author} {\bibfnamefont
  {S.}~\bibnamefont {Colom}}, \bibinfo {author} {\bibfnamefont
  {C.}~\bibnamefont {Davies}}, \bibinfo {author} {\bibfnamefont
  {J.}~\bibnamefont {Deal}}, \bibinfo {author} {\bibfnamefont {G.}~\bibnamefont
  {Decristoforo}}, \bibinfo {author} {\bibfnamefont {D.~A.}\ \bibnamefont
  {Diaz~Riega}}, \bibinfo {author} {\bibfnamefont {F.~M.}\ \bibnamefont
  {Dover}}, \bibinfo {author} {\bibfnamefont {D.}~\bibnamefont {Drozdov}},
  \bibinfo {author} {\bibfnamefont {T.}~\bibnamefont {Du}}, \bibinfo {author}
  {\bibfnamefont {L.}~\bibnamefont {Einhorn}}, \bibinfo {author} {\bibfnamefont
  {T.}~\bibnamefont {Ervin}}, \bibinfo {author} {\bibfnamefont
  {T.}~\bibnamefont {Fan}}, \bibinfo {author} {\bibfnamefont {S.~I.}\
  \bibnamefont {Farid}}, \bibinfo {author} {\bibfnamefont {M.}~\bibnamefont
  {Fischer}}, \bibinfo {author} {\bibfnamefont {B.}~\bibnamefont {Foo}},
  \bibinfo {author} {\bibfnamefont {H.-A.}\ \bibnamefont {Fütterer}}, \bibinfo
  {author} {\bibfnamefont {R.}~\bibnamefont {Gangadharan}}, \bibinfo {author}
  {\bibfnamefont {S.}~\bibnamefont {Gerow}}, \bibinfo {author} {\bibfnamefont
  {J.}~\bibnamefont {Gonzalez}}, \bibinfo {author} {\bibfnamefont
  {B.}~\bibnamefont {Goodall}}, \bibinfo {author} {\bibfnamefont
  {M.}~\bibnamefont {Gorelli}}, \bibinfo {author} {\bibfnamefont
  {S.}~\bibnamefont {Gordon-McKeon}}, \bibinfo {author} {\bibfnamefont
  {G.}~\bibnamefont {Goudeau}}, \bibinfo {author} {\bibfnamefont
  {S.}~\bibnamefont {Guidoni}}, \bibinfo {author} {\bibfnamefont
  {J.}~\bibnamefont {Guimiot}}, \bibinfo {author} {\bibfnamefont
  {C.}~\bibnamefont {Haggerty}}, \bibinfo {author} {\bibfnamefont {R.~S.}\
  \bibnamefont {Hansen}}, \bibinfo {author} {\bibfnamefont {M.}~\bibnamefont
  {Haque}}, \bibinfo {author} {\bibfnamefont {J.}~\bibnamefont {Hillairet}},
  \bibinfo {author} {\bibfnamefont {C.}~\bibnamefont {Hoang}}, \bibinfo
  {author} {\bibfnamefont {P.~Z.}\ \bibnamefont {How}}, \bibinfo {author}
  {\bibfnamefont {Y.-M.}\ \bibnamefont {Huang}}, \bibinfo {author}
  {\bibfnamefont {N.}~\bibnamefont {Humphrey}}, \bibinfo {author}
  {\bibfnamefont {M.}~\bibnamefont {Isupova}}, \bibinfo {author} {\bibfnamefont
  {A.}~\bibnamefont {Jeandet}}, \bibinfo {author} {\bibfnamefont
  {E.}~\bibnamefont {Jones}}, \bibinfo {author} {\bibfnamefont
  {M.}~\bibnamefont {Kastek}}, \bibinfo {author} {\bibfnamefont
  {J.}~\bibnamefont {Kent}}, \bibinfo {author} {\bibfnamefont {D.}~\bibnamefont
  {Klima}}, \bibinfo {author} {\bibfnamefont {A.}~\bibnamefont
  {Köhn-Seemann}}, \bibinfo {author} {\bibfnamefont {S.}~\bibnamefont
  {Kulshrestha}}, \bibinfo {author} {\bibfnamefont {S.}~\bibnamefont {Kumar}},
  \bibinfo {author} {\bibfnamefont {P.}~\bibnamefont {Kuszaj}}, \bibinfo
  {author} {\bibfnamefont {S.}~\bibnamefont {Langendorf}}, \bibinfo {author}
  {\bibfnamefont {A.}~\bibnamefont {Lanteri}}, \bibinfo {author} {\bibfnamefont
  {T.~T.}\ \bibnamefont {Lee}}, \bibinfo {author} {\bibfnamefont
  {D.}~\bibnamefont {Leonard}}, \bibinfo {author} {\bibfnamefont
  {N.}~\bibnamefont {Lequette}}, \bibinfo {author} {\bibfnamefont {P.~L.}\
  \bibnamefont {Lim}}, \bibinfo {author} {\bibfnamefont {A.}~\bibnamefont
  {Magarde}}, \bibinfo {author} {\bibfnamefont {J.~V.}\ \bibnamefont
  {Martinelli}}, \bibinfo {author} {\bibfnamefont {M.}~\bibnamefont {Masood}},
  \bibinfo {author} {\bibfnamefont {I.}~\bibnamefont {McHardy}}, \bibinfo
  {author} {\bibfnamefont {D.}~\bibnamefont {Modi}}, \bibinfo {author}
  {\bibfnamefont {K.}~\bibnamefont {Montes}}, \bibinfo {author} {\bibfnamefont
  {S.}~\bibnamefont {Mumford}}, \bibinfo {author} {\bibfnamefont
  {J.}~\bibnamefont {Munn}}, \bibinfo {author} {\bibfnamefont {L.}~\bibnamefont
  {Murphy}}, \bibinfo {author} {\bibfnamefont {S.}~\bibnamefont {Nie}},
  \bibinfo {author} {\bibfnamefont {C.}~\bibnamefont {Ortiz}}, \bibinfo
  {author} {\bibfnamefont {S.}~\bibnamefont {Panda}}, \bibinfo {author}
  {\bibfnamefont {M.}~\bibnamefont {Pannala}}, \bibinfo {author} {\bibfnamefont
  {T.}~\bibnamefont {Parashar}}, \bibinfo {author} {\bibfnamefont
  {N.}~\bibnamefont {Patel}}, \bibinfo {author} {\bibfnamefont {F.~S.}\
  \bibnamefont {Pavon}}, \bibinfo {author} {\bibfnamefont {R.~D.}\ \bibnamefont
  {Pérez}}, \bibinfo {author} {\bibfnamefont {P.}~\bibnamefont {Pitzer}},
  \bibinfo {author} {\bibfnamefont {J.}~\bibnamefont {Polak}}, \bibinfo
  {author} {\bibfnamefont {R.}~\bibnamefont {Qudsi}}, \bibinfo {author}
  {\bibfnamefont {R.}~\bibnamefont {Raj}}, \bibinfo {author} {\bibfnamefont
  {V.}~\bibnamefont {Rajashekar}}, \bibinfo {author} {\bibfnamefont
  {A.}~\bibnamefont {Rao}}, \bibinfo {author} {\bibfnamefont {J.}~\bibnamefont
  {Reep}}, \bibinfo {author} {\bibfnamefont {S.}~\bibnamefont {Richardson}},
  \bibinfo {author} {\bibfnamefont {J.}~\bibnamefont {Roberts}}, \bibinfo
  {author} {\bibfnamefont {S.}~\bibnamefont {Rodriguez}}, \bibinfo {author}
  {\bibfnamefont {R.}~\bibnamefont {Rojas~Zelaya}}, \bibinfo {author}
  {\bibfnamefont {A.}~\bibnamefont {Salcido}}, \bibinfo {author} {\bibfnamefont
  {A.}~\bibnamefont {Savcheva}}, \bibinfo {author} {\bibfnamefont
  {C.}~\bibnamefont {Schneck}}, \bibinfo {author} {\bibfnamefont
  {C.}~\bibnamefont {Shen}}, \bibinfo {author} {\bibfnamefont {A.}~\bibnamefont
  {Sheng}}, \bibinfo {author} {\bibfnamefont {D.~N.}\ \bibnamefont {Sherpa}},
  \bibinfo {author} {\bibfnamefont {L.}~\bibnamefont {Silvestri}}, \bibinfo
  {author} {\bibfnamefont {T.}~\bibnamefont {Simon}}, \bibinfo {author}
  {\bibfnamefont {A.}~\bibnamefont {Singh}}, \bibinfo {author} {\bibfnamefont
  {A.}~\bibnamefont {Singh}}, \bibinfo {author} {\bibfnamefont
  {B.}~\bibnamefont {Sipőcz}}, \bibinfo {author} {\bibfnamefont
  {C.}~\bibnamefont {Skinner}}, \bibinfo {author} {\bibfnamefont {T.~A.}\
  \bibnamefont {Skrzypczak}}, \bibinfo {author} {\bibfnamefont
  {N.}~\bibnamefont {Smirnov}}, \bibinfo {author} {\bibfnamefont
  {J.}~\bibnamefont {Smith}}, \bibinfo {author} {\bibfnamefont
  {S.}~\bibnamefont {Sobeske}}, \bibinfo {author} {\bibfnamefont
  {M.}~\bibnamefont {Spedicato}}, \bibinfo {author} {\bibfnamefont
  {D.}~\bibnamefont {Stansby}}, \bibinfo {author} {\bibfnamefont
  {T.}~\bibnamefont {Stinson}}, \bibinfo {author} {\bibfnamefont
  {S.}~\bibnamefont {Sugiharto}}, \bibinfo {author} {\bibfnamefont
  {M.}~\bibnamefont {Švancarová}}, \bibinfo {author} {\bibfnamefont
  {A.}~\bibnamefont {Tavant}}, \bibinfo {author} {\bibfnamefont
  {V.}~\bibnamefont {Tranquilino}}, \bibinfo {author} {\bibfnamefont
  {T.}~\bibnamefont {Ulrich}}, \bibinfo {author} {\bibfnamefont
  {M.}~\bibnamefont {Valle}}, \bibinfo {author} {\bibfnamefont
  {T.}~\bibnamefont {Varnish}}, \bibinfo {author} {\bibfnamefont
  {T.}~\bibnamefont {Vo}}, \bibinfo {author} {\bibfnamefont {T.}~\bibnamefont
  {Wu}}, \bibinfo {author} {\bibfnamefont {S.}~\bibnamefont {Xu}}, \bibinfo
  {author} {\bibfnamefont {C.~H.}\ \bibnamefont {Yip}},\ and\ \bibinfo {author}
  {\bibfnamefont {C.}~\bibnamefont {Zhang}},\ }\href
  {https://doi.org/10.5281/ZENODO.14010450} {\bibinfo {title} {Plasmapy}}
  (\bibinfo {year} {2024}),\ \bibinfo {note} {version 2024.10.0}\BibitemShut
  {NoStop}%
\bibitem [{\citenamefont {Brown}\ \emph {et~al.}(2018)\citenamefont {Brown},
  \citenamefont {Chadwick}, \citenamefont {Capote}, \citenamefont {Kahler},
  \citenamefont {Trkov}, \citenamefont {Herman}, \citenamefont {Sonzogni},
  \citenamefont {Danon}, \citenamefont {Carlson}, \citenamefont {Dunn},
  \citenamefont {Smith}, \citenamefont {Hale}, \citenamefont {Arbanas},
  \citenamefont {Arcilla}, \citenamefont {Bates}, \citenamefont {Beck},
  \citenamefont {Becker}, \citenamefont {Brown}, \citenamefont {Casperson},
  \citenamefont {Conlin}, \citenamefont {Cullen}, \citenamefont {Descalle},
  \citenamefont {Firestone}, \citenamefont {Gaines}, \citenamefont {Guber},
  \citenamefont {Hawari}, \citenamefont {Holmes}, \citenamefont {Johnson},
  \citenamefont {Kawano}, \citenamefont {Kiedrowski}, \citenamefont {Koning},
  \citenamefont {Kopecky}, \citenamefont {Leal}, \citenamefont {Lestone},
  \citenamefont {Lubitz}, \citenamefont {Márquez~Damián}, \citenamefont
  {Mattoon}, \citenamefont {McCutchan}, \citenamefont {Mughabghab},
  \citenamefont {Navratil}, \citenamefont {Neudecker}, \citenamefont {Nobre},
  \citenamefont {Noguere}, \citenamefont {Paris}, \citenamefont {Pigni},
  \citenamefont {Plompen}, \citenamefont {Pritychenko}, \citenamefont
  {Pronyaev}, \citenamefont {Roubtsov}, \citenamefont {Rochman}, \citenamefont
  {Romano}, \citenamefont {Schillebeeckx}, \citenamefont {Simakov},
  \citenamefont {Sin}, \citenamefont {Sirakov}, \citenamefont {Sleaford},
  \citenamefont {Sobes}, \citenamefont {Soukhovitskii}, \citenamefont {Stetcu},
  \citenamefont {Talou}, \citenamefont {Thompson}, \citenamefont {van~der
  Marck}, \citenamefont {Welser-Sherrill}, \citenamefont {Wiarda},
  \citenamefont {White}, \citenamefont {Wormald}, \citenamefont {Wright},
  \citenamefont {Zerkle}, \citenamefont {Žerovnik},\ and\ \citenamefont
  {Zhu}}]{Brown2018ENDF}%
  \BibitemOpen
  \bibfield  {author} {\bibinfo {author} {\bibfnamefont {D.}~\bibnamefont
  {Brown}}, \bibinfo {author} {\bibfnamefont {M.}~\bibnamefont {Chadwick}},
  \bibinfo {author} {\bibfnamefont {R.}~\bibnamefont {Capote}}, \bibinfo
  {author} {\bibfnamefont {A.}~\bibnamefont {Kahler}}, \bibinfo {author}
  {\bibfnamefont {A.}~\bibnamefont {Trkov}}, \bibinfo {author} {\bibfnamefont
  {M.}~\bibnamefont {Herman}}, \bibinfo {author} {\bibfnamefont
  {A.}~\bibnamefont {Sonzogni}}, \bibinfo {author} {\bibfnamefont
  {Y.}~\bibnamefont {Danon}}, \bibinfo {author} {\bibfnamefont
  {A.}~\bibnamefont {Carlson}}, \bibinfo {author} {\bibfnamefont
  {M.}~\bibnamefont {Dunn}}, \bibinfo {author} {\bibfnamefont {D.}~\bibnamefont
  {Smith}}, \bibinfo {author} {\bibfnamefont {G.}~\bibnamefont {Hale}},
  \bibinfo {author} {\bibfnamefont {G.}~\bibnamefont {Arbanas}}, \bibinfo
  {author} {\bibfnamefont {R.}~\bibnamefont {Arcilla}}, \bibinfo {author}
  {\bibfnamefont {C.}~\bibnamefont {Bates}}, \bibinfo {author} {\bibfnamefont
  {B.}~\bibnamefont {Beck}}, \bibinfo {author} {\bibfnamefont {B.}~\bibnamefont
  {Becker}}, \bibinfo {author} {\bibfnamefont {F.}~\bibnamefont {Brown}},
  \bibinfo {author} {\bibfnamefont {R.}~\bibnamefont {Casperson}}, \bibinfo
  {author} {\bibfnamefont {J.}~\bibnamefont {Conlin}}, \bibinfo {author}
  {\bibfnamefont {D.}~\bibnamefont {Cullen}}, \bibinfo {author} {\bibfnamefont
  {M.-A.}\ \bibnamefont {Descalle}}, \bibinfo {author} {\bibfnamefont
  {R.}~\bibnamefont {Firestone}}, \bibinfo {author} {\bibfnamefont
  {T.}~\bibnamefont {Gaines}}, \bibinfo {author} {\bibfnamefont
  {K.}~\bibnamefont {Guber}}, \bibinfo {author} {\bibfnamefont
  {A.}~\bibnamefont {Hawari}}, \bibinfo {author} {\bibfnamefont
  {J.}~\bibnamefont {Holmes}}, \bibinfo {author} {\bibfnamefont
  {T.}~\bibnamefont {Johnson}}, \bibinfo {author} {\bibfnamefont
  {T.}~\bibnamefont {Kawano}}, \bibinfo {author} {\bibfnamefont
  {B.}~\bibnamefont {Kiedrowski}}, \bibinfo {author} {\bibfnamefont
  {A.}~\bibnamefont {Koning}}, \bibinfo {author} {\bibfnamefont
  {S.}~\bibnamefont {Kopecky}}, \bibinfo {author} {\bibfnamefont
  {L.}~\bibnamefont {Leal}}, \bibinfo {author} {\bibfnamefont {J.}~\bibnamefont
  {Lestone}}, \bibinfo {author} {\bibfnamefont {C.}~\bibnamefont {Lubitz}},
  \bibinfo {author} {\bibfnamefont {J.}~\bibnamefont {Márquez~Damián}},
  \bibinfo {author} {\bibfnamefont {C.}~\bibnamefont {Mattoon}}, \bibinfo
  {author} {\bibfnamefont {E.}~\bibnamefont {McCutchan}}, \bibinfo {author}
  {\bibfnamefont {S.}~\bibnamefont {Mughabghab}}, \bibinfo {author}
  {\bibfnamefont {P.}~\bibnamefont {Navratil}}, \bibinfo {author}
  {\bibfnamefont {D.}~\bibnamefont {Neudecker}}, \bibinfo {author}
  {\bibfnamefont {G.}~\bibnamefont {Nobre}}, \bibinfo {author} {\bibfnamefont
  {G.}~\bibnamefont {Noguere}}, \bibinfo {author} {\bibfnamefont
  {M.}~\bibnamefont {Paris}}, \bibinfo {author} {\bibfnamefont
  {M.}~\bibnamefont {Pigni}}, \bibinfo {author} {\bibfnamefont
  {A.}~\bibnamefont {Plompen}}, \bibinfo {author} {\bibfnamefont
  {B.}~\bibnamefont {Pritychenko}}, \bibinfo {author} {\bibfnamefont
  {V.}~\bibnamefont {Pronyaev}}, \bibinfo {author} {\bibfnamefont
  {D.}~\bibnamefont {Roubtsov}}, \bibinfo {author} {\bibfnamefont
  {D.}~\bibnamefont {Rochman}}, \bibinfo {author} {\bibfnamefont
  {P.}~\bibnamefont {Romano}}, \bibinfo {author} {\bibfnamefont
  {P.}~\bibnamefont {Schillebeeckx}}, \bibinfo {author} {\bibfnamefont
  {S.}~\bibnamefont {Simakov}}, \bibinfo {author} {\bibfnamefont
  {M.}~\bibnamefont {Sin}}, \bibinfo {author} {\bibfnamefont {I.}~\bibnamefont
  {Sirakov}}, \bibinfo {author} {\bibfnamefont {B.}~\bibnamefont {Sleaford}},
  \bibinfo {author} {\bibfnamefont {V.}~\bibnamefont {Sobes}}, \bibinfo
  {author} {\bibfnamefont {E.}~\bibnamefont {Soukhovitskii}}, \bibinfo {author}
  {\bibfnamefont {I.}~\bibnamefont {Stetcu}}, \bibinfo {author} {\bibfnamefont
  {P.}~\bibnamefont {Talou}}, \bibinfo {author} {\bibfnamefont
  {I.}~\bibnamefont {Thompson}}, \bibinfo {author} {\bibfnamefont
  {S.}~\bibnamefont {van~der Marck}}, \bibinfo {author} {\bibfnamefont
  {L.}~\bibnamefont {Welser-Sherrill}}, \bibinfo {author} {\bibfnamefont
  {D.}~\bibnamefont {Wiarda}}, \bibinfo {author} {\bibfnamefont
  {M.}~\bibnamefont {White}}, \bibinfo {author} {\bibfnamefont
  {J.}~\bibnamefont {Wormald}}, \bibinfo {author} {\bibfnamefont
  {R.}~\bibnamefont {Wright}}, \bibinfo {author} {\bibfnamefont
  {M.}~\bibnamefont {Zerkle}}, \bibinfo {author} {\bibfnamefont
  {G.}~\bibnamefont {Žerovnik}},\ and\ \bibinfo {author} {\bibfnamefont
  {Y.}~\bibnamefont {Zhu}},\ }\bibfield  {title} {\bibinfo {title}
  {Endf/b-viii.0: The 8 th major release of the nuclear reaction data library
  with cielo-project cross sections, new standards and thermal scattering
  data},\ }\href {https://doi.org/10.1016/j.nds.2018.02.001} {\bibfield
  {journal} {\bibinfo  {journal} {Nuclear Data Sheets}\ }\textbf {\bibinfo
  {volume} {148}},\ \bibinfo {pages} {1} (\bibinfo {year} {2018})}\BibitemShut
  {NoStop}%
\bibitem [{\citenamefont {Ziegler}(2013)}]{SRIM}%
  \BibitemOpen
  \bibfield  {author} {\bibinfo {author} {\bibfnamefont {J.~F.}\ \bibnamefont
  {Ziegler}},\ }\href {http://www.srim.org/} {\bibinfo {title} {{SRIM} - {T}he
  {S}topping and {R}anging of {I}ons in {M}atter}} (\bibinfo {year} {2013}),\
  \bibinfo {note} {http://www.srim.org/}\BibitemShut {NoStop}%
\bibitem [{\citenamefont {Rygg}\ \emph {et~al.}(2008)\citenamefont {Rygg},
  \citenamefont {Seguin}, \citenamefont {Li}, \citenamefont {Frenje},
  \citenamefont {Manuel}, \citenamefont {Petrasso}, \citenamefont {Betti},
  \citenamefont {Delettrez}, \citenamefont {Gotchev}, \citenamefont {Knauer},
  \citenamefont {Meyerhofer}, \citenamefont {Marshall}, \citenamefont
  {Stoeckl},\ and\ \citenamefont {Theobald}}]{Rygg2008proton}%
  \BibitemOpen
  \bibfield  {author} {\bibinfo {author} {\bibfnamefont {J.~R.}\ \bibnamefont
  {Rygg}}, \bibinfo {author} {\bibfnamefont {F.~H.}\ \bibnamefont {Seguin}},
  \bibinfo {author} {\bibfnamefont {C.~K.}\ \bibnamefont {Li}}, \bibinfo
  {author} {\bibfnamefont {J.~A.}\ \bibnamefont {Frenje}}, \bibinfo {author}
  {\bibfnamefont {M.~J.-E.}\ \bibnamefont {Manuel}}, \bibinfo {author}
  {\bibfnamefont {R.~D.}\ \bibnamefont {Petrasso}}, \bibinfo {author}
  {\bibfnamefont {R.}~\bibnamefont {Betti}}, \bibinfo {author} {\bibfnamefont
  {J.~A.}\ \bibnamefont {Delettrez}}, \bibinfo {author} {\bibfnamefont {O.~V.}\
  \bibnamefont {Gotchev}}, \bibinfo {author} {\bibfnamefont {J.~P.}\
  \bibnamefont {Knauer}}, \bibinfo {author} {\bibfnamefont {D.~D.}\
  \bibnamefont {Meyerhofer}}, \bibinfo {author} {\bibfnamefont {F.~J.}\
  \bibnamefont {Marshall}}, \bibinfo {author} {\bibfnamefont {C.}~\bibnamefont
  {Stoeckl}},\ and\ \bibinfo {author} {\bibfnamefont {W.}~\bibnamefont
  {Theobald}},\ }\bibfield  {title} {\bibinfo {title} {Proton radiography of
  inertial fusion implosions},\ }\href
  {https://doi.org/10.1126/science.1152640} {\bibfield  {journal} {\bibinfo
  {journal} {Science}\ }\textbf {\bibinfo {volume} {319}},\ \bibinfo {pages}
  {1223} (\bibinfo {year} {2008})}\BibitemShut {NoStop}%
\bibitem [{\citenamefont {S{\'{e}}guin}\ \emph {et~al.}(2012)\citenamefont
  {S{\'{e}}guin}, \citenamefont {Li}, \citenamefont {Manuel}, \citenamefont
  {Rinderknecht}, \citenamefont {Sinenian}, \citenamefont {Frenje},
  \citenamefont {Rygg}, \citenamefont {Hicks}, \citenamefont {Petrasso},
  \citenamefont {Delettrez}, \citenamefont {Betti}, \citenamefont {Marshall},\
  and\ \citenamefont {Smalyuk}}]{Seguin2012time}%
  \BibitemOpen
  \bibfield  {author} {\bibinfo {author} {\bibfnamefont {F.~H.}\ \bibnamefont
  {S{\'{e}}guin}}, \bibinfo {author} {\bibfnamefont {C.~K.}\ \bibnamefont
  {Li}}, \bibinfo {author} {\bibfnamefont {M.~J.-E.}\ \bibnamefont {Manuel}},
  \bibinfo {author} {\bibfnamefont {H.~G.}\ \bibnamefont {Rinderknecht}},
  \bibinfo {author} {\bibfnamefont {N.}~\bibnamefont {Sinenian}}, \bibinfo
  {author} {\bibfnamefont {J.~A.}\ \bibnamefont {Frenje}}, \bibinfo {author}
  {\bibfnamefont {J.~R.}\ \bibnamefont {Rygg}}, \bibinfo {author}
  {\bibfnamefont {D.~G.}\ \bibnamefont {Hicks}}, \bibinfo {author}
  {\bibfnamefont {R.~D.}\ \bibnamefont {Petrasso}}, \bibinfo {author}
  {\bibfnamefont {J.}~\bibnamefont {Delettrez}}, \bibinfo {author}
  {\bibfnamefont {R.}~\bibnamefont {Betti}}, \bibinfo {author} {\bibfnamefont
  {F.~J.}\ \bibnamefont {Marshall}},\ and\ \bibinfo {author} {\bibfnamefont
  {V.~A.}\ \bibnamefont {Smalyuk}},\ }\bibfield  {title} {\bibinfo {title}
  {Time evolution of filamentation and self-generated fields in the coronae of
  directly driven inertial-confinement fusion capsules},\ }\href
  {https://doi.org/10.1063/1.3671908} {\bibfield  {journal} {\bibinfo
  {journal} {Phys. Plasmas}\ }\textbf {\bibinfo {volume} {19}},\ \bibinfo
  {pages} {012701} (\bibinfo {year} {2012})}\BibitemShut {NoStop}%
\bibitem [{\citenamefont {Zylstra}\ \emph {et~al.}(2016)\citenamefont
  {Zylstra}, \citenamefont {Frenje}, \citenamefont {Grabowski}, \citenamefont
  {Li}, \citenamefont {Collins}, \citenamefont {Fitzsimmons}, \citenamefont
  {Glenzer}, \citenamefont {Graziani}, \citenamefont {Hansen}, \citenamefont
  {Hu}, \citenamefont {Johnson}, \citenamefont {Keiter}, \citenamefont
  {Reynolds}, \citenamefont {Rygg}, \citenamefont {Séguin},\ and\
  \citenamefont {Petrasso}}]{Zylstra2016development}%
  \BibitemOpen
  \bibfield  {author} {\bibinfo {author} {\bibfnamefont {A.~B.}\ \bibnamefont
  {Zylstra}}, \bibinfo {author} {\bibfnamefont {J.~A.}\ \bibnamefont {Frenje}},
  \bibinfo {author} {\bibfnamefont {P.~E.}\ \bibnamefont {Grabowski}}, \bibinfo
  {author} {\bibfnamefont {C.~K.}\ \bibnamefont {Li}}, \bibinfo {author}
  {\bibfnamefont {G.~W.}\ \bibnamefont {Collins}}, \bibinfo {author}
  {\bibfnamefont {P.}~\bibnamefont {Fitzsimmons}}, \bibinfo {author}
  {\bibfnamefont {S.}~\bibnamefont {Glenzer}}, \bibinfo {author} {\bibfnamefont
  {F.}~\bibnamefont {Graziani}}, \bibinfo {author} {\bibfnamefont {S.~B.}\
  \bibnamefont {Hansen}}, \bibinfo {author} {\bibfnamefont {S.~X.}\
  \bibnamefont {Hu}}, \bibinfo {author} {\bibfnamefont {M.~G.}\ \bibnamefont
  {Johnson}}, \bibinfo {author} {\bibfnamefont {P.}~\bibnamefont {Keiter}},
  \bibinfo {author} {\bibfnamefont {H.}~\bibnamefont {Reynolds}}, \bibinfo
  {author} {\bibfnamefont {J.~R.}\ \bibnamefont {Rygg}}, \bibinfo {author}
  {\bibfnamefont {F.~H.}\ \bibnamefont {Séguin}},\ and\ \bibinfo {author}
  {\bibfnamefont {R.~D.}\ \bibnamefont {Petrasso}},\ }\bibfield  {title}
  {\bibinfo {title} {Development of a wdm platform for charged-particle
  stopping experiments},\ }\href
  {https://doi.org/10.1088/1742-6596/717/1/012118} {\bibfield  {journal}
  {\bibinfo  {journal} {Journal of Physics: Conference Series}\ }\textbf
  {\bibinfo {volume} {717}},\ \bibinfo {pages} {012118} (\bibinfo {year}
  {2016})}\BibitemShut {NoStop}%
\bibitem [{\citenamefont {Heuer}\ \emph {et~al.}(2022)\citenamefont {Heuer},
  \citenamefont {Leal}, \citenamefont {Davies}, \citenamefont {Hansen},
  \citenamefont {Barnak}, \citenamefont {Peebles}, \citenamefont
  {Garc{\'{\i}}a-Rubio}, \citenamefont {Pollock}, \citenamefont {Moody},
  \citenamefont {Birkel},\ and\ \citenamefont {Seguin}}]{Heuer2022diagnosing}%
  \BibitemOpen
  \bibfield  {author} {\bibinfo {author} {\bibfnamefont {P.~V.}\ \bibnamefont
  {Heuer}}, \bibinfo {author} {\bibfnamefont {L.~S.}\ \bibnamefont {Leal}},
  \bibinfo {author} {\bibfnamefont {J.~R.}\ \bibnamefont {Davies}}, \bibinfo
  {author} {\bibfnamefont {E.~C.}\ \bibnamefont {Hansen}}, \bibinfo {author}
  {\bibfnamefont {D.~H.}\ \bibnamefont {Barnak}}, \bibinfo {author}
  {\bibfnamefont {J.~L.}\ \bibnamefont {Peebles}}, \bibinfo {author}
  {\bibfnamefont {F.}~\bibnamefont {Garc{\'{\i}}a-Rubio}}, \bibinfo {author}
  {\bibfnamefont {B.}~\bibnamefont {Pollock}}, \bibinfo {author} {\bibfnamefont
  {J.}~\bibnamefont {Moody}}, \bibinfo {author} {\bibfnamefont
  {A.}~\bibnamefont {Birkel}},\ and\ \bibinfo {author} {\bibfnamefont {F.~H.}\
  \bibnamefont {Seguin}},\ }\bibfield  {title} {\bibinfo {title} {Diagnosing
  magnetic fields in cylindrical implosions with oblique proton radiography},\
  }\href {https://doi.org/10.1063/5.0092652} {\bibfield  {journal} {\bibinfo
  {journal} {Phys. Plasmas}\ }\textbf {\bibinfo {volume} {29}},\ \bibinfo
  {pages} {072708} (\bibinfo {year} {2022})}\BibitemShut {NoStop}%
\bibitem [{\citenamefont {Zylstra}\ \emph {et~al.}(2012)\citenamefont
  {Zylstra}, \citenamefont {Li}, \citenamefont {Rinderknecht}, \citenamefont
  {S{\'{e}}guin}, \citenamefont {Petrasso}, \citenamefont {Stoeckl},
  \citenamefont {Meyerhofer}, \citenamefont {Nilson}, \citenamefont {Sangster},
  \citenamefont {Pape}, \citenamefont {Mackinnon},\ and\ \citenamefont
  {Patel}}]{Zylstra2012using}%
  \BibitemOpen
  \bibfield  {author} {\bibinfo {author} {\bibfnamefont {A.~B.}\ \bibnamefont
  {Zylstra}}, \bibinfo {author} {\bibfnamefont {C.~K.}\ \bibnamefont {Li}},
  \bibinfo {author} {\bibfnamefont {H.~G.}\ \bibnamefont {Rinderknecht}},
  \bibinfo {author} {\bibfnamefont {F.~H.}\ \bibnamefont {S{\'{e}}guin}},
  \bibinfo {author} {\bibfnamefont {R.~D.}\ \bibnamefont {Petrasso}}, \bibinfo
  {author} {\bibfnamefont {C.}~\bibnamefont {Stoeckl}}, \bibinfo {author}
  {\bibfnamefont {D.~D.}\ \bibnamefont {Meyerhofer}}, \bibinfo {author}
  {\bibfnamefont {P.}~\bibnamefont {Nilson}}, \bibinfo {author} {\bibfnamefont
  {T.~C.}\ \bibnamefont {Sangster}}, \bibinfo {author} {\bibfnamefont {S.~L.}\
  \bibnamefont {Pape}}, \bibinfo {author} {\bibfnamefont {A.}~\bibnamefont
  {Mackinnon}},\ and\ \bibinfo {author} {\bibfnamefont {P.}~\bibnamefont
  {Patel}},\ }\bibfield  {title} {\bibinfo {title} {Using high-intensity
  laser-generated energetic protons to radiograph directly driven implosions},\
  }\href {https://doi.org/10.1063/1.3680110} {\bibfield  {journal} {\bibinfo
  {journal} {Rev. Sci. Instrum.}\ }\textbf {\bibinfo {volume} {83}},\ \bibinfo
  {pages} {013511} (\bibinfo {year} {2012})}\BibitemShut {NoStop}%
\bibitem [{\citenamefont {Sutcliffe}\ \emph {et~al.}(2022)\citenamefont
  {Sutcliffe}, \citenamefont {Adrian}, \citenamefont {Pearcy}, \citenamefont
  {Johnson}, \citenamefont {Kunimune}, \citenamefont {Pollock}, \citenamefont
  {Moody}, \citenamefont {Loureiro},\ and\ \citenamefont
  {Li}}]{Sutcliffe2022observation}%
  \BibitemOpen
  \bibfield  {author} {\bibinfo {author} {\bibfnamefont {G.~D.}\ \bibnamefont
  {Sutcliffe}}, \bibinfo {author} {\bibfnamefont {P.~J.}\ \bibnamefont
  {Adrian}}, \bibinfo {author} {\bibfnamefont {J.~A.}\ \bibnamefont {Pearcy}},
  \bibinfo {author} {\bibfnamefont {T.~M.}\ \bibnamefont {Johnson}}, \bibinfo
  {author} {\bibfnamefont {J.}~\bibnamefont {Kunimune}}, \bibinfo {author}
  {\bibfnamefont {B.}~\bibnamefont {Pollock}}, \bibinfo {author} {\bibfnamefont
  {J.~D.}\ \bibnamefont {Moody}}, \bibinfo {author} {\bibfnamefont {N.~F.}\
  \bibnamefont {Loureiro}},\ and\ \bibinfo {author} {\bibfnamefont {C.~K.}\
  \bibnamefont {Li}},\ }\bibfield  {title} {\bibinfo {title} {Observation of
  electromagnetic filamentary structures produced by the weibel instability in
  laser-driven plasmas},\ }\href@noop {} {\bibfield  {journal} {\bibinfo
  {journal} {arXiv}\ } (\bibinfo {year} {2022})},\ \Eprint
  {https://arxiv.org/abs/2209.02565} {arXiv:2209.02565 [physics.plasm-ph]}
  \BibitemShut {NoStop}%
\bibitem [{\citenamefont {Birdsall}\ and\ \citenamefont
  {Langdon}(2004)}]{Birdsall2004plasma}%
  \BibitemOpen
  \bibfield  {author} {\bibinfo {author} {\bibfnamefont {C.}~\bibnamefont
  {Birdsall}}\ and\ \bibinfo {author} {\bibfnamefont {A.~B.}\ \bibnamefont
  {Langdon}},\ }\href@noop {} {\emph {\bibinfo {title} {Plasma Physics via
  Computer Simulation}}},\ \bibinfo {edition} {1st}\ ed.,\ Plasma Physics
  Series\ (\bibinfo  {publisher} {CRC Press},\ \bibinfo {year}
  {2004})\BibitemShut {NoStop}%
\bibitem [{\citenamefont {Lahmann}\ \emph {et~al.}(2023)\citenamefont
  {Lahmann}, \citenamefont {Saunders}, \citenamefont {Döppner}, \citenamefont
  {Frenje}, \citenamefont {Glenzer}, \citenamefont {Gatu-Johnson},
  \citenamefont {Sutcliffe}, \citenamefont {Zylstra},\ and\ \citenamefont
  {Petrasso}}]{Lahmann2023measuring}%
  \BibitemOpen
  \bibfield  {author} {\bibinfo {author} {\bibfnamefont {B.}~\bibnamefont
  {Lahmann}}, \bibinfo {author} {\bibfnamefont {A.~M.}\ \bibnamefont
  {Saunders}}, \bibinfo {author} {\bibfnamefont {T.}~\bibnamefont {Döppner}},
  \bibinfo {author} {\bibfnamefont {J.~A.}\ \bibnamefont {Frenje}}, \bibinfo
  {author} {\bibfnamefont {S.~H.}\ \bibnamefont {Glenzer}}, \bibinfo {author}
  {\bibfnamefont {M.}~\bibnamefont {Gatu-Johnson}}, \bibinfo {author}
  {\bibfnamefont {G.}~\bibnamefont {Sutcliffe}}, \bibinfo {author}
  {\bibfnamefont {A.~B.}\ \bibnamefont {Zylstra}},\ and\ \bibinfo {author}
  {\bibfnamefont {R.~D.}\ \bibnamefont {Petrasso}},\ }\bibfield  {title}
  {\bibinfo {title} {Measuring stopping power in warm dense matter plasmas at
  omega},\ }\href {https://doi.org/10.1088/1361-6587/ace4f2} {\bibfield
  {journal} {\bibinfo  {journal} {Plasma Physics and Controlled Fusion}\
  }\textbf {\bibinfo {volume} {65}},\ \bibinfo {pages} {095002} (\bibinfo
  {year} {2023})}\BibitemShut {NoStop}%
\end{thebibliography}

%

\end{document}